\documentclass[a4paper,11pt]{report}
\usepackage[english]{babel}
\usepackage[latin1]{inputenc}
\usepackage[T1]{fontenc}
\usepackage{a4wide}
\usepackage{times}
\usepackage{latexsym,verbatim}
\usepackage{amsmath,amssymb,amsfonts}
\usepackage[all,arc,curve,color,frame,knot]{xy}
\newxyColor{grey}{0.7,0.7,0.7}{rgb}{}
\usepackage{color}
\usepackage[table]{xcolor}
\definecolor{bl}{cmyk}{1,1,0.3,0}
\definecolor{gr}{rgb}{0,0.35,0.1}
\definecolor{wh}{rgb}{1,1,1}
\definecolor{orange}{rgb}{1,0.4,0.2}
\definecolor{new}{rgb}{0.21,0.43,0.5}
\definecolor{ulb}{rgb}{0,0.27,0.55}

\newcommand{\red}[1]{{\color{red}{#1}}}
\newcommand{\white}[1]{{\color{wh}{#1}}}
\usepackage[pdfauthor={Michael V. Bebronne},pdftitle={Theoretical and phenomenological aspects of theories with massive gravitons},pdftex]{hyperref}
\hypersetup{colorlinks=true,urlcolor=new,linkcolor=new,citecolor=new}
\usepackage{graphpap,graphicx}
\usepackage{chngpage}
\usepackage{rotating}
\newcommand{\dif}{\textrm{d}}
\newcommand{\tx}[1]{\textrm{#1}}
\newcommand{\tr}{\textrm{Tr}\,}
\newcommand{\Mpl}{\textrm{M}_{\textrm{pl}}^2}

\newcommand{\act}{\mathcal{S}}
\newcommand{\lag}{\mathcal{L}}
\newcommand{\func}{\mathcal{F}}
\newcommand{\ricci}{\mathcal{R}}
\newcommand{\ex}[1]{\textrm{e}^{#1}}

\newcommand{\refp}[1]{(\ref{#1})}
\newcommand{\sm}[1]{{\scriptscriptstyle{#1}}}
\newcommand{\C}[3]{\Gamma_{#1#2}^{\phantom{#1#2}#3}}

\newcommand{\cod}[3]{{#1}_{#2}^{\phantom{#2}#3}}
\title{Theoretical and phenomenological aspects of theories with massive gravitons}
\author{Michael V. Bebronne}
\date{October 2009}

\makeatletter
\def\@makechapterhead#1{%
  \vspace*{-30\p@}%
  {\parindent \z@ \raggedleft \normalfont
    \interlinepenalty\@M
    \ifnum \c@secnumdepth >\m@ne
 \bfseries {\fontsize{76}{80}\usefont{OT1}{pzc}{m}{n}\selectfont \thechapter}\quad
    \fi \\
    \Huge \bfseries {\usefont{OT1}{pzc}{m}{n}\selectfont #1}\par\nobreak
    \vskip 50\p@
\medskip{\color{gray}{\hrule}}
    \vskip 30\p@
  }}

\def\@makeschapterhead#1{%
  \vspace*{50\p@}%
  {\parindent \z@ \raggedright
    \normalfont
    \interlinepenalty\@M
    \Huge \bfseries  #1\par\nobreak
    \vskip 40\p@
  }}
\makeatother

\begin{document}


\pagenumbering{alph}

\changepage{2cm}{}{}{}{}{-2cm}{}{}{}

\thispagestyle{empty}

\begin{adjustwidth}[]{-0.7cm}{-0.7cm}

\begin{minipage}[]{2cm}
\includegraphics[angle=0,width=2cm]{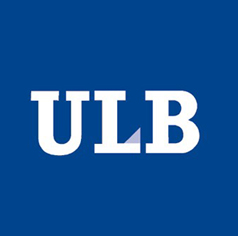}
\end{minipage}
\hspace{1cm}
\begin{minipage}[]{13cm}
{\color{ulb}{\Large \textbf{\textsc{U n i v e r s i t \'e \,\,\, L i b r e \,\,\, d e \,\,\, B r u x e l l e s}}}}
\vspace{0.4cm}\\
{ \color{ulb}{\textsc{S e r v i c e \,\,\, d e \,\,\, P h y s i q u e \,\,\, T h \'e o r i q u e}}}
\end{minipage}

\begin{minipage}{8cm}
\vspace{8cm}
\includegraphics[angle=0,width=8cm]{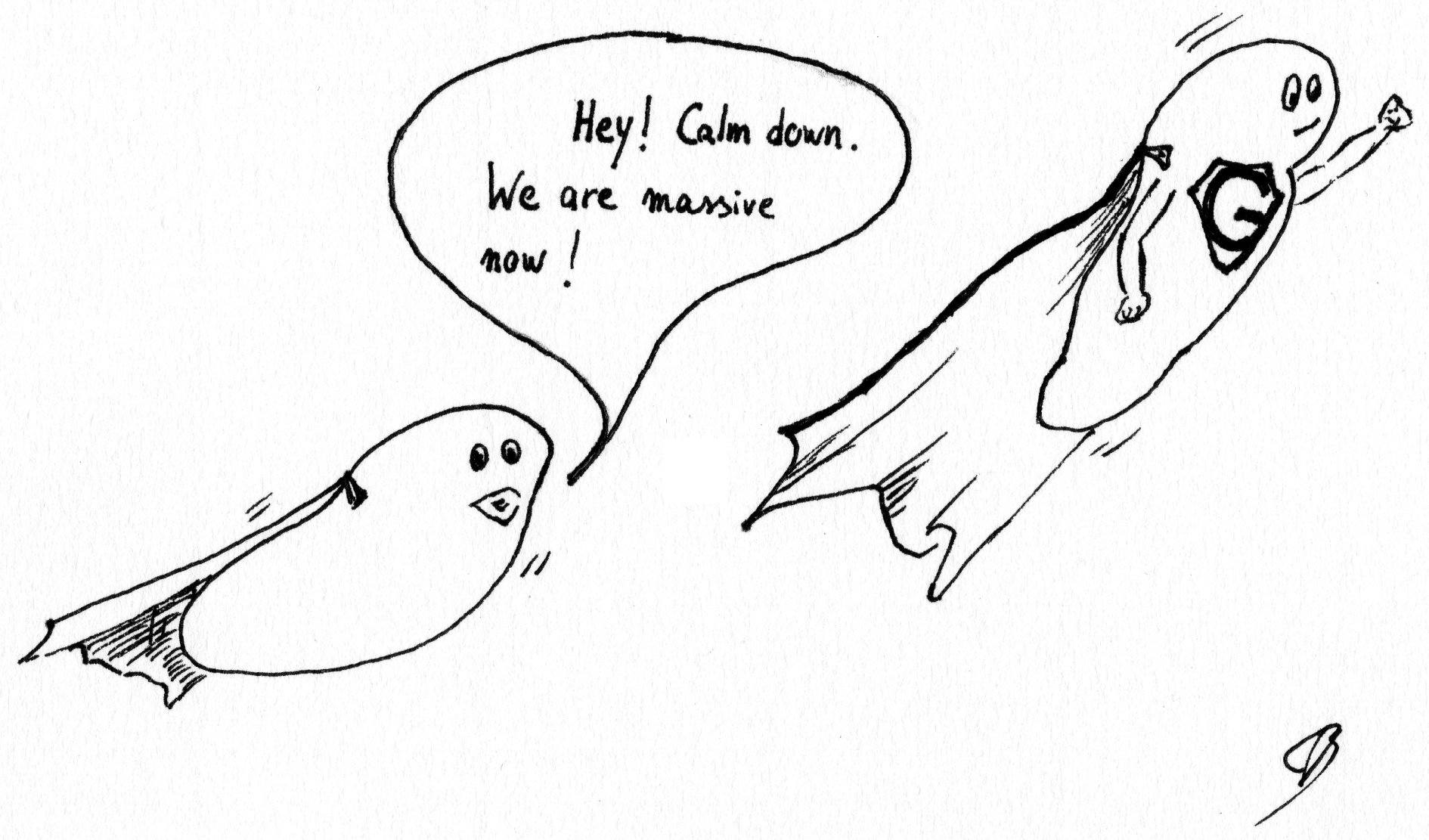}
\vspace{2cm}\\
{\Large\usefont{OT1}{pzc}{m}{n}\selectfont Michael V{\small .} Bebronne}
\end{minipage}
\hspace{3cm}
\begin{minipage}{5cm}
\vspace{1cm}
\begin{tabular}{cccccc}
& \rotatebox{90}{{\Huge \textsc{t h e o r e t i c a l \,\,\, a n d}}} &
&\rotatebox{90}{{\Huge \textsc{p h e n o m e n o l o g i c a l \,\,\, a s p e c t s \,\,\, o f}}} &
& \rotatebox{90}{{\Huge \textsc{t h e o r i e s \,\,\, w i t h \,\,\, m a s s i v e \,\,\, g r a v i t o n s}}}
\end{tabular}

\end{minipage}

\end{adjustwidth}

\newpage

\changepage{-2cm}{}{}{}{}{2cm}{}{}{}
\setcounter{page}{1}
\renewcommand{\thepage}{\roman{page}}
\begin{center}
\vspace*{4cm}
\textsc{{\Large Theoretical and phenomenological aspects of \\ theories with massive gravitons}}
\vspace{1cm}\\ Michael V. Bebronne \vspace{1cm}\\
{\large PhD made under the supervision of Professor Peter Tinyakov}
\end{center}
\vfill
\begin{minipage}{8cm}
\begin{flushleft}
Deposit: 23 September 2009\\
Public defense: 15 October 2009 \\
Last modifications: 16 October 2009
\vspace{0.5cm}\\
This work has been supported by the Belgian\\ \emph{Fond pour la Formation \`a la
Recherche dans \\ l'Industrie et dans l'Agriculture (FRIA)} \\

[F 3/5/5 - MCF/ROI/FC - 20990 ($1^{\tx{\`ere}}$ bourse)]

[F 3/5/5 - MCF/ROI/BC - 22945 ($2^{\tx{\`eme}}$ bourse)].
\end{flushleft}
\end{minipage}

\newpage
\emph{}
\newpage

\vspace*{3cm}
\begin{flushright}
\begin{minipage}{5cm}
The important thing is not to stop questioning. Curiosity has its own reason for existing.
\vspace{0.3cm}\\
\emph{Albert Einstein}
\end{minipage}

\end{flushright}

\newpage

\chapter*{\emph{Et voil\`a ... c'est fait !}}

\emph{Vous avez été nombreux, ces dernières années, à vous poser la même question. Certains d'entre vous allèrent même jusqu'à la prononcer à haute et intelligible voix. Parfois avec malice, mais le plus souvent avec compassion. L'inquiétude était même perceptible dans la voix de certains. Pour être franc, je me la suis souvent posé moi-même, cette question ...
\begin{center}
Et alors cette thèse, ça avance ?
\end{center}
Je sais, je n'ai que trop rarement répondu à vos interrogations. La raison en était très simple: la seule chose que je savais, c'est qu'il faudrait de toute façon qu'elle soit terminée à temps. Mais voilà, le temps passe et avec lui, les années défilent. Le travail accompli prend forme. La dernière année de thèse pointe le bout de son nez avec, dans ses bagages, la rédaction tant redoutée. L'hiver est rude, le printemps accueilli à bras ouvert. Mais celui-ci cède déjà sa place à l'été qui lui-même se termine. Cette thèse qui a suscité tant d'interrogations prend forme. Petit à petit, les corrections et modifications défilent. Jusqu'à ce que ... Aujourd'hui, je suis enfin prêt à répondre à cette question. Alors, n'hésitez pas !}
\vspace{0.5cm}

\emph{Vous avez été nombreux à contribuer d'une manière ou d'une autre à l'élaboration de cette thèse. Certains directement, d'autres par leur seul présence. Il ne me reste plus qu'à trouver les mots pour vous remercier tous.}
\vspace{0.5cm}

\emph{La première fois que j'ai franchi la porte du bureau de Peter, je cherchais un sujet de mémoire. J'ignorais à l'époque qu'entrer dans ce bureau me conduirait à terminer une thèse cinq ans plus tard. Mais a posteriori, si tout était à refaire, je franchirais à nouveau cette porte! Durant toutes ces années, Peter a toujours été disponible, que ce soit pour répondre à mes questions toujours plus nombreuses (et pas forcement bien formulées, voire même parfois dénuées de sens!) ou pour discuter de nouvelles idées à explorer. Même s'il m'est souvent arrivé d'être encore plus perplexe après avoir discuté avec lui, je pense avoir beaucoup plus appris auprès de lui que durant n'importe quelle autre formation que j'ai pu suivre. Je lui exprime dès lors toute ma gratitude.}
\vspace{0.5cm}

\emph{Comme toute thèse, celle-ci a été réalisée au sein d'un groupe de recherches, le Service de Physique Théorique de l'Université Libre de Bruxelles. Que de bons moments passés durant ces quatre années à discuter avec ses membres (durant les breaks bien sûr) ou à faire les cent pas dans ce fameux couloir du NO7. Merci à vous tous, Chiara, François-Xavier, Gilles, Hylke, Isabelle, Jean-Marie, Jonathan, Manu, Paola et Thomas. Je souhaiterais tout particulièrement remercier Michel pour avoir été mon promoteur « officiel » pendant trois ans, avec tout le travail administratif que cette charge implique, et Laura pour ses conseils qui tombèrent toujours à point ainsi que pour sa patience sur les pistes de ski (et oui, le travail c'est dur!). Merci également à Josep dont j'ai partagé le bureau pendant près d'un an et qui m'a sorti à plusieurs reprises de mes problèmes sous Linux. Merci à vous, Quentin et Sabrina, pour tous ces fous rires qu'on a partagés dans ce même bureau et qui m'ont permis de m'évader un instant. Merci Fu-Sin pour ces quelques breaks au Soleil qui furent trop peu nombreux mais qui, j'en suis sûr, seront plus fréquents à l'avenir.}
\vspace{0.5cm}

\emph{Ces huit années passées à l'Université Libre de Bruxelles furent une expérience unique. Je souhaiterais dès lors remercier l'ensemble du corps professoral du département de physique pour l'enseignement reçu ainsi que les étudiants qui m'ont accompagné dans ce qui s'apparentait parfois à un véritable parcours du combattant. Merci tout particulièrement à Nassiba et Vincent.}
\vspace{0.5cm}

\emph{Cette thèse, je la dois en grande partie aux professeurs du Collège Don Bosco de Woluwé-Saint-Lambert qui m'ont enseigné des disciplines aussi variées que l'histoire et les mathématiques. Je voudrais remercier en particulier pour leur passion communicative Albert Demelenne, Ingrid t'Kint, Yves Monin, Philippe Moreau, Véronique Bouquelle ainsi que Françoise Damien à qui je dois mes premières observations de M 31, Jupiter et Saturne.}
\vspace{0.5cm}

\emph{J'adresse également mes remerciements à tous les membres de ma famille pour m'avoir toujours soutenu à chaque étape de mon existence. Je remercie en particulier ma « petite » soeur pour l'illustration placée sur la couverture de cette thèse.}
\vspace{0.5cm}

\emph{Je ne sais pas de quoi demain sera fait. Mais je sais d'où je viens et à qui je dois d'être ce que je suis. Pour vos valeurs que vous m'avez enseignées, pour votre patience, pour m'avoir toujours soutenu, pour tout cet amour que vous m'avez prodigué et pour tous ces moments de joie que vous m'avez offerts, Papa, Maman, merci.}
\vspace{0.5cm}

\emph{Ces remerciements ne seraient pas complets si j'oubliais celle qui supporte mes humeurs et mes interrogations depuis si longtemps. Merci pour ton amour et ton soutien sans failles.}
\vspace{1cm}
\vspace{1cm}
\begin{flushright}
Michael, \\ Septembre 2009.
\end{flushright}


\tableofcontents

\newpage
\section*{\Huge\usefont{OT1}{pzc}{m}{n}\selectfont Abbreviations \& Conventions} \addcontentsline{toc}{chapter}{Abbreviations \& Conventions}

\vspace*{1cm}

Throughout this thesis, we will work in ``natural unites'' for which the speed of light and the Planck constant are one, i.e. $c = h = 1$. Greek indices will be used for space-time coordinates ($\mu = 0, \ldots, 3$) and Latin indices for space coordinates ($i = 1, 2, 3$). The metric ``sign'' convention will be the one used by Landau \& Lifchitz in \emph{The classical theory of field}, for which the space-time interval is given by
\begin{eqnarray*}
\dif s^2 = \dif t^2 - \dif l^2 + \ldots \, .
\end{eqnarray*}
Therefore, the metric of Minkowski describing flat space-time is given in Cartesian coordinates by
\begin{eqnarray*}
\eta_{\mu\nu} = \textrm{diag.} \left( 1, -1, -1, -1 \right) .
\end{eqnarray*}
Finally, the following abbreviations and conventions will be used throughout this thesis :
\vspace*{0.1cm}
\hrule
\vspace{0.3cm}
\begin{tabular}{l|l}
CMB & Cosmic Microwave Background \\
EoM & Equation(s) of motion \\
FLRW & Friedmann-Lema\^{\i}tre-Robertson-Walker \\
FP & Fierz-Pauli (theory of) \\
GR & General Relativity \\
$\Phi$ & Newton's potential \\
$G = \left( 8 \pi \Mpl \right)^{-1}$ & Newton's constant \\
$g_{\mu\nu}$ & Curved metric \\
$g$ & Determinant of the metric \\
$\mathcal{G}_{\mu\nu}$ & Einstein tensor \\
$k = \sqrt{k_i^2}$ & Three-dimensional momentum \\
$\Lambda_{c}$ & Cosmological constant \\
$\Mpl$ & Planck Mass \\
$\mathcal{R}$ & Ricci scalar \\
$\mathcal{R}_{\mu\nu}$ & Ricci tensor
\end{tabular}

\newpage
\renewcommand{\thepage}{\arabic{page}}
\setcounter{page}{1}


\part[Introduction]{\usefont{OT1}{pzc}{m}{n}\selectfont Introduction}

\chapter{Motivations \& outline}

When General Relativity (GR) was formulated in the beginning of the twentieth century, the anomalous perihelion advance of Mercury was an unsolved problem. When it was clear that Einstein's new theory predicted the same perihelion advance as the one observed, GR became the standard theory for the gravitational interaction. Its popularity, firstly motivated by its mathematical elegance and consistence, was reinforced by its prediction of the bending of light by a massive source, which was first observed in 1919. Since then, GR has been tested with an ever-growing precision \cite{Shapiro:1999fn,Bertotti:2003rm,Long:2003ta,Long:2003dx,Hoyle:2004cw,Will:2005va,Turyshev:2008dr}. However, gravity remains little-known mainly because of its particularly low coupling to matter\footnote{The gravitational interaction between two protons is 38 orders of magnitude smaller than the Coulomb interaction between them.}.

GR predicts \cite{Weinberg:1972,Wald:1984rg} the existence of gravitational waves, in analogy with Maxwell's theory of electrodynamics. There are many similarities between gravity and electromagnetism, but also striking differences. For instance, while the existence of photon is well established, gravitational waves have still not been directly observed despite the huge efforts invested recently in developing gravitational waves detection experiments \cite{Acernese:2008zz,Abramovici:1992ah}. At present, the best evidence one has for the existence of such waves comes from timing observations of binary pulsars \cite{Damour:1991rd,Stairs:2003eg}; the secular decrease of orbital period of the system PSR 1913+16 has been shown to be compatible with the emission of gravitational waves predicted by GR \cite{1979Natur.277..437T,Taylor:1982zz}. Hence, gravitational waves are in a situation similar to neutrinos after Pauli's proposal in 1930, when their emission could be inferred from energy and momentum conservation in beta decays without being observed in absorption experiments.

According to GR, gravitational waves have zero rest mass, just as electrodynamic waves. If gravitons were to have a small, non-zero mass $m$, one would expect that all effects arising because of their mass would be small corrections to Einstein's theory of gravity, completely negligible at distances smaller than the graviton's Compton wavelength $\lambda \sim m^{-1}$. Moreover, it is well known in fields theory that the very concept of a field rest mass is closely related to the range of the interaction it mediates. If the gravitons were to have a non-zero mass, one would expect the potential of a static source to have a Yukawa shape $e^{-mr}/r$ effectively cutting off the gravitational interaction at distances larger than $m^{-1}$. For this reason, much attention has been paid to the apparent absence of such cutoff \cite{Goldhaber:1974wg,Talmadge:1988qz,Smullin:2005iv}. The dynamic of gravitational waves allows for independent constraints on the graviton rest mass through the modification of their dispersion relation, which implies that these waves propagate at velocity
\begin{eqnarray*}
v_{\tx{group}} \left( \omega \right) \equiv \dfrac{\partial \omega}{\partial k} = \sqrt{1 - \dfrac{m^2}{\omega^2}} .
\end{eqnarray*}
This relation implies that massive gravitational waves of different frequence should propagate at different speed. Direct observations of gravitational waves could then be used to set a bound on the graviton mass, for instance from observations of compact binary systems which slowly spiral together due to the emission of gravitational waves \cite{Will:1997bb,Larson:1999kg,Cutler:2002ef,Yagi:2009zm,Arun:2009pq,Finn:2001qi}. Other bounds on the graviton rest mass could be set through the delay between electromagnetic and gravitational waves \cite{Cooray:2003cv} or through the relativistic time delay \cite{Will:2003yj} (for a recent review on graviton mass limits, see \cite{Goldhaber:2008xy}).

Although observations made so far seem compatible with massless gravitational waves, a very small graviton rest mass cannot be excluded. Hence, one may wonder whether it is possible to modify Einstein's theory of gravity as to describe massive gravitons. Giving a mass to gravitons is a challenging problem \cite{Rubakov:2008nh}, which has motivated many studies since the original work of W. Pauli and M. Fierz \cite{Fierz:1939ix}. Because of the intricate structure of GR, it is much more complicated to give a mass to gravitons than imitating Proca's work on massive photons, with consequence that no clear consensus has emerged on the subject.

Beside the theoretical interest of such question, theories with massive gravitons also have more pragmatic motivations. Surprisingly, one of them comes from Quantum Chromodynamics (QCD) \cite{'tHooft:2007bf}. Indeed, it is believed that QCD can be approximated by some sort of string theory. However, string theories often have massless spin-2 modes, while QCD does not. This apparent contradiction may be an illusion if it turns out that these modes can be removed from the massless sector of string theory by giving them a mass, a task which may seem very similar to giving a mass to gravitons. Yet, there are other original motivations for theories with massive gravitons. For instance, massless scalar fields only interacting with gravity could exist in principle. If so, they could be responsible for a non-zero graviton rest mass, just because of their interaction with gravity. The class of models which are the subject of this thesis makes use of this possibility.

All these considerations legitimate the efforts invested in the attempt of building a consistent theory with massive gravitons. Yet, the main motivation for such theories is elsewhere. Generalizations of Einstein's theory of gravity with a small, non-zero, graviton mass may give rise to large scale (or infrared) modifications of GR. Despite that Einstein's theory has been directly tested from scales of a fraction of millimeters up to Solar System scales, there is no direct confirmation of it on larger distances, with consequence that infrared deviations from GR cannot be excluded. The recent developments in theories describing massive gravitons were parts of this revival of interest for infrared modification of GR.

Theories that predict infrared modifications of the gravitational interaction have two main motivations. The first one is to be found in theories with brane-worlds and extra-dimensions of large or infinite size (see \cite{Rubakov:2001kp} for a review). In these models, the usual matter is supposed to reside on a 3-brane embedded in higher-dimensional space, while gravity may propagate in all dimensions \cite{Dvali:2000hr,Gregory:2000jc}. Hence, they generally predict that gravity is modified at very large scales while being conventional at smaller ones. A notable example is given by the Dvali-Gabadadze-Porrati (DGP) model \cite{Dvali:2000hr}, for which gravitons may appear as a massive resonance \cite{Dvali:2000rv,Dvali:2008em} to observers trapped in the 3-brane. The second motivation for large scale modifications of GR is to be found in cosmological observations.

\subsubsection{The dark sides of the Universe}

Because of the non-renormalizability of Einstein's theory of gravity, it is commonly believed that GR gives way to a quantum theory of gravity \cite{Donoghue:1995cz,Burgess:2003jk,Goldberger:2007hy,Burgess:2007pt} at high enough energies. For dimensional reasons, it seems natural to identify this energy scale with the Planck mass $\Mpl \equiv \left( 8 \pi G \right)^{-1}$, where $G$ is Newton's constant. The standard assumption is that GR is valid as an effective field theory\footnote{For reviews on effective field theory, see \cite{Donoghue:1995cz,Burgess:2003jk,Goldberger:2007hy,Burgess:2007pt}.} for energies below $\tx{M}_\tx{pl}$, and therefore for length scales larger than  $\lambda_\tx{pl} = \tx{M}_\tx{pl}^{-1}$. Nevertheless, at very low energies GR also faces some troubles in explaining observations made from galactic up to cosmological scales; the agreement is only achieved after the introduction of the otherwise undetected \emph{dark matter} and \emph{dark energy}. However, with some assumptions about these dark components, a picture emerges which describes with an impressive precision the whole bulk of cosmological data \cite{Astier:2005qq,Seljak:2006bg,AdelmanMcCarthy:2007wu,Dunkley:2008ie}. The question arises to what extent one should consider this agreement as a confirmation of GR itself. It is not inconceivable that large scale modifications of GR, such as theories with a small graviton mass, may account for these observations without need of dark hypothesis. Over the last years, there have been several attempts \cite{Bekenstein:2004ne,Carroll:2003wy,Kogan:2000vb,Damour:2002ws,Arkani-Hamed:2003uy,Gabadadze:2003ii} in constructing self-consistent theories which deviate from GR at large distances and time scale\footnote{The general philosophy of those works is to view these models as the low energy limits of an unknown fundamental theory, without worrying about issues like renormalizability or embedding into an ultraviolet-complete theory.}, most of them being motivated by these \textquotedblleft dark paradigms\textquotedblright.

The existence of dark matter in the form of an undetected planet was already assumed in the 19th century to reconcile astronomical observations and Newton's theory of gravity; the astronomers Jean Joseph le Verrier and John Couch Adams both predicted the existence of a new, unseen planet by studying the anomalies in the orbit of Uranus. Although Neptune was discovered in 1846 at the precise location predicted by these astronomers, this \textquotedblleft dark matter\textquotedblright problem also had an alternative explanation as modification of gravity. Soon after, le Verrier started a complete study of orbital motions of the other planets. Motivated by his previous achievement, he postulated the existence of another undetected planet, Vulcan, between Mercury and the Sun to explain the perihelion advance of Mercury. However, this time the solution was not a missing dark matter, and the problems in the motion of Mercury were solved by Einstein's new theory of gravity. Nowadays, both dark matter and dark energy are needed to reconcile theory and observations.

Modern dark matter was first advocated by Jan Henrik Oort \cite{1932BAN.....6..249O} and Fritz Zwicky \cite{1933AcHPh...6..110Z} in the early 1930s to explain vertical motions of stars in the Milky Way and radial velocities of galaxies in the Coma cluster, respectively. Since then, the number of observational evidences \cite{Bergstrom:2000pn} for non-baryonic dark matter has burst out (see \cite{Sahni:2004ai,Bertone:2004pz,Einasto:2009zd} for reviews): dark matter is needed from galaxy rotation curves \cite{Corbelli:1999af,Sofue:2000jx} up to structure formation \cite{Blumenthal:1984bp}. Moreover, there were recently claims for direct evidences for the existence of dark matter from two colliding clusters of galaxies known as the \emph{Bullet Cluster} \cite{Clowe:2006eq}.

The problem of modern dark matter is very similar to the old problem of unseen planets. One may either resolve it by assuming the existence of a large amount of unseen, dark, matter existing on system with size ranging from galactic to cosmological scale, or by assuming deviations from the standard laws of gravity at large distances and time scales \cite{Mannheim:2005bfa}. Both approaches are popular. An example of the first approach is given by \emph{weakly interacting massive particles} (or WIMP's) \cite{Bertone:2004pz}. The idea is to suppose the existence of a stable, weakly interacting, massive particles which has been produced in a huge amount in the early Universe and which could explain the discrepancy between theory and observations through its interaction with gravity. The second approach, often illustrated by \emph{Modified Newtonian Dynamics} (MOND) \cite{Milgrom:1983pn} and theories attempting to incorporate it in a full relativistic description \cite{Bekenstein:2009bd}, consist in searching for possible alternatives to GR which are often designed to explain rotation curves of galaxies without dark matter.

Modern physicists face more serious troubles when trying to understand the nature of dark energy (for reviews on the subject, see \cite{RevModPhys.61.1,Carroll:1991mt}). First evidence for dark energy dates back to the early 1990s. At that time, it become clear from the analyze of the Cosmic Microwave Background (CMB) power spectrum that the Universe is nearly flat \cite{Melchiorri:1999br,Balbi:2000tg}. Yet, the fractional energy density of all non-relativistic matter (baryons and dark matter) was bounded by $\Omega_{\tx{Matter}} \leq 0.3$ \cite{1986Natur.321...27P,Carlberg:1995aq} so that something was missing in order for the Universe to be flat. Nowadays, the existence of dark energy is inferred from many cosmological observations \cite{Perlmutter:1998np,Riess:1998cb} all indicating that the Universe is currently accelerating. In the framework of GR, this is only possible if the Universe is dominated by a fluid with negative effective pressure
\begin{eqnarray*}
\dfrac{p_{\Lambda}}{\rho_{\Lambda}} \equiv w_{\Lambda} < - \dfrac{1}{3} ,
\end{eqnarray*}
called dark energy and which is some kind of effective version of Einstein's cosmological constant. Dark energy is assumed to be responsible for roughly $70 \%$ of the total energy content of the Universe, $\Omega_{\Lambda} = 0.742 \pm 0.030$ \cite{Dunkley:2008ie}\footnote{Standard baryonic matter and cold dark matter amount only to $\Omega_{\tx{Baryons}} \sim 0.04$ and $\Omega_{\tx{CDM}} = \Omega_{\tx{Matter}} - \Omega_{\tx{Baryons}} \sim 0.22$, respectively.}, with an energy density given by
\begin{eqnarray*}
\rho_{\Lambda} \equiv 3 H_{0}^2 \Mpl \Omega_{\Lambda} \sim 4 \cdot 10^{-6} \,\, \dfrac{\tx{Gev}}{\tx{cm}^3} \sim 10^{-29} \dfrac{\tx{g}}{\tx{cm}^3} .
\end{eqnarray*}
This energy scale is much smaller than the energy scale one would associate, on dimensional grounds, to the three known interactions
\begin{eqnarray*}
\rho_{\Lambda} \sim 10^{-46} \rho_{\tx{QCD}} \sim 10^{-54} \rho_{\tx{EW}} \sim 10^{-123} \rho_{\tx{gravity}} .
\end{eqnarray*}
Its seems then really difficult to understand the origin of dark energy. The unnatural smallness of $\rho_{\Lambda}$ is a huge problem. It implies that the vacuum energy density\footnote{Gravity couples universally to all forms of energy, including the vacuum energy which acts like a cosmological constant in Einstein's equations.} is canceled in the gravitational field equations to a very small value by a mechanism which does not seem to be related to any of the three known interactions. In fact, this problem could be divided into two parts. The first part consists in understanding why is dark energy not as high as it should be to agree with estimations of the vacuum energy density \cite{RevModPhys.61.1}; in other words, why is $\rho_{\Lambda}$ essentially zero. The second part of the problem consist in understanding why $\rho_{\Lambda}$ is in fact different from zero.

Although it is not inconceivable that the first part of this problem may be solved by a mechanism which drives the cosmological constant to zero in the early Universe \cite{Dolgov:2006xi}, its second part requires to understand what is hidden behind $\rho_{\Lambda}$. There are various hypotheses as for the nature of dark energy \cite{Sahni:1999gb,Peebles:2002gy}:
\begin{itemize}
\item In order to fit the cosmological observations, assumptions are made about our Universe. In particular, the Universe is assumed to be homogeneous and isotropic, although this is clearly not the case below scales of $\sim 100-200\,\tx{Mpc}$. The acceleration of the Universe's expansion could be an illusion if one actually lives in a kind of void, where the matter density is smaller than the spatial average density \cite{PascualSanchez:1999zr,Celerier:1999hp,Alexander:2007xx} (see \cite{Celerier:2007jc} for a review).
\item Models with weakly interacting, time-varying, negative pressure fluids such as quintessence \cite{Ratra:1987rm,Caldwell:1997ii,Zhang:2007nk} are perhaps the most popular candidates for solving the dark energy problem (see \cite{Caldwell:2000wt} for an introduction to quintessence) since such fields naturally imply an acceleration of the Universe's expansion at late time.
\item The weak anthropic principle \cite{Carter:2006gy} could also somehow explain the observed value of the vacuum energy density \cite{RevModPhys.61.1,Linde:2002gj}. Indeed, if the Universe is much larger than its visible part, and if the cosmological constant take different values in different regions of cosmological size, then we happen to live in a region where the effective vacuum energy density is small enough to allow galaxies and stars to form. In other regions where the effective cosmological constant is not that small, there is simply nobody to worry about this issue.
\end{itemize}
This list is not exhaustive. Independently of the relevance of all these hypothesis, large scale deviation from GR may also account for the unnatural smallness of $\rho_{\Lambda}$ \cite{Nojiri:2006ri,Bertschinger:2008zb}. For instance, theories based on an arbitrary function of the Ricci scalar \cite{Sotiriou:2008rp} are designed to fit the cosmological observations without need of dark energy. But what is even more interesting is that theories with massive gravitons may also contribute to the cosmological constant and therefore shed light on the dark energy paradigm (see chapter \ref{ch:cosmo}).

\subsubsection{Massive gravitons}

It is an open question whether all current observations could naturally fit into a unique theory of gravity without need of dark matter and, or, dark energy. In order to address this question an alternative model is needed whose predictions can be compared to those of GR. Theories with massive gravitons are candidates for such alternatives. They are interesting from both theoretical and phenomenological point of view, independently of the answer to this question.

As already mentioned, the first attempts trying to describe massive gravitons were based on the original work of W. Pauli and M. Fierz \cite{Fierz:1939ix}. These attempts were concerning a Lorentz-invariant model, in the linear regime, which was supposed to generalize Einstein theory of gravity with massive gravitons. This model, known as the Fierz-Pauli theory, is the only one which is both Lorentz-invariant and free of instabilities in Minkowski space-time. Unfortunately, it turned out that it could hardly be considered as a coherent candidate for a model of massive gravity (i.e., with massive gravitons). Although some of its problems may be solved by a non-linear completion, any theory with Lorentz-invariant massive gravitons of the Fierz-Pauli type suffers from strong coupling at very low energy scales \cite{Arkani-Hamed:2002sp}. Therefore, such theory requires an UV completion to describe the gravitational interaction at scales where GR has already been tested.

It is not clear how to build consistently a theory of massive gravity. Yet, the example given by the Fierz-Pauli model enables one to understand that it is essential to have a full non-linear formulation for both theoretical and observational considerations. Because of the success of Einstein's theory of gravity, it seems natural to add to the action of GR a term which will, in the linearized approximation, give a mass to gravitons without modifying the kinetic terms coming from GR. However, no scalar term can be build out of the metric without derivative. Therefore, either one gives up the requirement for the invariance under the whole group of diffeomorphisms, or one adds dynamical fields to the theory. In fact, both approaches may appear equivalent through the St\"{u}ckelberg formalism \cite{Stuckelberg:1938}. Indeed, any action containing the Einstein-Hilbert term plus a non-gauge-invariant function of the metric can be thought of as the unitary gauge description of some gauge-invariant theory containing one scalar field for each broken diffeomorphism. Theories of massive gravity discussed in this thesis are of this kind; they are based on a gauge-invariant action containing a function of the metric derivatively coupled to four scalar fields \cite{Dubovsky:2004sg}, along with the usual Einstein-Hilbert term. Gravitons acquire a mass due to a mechanism which may be thought of as the low energy limit of the \emph{Brout-Englert-Higgs} mechanism for gravity \cite{'tHooft:2007bf}: the vacuum expectation value of each scalar field may break one coordinate reparametrization invariance, and gravitons acquire a mass. One should note here that adding dynamical scalar fields is not the only possibility to restore the gauge-invariance of the action. For instance, bigravity theories \cite{Damour:2002ws,Blas:2007zz,Berezhiani:2007zf} assume two dynamical metrics: their action is made of two Einstein-Hilbert term along with a gauge-invariant coupling between the metric.

Instead of breaking the whole diffeomorphism group of GR, one may require the vacuum to break only the invariance under the boosts of the Lorentz group. Indeed, one already knows that if the Lorentz-invariance is not broken, either the vacuum is unstable (for non-Fierz-Pauly mass terms) or there is a strong coupling problem (for Fierz-Pauly mass terms). It has recently been argued that all the standard problems of massive gravity could be avoided just by renouncing the Lorentz-invariance of the background. Consequently, theories of massive gravity with spontaneous breaking of the Lorentz symmetry have been considered intensively. Their study started a few years ago (for a review, see \cite{Rubakov:2008nh}) with discussions about their stability in Minkowski space-time \cite{Dubovsky:2004sg,Rubakov:2004eb}. Because they are free of the usual problems encounter in the study of Lorentz-invariant models, they are candidates for healthy theories with massive gravitons. They may be thought of as generalizations of a model known as the Ghost Condensate \cite{Arkani-Hamed:2003uy} (see chapter \ref{ch:mg}) with which they share several properties, although gravitons are massless in the latter.

For the sake of completeness, one should mention that there are other directions of search. One of them is based on topological massive gravity in $2 + 1$ dimensions \cite{Deser:1981wh,Bergshoeff:2009hq}, whose main motivation comes from quantum gravity considerations (one expects less severe short-distance behavior in  $2 + 1$ than in  $3 + 1$ dimensions). Theories with fourth-order derivatives have also been studied \cite{Stelle:1977ry,Nojiri:2006ri}. In these models, there is a massive spin two graviton along with the massless one. Unfortunately, such models are plagued by ghost instabilities. Another possibility is provided by models where the four dimensional graviton is a metastable resonance with a finite lifetime \cite{Dvali:2000rv,Dvali:2008em,Csaki:2000pp}. Such particular graviton state are found in brane theories with extra dimensions, as for example in the DGP model \cite{Dvali:2000hr}.

\subsubsection{Outline of this thesis}

Up to now, several aspects of theories of massive gravity with spontaneous breaking of the Lorentz symmetry have been studied, most of them concerning a minimal class of massive gravity models \cite{Dubovsky:2004ud} for which there is no modifications of Newton's potential. In this thesis, we discuss some other theoretical and phenomenological aspects of this minimal class of models. The first part of this thesis is merely introductory. It contains two chapters in addition to the present one:
\begin{itemize}
\item Chapter \ref{ch:mg} consists of two parts. The first part overviews the ins and outs of Lorentz-invariant theories of massive gravity, with the aim of illustrating the problems which arise when trying to give a mass to gravitons. Although this discussion is not essential for the understanding of the original part of this thesis, it enables one to comprehend why people have started studying models with Lorentz symmetry breaking. In the second part of chapter \ref{ch:mg}, we introduce models of massive gravity with spontaneous breaking of Lorentz symmetry. This second discussion gives some details about the stability issues in these models, which lead to a minimal class of massive gravity theories possessing only two massive modes with helecities $\pm 2$. It is this minimal class of models that will be considered in the rest of the thesis.
\item Chapter \ref{ch:weak_field} is devoted to the weak field limit of the massive gravitational field. This discussion has two aims. The first one is to introduce a sub-class of minimal models for which there is no modifications of Newton's potential. The second one is to provides the equations which will be the starting point of chapter \ref{ch:inst_int}.
\end{itemize}
The second part of this thesis contains all the original contributions to the study of massive gravity theories with spontaneous breaking of the Lorentz-invariance. It contains four chapters:
\begin{itemize}
\item In chapter \ref{ch:inst_int}, we discuss the first original contribution of this thesis, consisting in the study of a physical instantaneous interaction present in massive gravity theories. In GR, the gravitational potentials are all instantaneous potentials, as in classical electrodynamics. However, these instantaneous contributions cancel each other in observables, leaving the theory free of physical instantaneous interactions. For the models considered in this thesis, these subtle cancellations are spoiled by the presence of four scalar fields which break Lorentz-invariance. One then expects these models to have physical instantaneous interactions. The existence of such interactions is related to the presence of a mode with the dispersion relation $k^2 = 0$ which could be interpreted as a mode with an infinite propagation velocity.

Beside these theoretical expectations, the presence of a superluminal interaction is required by a crucial feature of black holes: it has been shown that black holes do possess \emph{hair} in massive gravity models with Lorentz-symmetry breaking \cite{Dubovsky:2007zi}. This violation of the \textquotedblleft no-hair\textquotedblright theorem of GR is supposed to be the consequence of the presence of physical instantaneous interactions who could carry informations outside of the black hole horizon.

The aim of this chapter is to demonstrate the existence of a physical instantaneous interaction in massive gravity theories with spontaneous breaking of Lorentz-invariance. A concrete example will be studied, consisting in an instantaneous frequency shift of a light beam by a distant gravitational source.
\item In chapter \ref{ch:bh}, we obtain the exact static vacuum spherically symmetric solutions of massive gravity equations. In GR, this solution plays a crucial role. First, this solution describes the metric outside of spherical non-rotating bodies and gives rise, in the weak field limit, to the Newtonian gravity. It provides therefore a useful approximation in many astrophysical situations. Second, the Schwarzschild solution describes the result of a gravitational collapse, the black hole.

Black holes are perhaps the most interesting objects to constrain alternative models of gravity, since by reconstructing the metric around an astrophysical black hole one should be able to test whether it has the Schwarzschild or Kerr form, and therefore probe GR in the strong field limit. In massive gravity theories with breaking of the Lorentz-invariance, the properties of black holes are expected to be different. In particular, rotating black holes are certainly modified, and, more generally, black holes are expected to have hair \cite{Dubovsky:2007zi}. The possible existence of black hole hair in massive gravity models suggests that there might exist spherically symmetric solutions other than the Schwarzschild one.

The aim of this chapter is to present a new class of vacuum spherically-symmetric solutions in massive gravity theories. These solutions constitute the second original contribution to this thesis. One will see that, in addition to the Schwarzschild radius, these solutions depend on one more parameter, called the \emph{scalar charge} $S$. At zero value of this parameter the standard Schwarzschild solution is recovered, while at non-zero values of $S$ the Schwarzschild metric gets modified. The modified solutions are non-linear at all distances; they cannot be obtained in the linear approximation. These new solutions may have event horizons and are, therefore, candidates for modified black holes. Both analytical and numerical examples of such modified black holes are discussed in this chapter.
\item Chapter \ref{ch:cosmo} contains the third and last original contribution of this thesis. Given that massive gravity models with spontaneous breaking of Lorentz-invariance passes the most obvious constraints, one may wonder if it reproduces correctly more subtle parts of modern cosmology, in particular, the theory of structure formation. This is the question which is addressed in this chapter. The answer is not obvious {\em a priori} since the vacuum in these models contains the condensates of four scalar fields whose perturbations mix with the matter density perturbations.

Cosmological perturbations are studied in the minimal class of massive gravity models. They consist of two parts. The first part behaves
identically to the perturbations in GR while the second, \textquotedblleft anomalous\textquotedblright part is proportional to an unknown function $\Psi_0(x^i)$ of the space coordinates which arises as an integration constant. The growth of the \textquotedblleft anomalous\textquotedblright perturbations depends on the value of a single parameter $\gamma$. For $- 1 < \gamma < 0$ they grow slower than the standard ones, so that the latter dominate, while at $\gamma = 1$ they
cancel out. Thus, at least in these two cases the perturbations behave in a standard way and massive gravity models are consistent with the formation of structures. This is the main result of this chapter.
\item Finally, chapter \ref{ch:conclusion} contains the general conclusion of this thesis.
\end{itemize}

\chapter{Massive gravity} \label{ch:mg}

One may wonder whether it is possible to construct a well-defined theory describing massive gravitons. If such theory exists and if the mass $m$ of the gravitons is small enough, or equivalently if its Compton wavelength $\lambda \sim m^{-1}$ is large enough, one may not be able to distinguish this theory from GR in the current observations of the gravitational field. All effects arising because of the mass $m$ would be small corrections to Einstein's theory. Unfortunately, things are much more complicated.

In this chapter, one review the problems and issues of theories with massive gravitons. In section \ref{sc:LIMG}, one discuses Lorentz-invariant models in their linear regime with the aim of demonstrating that they cannot lead to satisfying theory of massive gravity. Although this discussion is not essential for the understanding of the original part of this thesis, it enables one to comprehend why people have started studying models where the Lorentz symmetry is dynamically broken by the vacuum expectation value of four scalar fields, models who are introduce in section \ref{sc:LVMG}. This second discussion gives the details of the models which will be considered in the rest of the thesis. The original part of this thesis concern a minimal class of massive gravity models who will be introduced at the end of this chapter.

\section{Lorentz-invariant massive gravity} \label{sc:LIMG}

Theories of gravity are in general non-linear as GR, since the gravitational field is a source of gravity itself. Such theories are complicated and it is difficult to find analytical solutions to them (we only have a few analytical solutions to GR equations). Therefore, when a solution is known, it is common to analyze small perturbations around this solution. For example, the Solar System is well described in GR by considering small linear perturbations around Minkowski background (this will be discussed in details in section \ref{sc:FP_vainst}). For this reason and because linear approximation reveals the particle content of a theory, massive gravity where first considered in the linear approximation, just by adding mass terms for the perturbations around Minkowski vacuum.

First attempts in constructing a model describing massive graviton were related to a Lorentz-invariant generalization of Einstein theory known as the Fierz-Pauli (FP) theory of massive gravity \cite{Fierz:1939ix}. But before discussing the FP theory, let us first have a look at the most general theory of Lorentz-invariant massive gravity in order to prove that, at the linearized level, the FP theory is the only healthy Lorentz-invariant theory describing massive gravitational waves. For this purpose, consider small fluctuations about Minkowski space-time parameterized as follow
\begin{eqnarray*}
g_{\mu\nu} = \eta_{\mu\nu} + h_{\mu\nu} , & & | h_{\mu\nu} | \sim \epsilon \ll 1 .
\end{eqnarray*}
At the quadratic level in these small fluctuations, the action for Lorentz-invariant massive gravitational waves is simply given by the Einstein-Hilbert action plus the most general Lorentz-invariant mass terms
\begin{eqnarray*}
\act = \Mpl \int \dif^4 x \mathcal{L} , & & \mathcal{L} = \mathcal{L}_{\tx{EH}} - \dfrac{m^2}{4} \left( h_{\mu\nu} h^{\mu\nu} + \alpha h^2 \right) ,
\end{eqnarray*}
where $\alpha$ is a dimensionless constant and $h \equiv h_{\mu}^{\mu}$. Indices are raised and lowered with the Minkowski metric. In what follows the Lagrangian densities will be defined up to an overall $\Mpl$ factor. $\mathcal{L}_{\tx{EH}}$ includes the gauge invariant kinetic terms for the metric fluctuations (see appendix \ref{app:sc:line_EH} for details about the linearization procedure)
\begin{eqnarray*}
\mathcal{L}_{\tx{EH}} = \dfrac{1}{4} \left[ \partial_{\alpha} h^{\mu\nu} \partial^{\alpha} h_{\mu\nu} - 2 \partial_{\mu} h^{\mu\nu} \partial^{\alpha} h_{\nu\alpha} + h \left( \partial^{\mu} \partial_{\mu} h - 2 \partial^{\mu} \partial^{\nu} h_{\mu\nu} \right) \right] .
\end{eqnarray*}
Since the gravitational field is assumed to be zero on the border of the integration domain, terms with total derivatives are neglected. The mass terms break the gauge invariance of the theory. Hence, it is tempting to enlarge the field content of the previous action so as to restore the gauge invariance. This can be done through the St\"{u}ckelberg formalism \cite{Stuckelberg:1938} (for a recent review on the subject, see \cite{Ruegg:2003ps}). Let us introduce four new scalar fields $\xi^\mu$ and replace $h_{\mu\nu}$ in the Lagrangian by the combination
\begin{eqnarray*}
h_{\mu\nu} = \tilde{h}_{\mu\nu} - \partial_{\mu} \xi_{\nu} - \partial_{\nu} \xi_{\mu} .
\end{eqnarray*}
With this definition, the Lagrangian for the massive gravitational waves becomes
\begin{eqnarray} \label{eq:FP_LIMG_action}
\lag &=& \mathcal{L}_{\tx{EH}} - \dfrac{m^2}{4} \left( \tilde{h}_{\mu\nu} \tilde{h}^{\mu\nu} + \alpha \tilde{h}^2 \right) + m^2 \left( \tilde{h}^{\mu\nu} \partial_{\mu} \xi_{\nu} + \alpha \tilde{h} \partial_{\mu} \xi^{\mu} \right) \nonumber \\
& & - \dfrac{m^2}{2} \left( \partial^{\mu} \xi^{\nu} \partial_{\mu} \xi_{\nu} + \left( 1 + 2 \alpha \right) \left( \partial^{\mu} \xi_{\mu} \right)^2 \right) .
\end{eqnarray}
Now, it is straightforward to see that the theory is invariant under the following gauge transformations
\begin{eqnarray} \label{eq:Inf_gauge}
\tilde{h}_{\mu\nu} \rightarrow \tilde{h}_{\mu\nu} + \partial_{\mu} \varsigma_{\nu} + \partial_{\nu} \varsigma_{\mu} , & & \xi_{\mu} \rightarrow \xi_{\mu} + \varsigma_{\mu} ,
\end{eqnarray}
which correspond to an infinitesimal coordinate transformations $x^\mu \rightarrow x^{\mu} - \varsigma^{\mu}$. This gauge-invariant formulation is completely equivalent to the previous one, since it is always possible to go to the unitary gauge $\varsigma^{\mu} = - \xi^{\mu}$ where there is no scalar fields $\xi^\mu = 0$. Hence, despite the presence of the scalar fields, this St\"{u}ckelberg trick do not introduce new degrees of freedom. These scalar fields are known in the literature as \textquotedblleft Goldstone fields\textquotedblright.

Quadratic Lagrangian reveals the particle content of a theory. In GR,  there are only two propagating modes because of the gauge-invariance. These degrees of freedom correspond to the massless spin-2 graviton, with the helicity states $\pm 2$. Since mass terms break the gauge invariance in the metric sector of the massive theory \refp{eq:FP_LIMG_action}, extra propagating degrees of freedom are expected to emerge. The massless graviton of GR should turn into a massive spin-2 field which has five helicity states: two tensor modes with helicities $\pm 2$, two vector modes with helicities $\pm 1$ and one scalar mode with helicity $0$, according to the irreducible representations of the Poincare symmetry group of massive spin-2 fields. Therefore, besides the usual tensor modes describing helicities $\pm 2$ states, the massive theory should posses at least three more degrees of freedom corresponding to the helicities $0, \pm 1$. However, the situation is more complicated because of the scalar sector of the theory.

One of the greatest problem when dealing with a theory of gravity which has more dynamical modes than GR is to guarantee that those modes do not imply instabilities, i.e. the vacuum should be stable against small perturbations. There are basically two kinds of highly dangerous instabilities. The first one is the classical instability characterized by $\omega^2 = - k^2 + m^2$. Modes with such dispersion relation are stable at low momenta $0 < k^2 < m^2$. However, they grow exponentially at arbitrarily high momenta $m^2 < k^2$ and are therefore unacceptable. The second type of pathological behavior is known as ghost. Ghosts have healthy dispersion relations ($\omega^2 > 0$), but they lead to an instability because they have the wrong sign in front of their kinetic terms, and hence negative energy. Therefore, it is possible to endlessly create ghost fields along with \textquotedblleft healthy\textquotedblright fields without contribution of external energy \cite{Cline:2003gs,Holdom:2004yx}. In order to clarify the particle content of the quadratic Lagrangian, let us introduce the transverse and longitudinal decomposition.

\subsection{Transverse and longitudinal decomposition} \label{sc:Pert_Mink}

A convenient way to study the fluctuations about Minkowski space-time in metric theories of gravity consist in decomposing the metric perturbations into transverse and longitudinal fields \cite{Mukhanov:1990me}. In massive gravity, this decomposition is generalized to include the four Goldstone fields
\begin{eqnarray} \label{eq:Pert_Mk}
\left. \begin{array}{l}
h_{00} = 2 \varphi, \\
h_{0i} = S_{i} + \partial_{i} B, \\
h_{ij} = 2 \Psi \delta_{ij} - 2 \partial_{i} \partial_{j} E - \partial_{i} F_{j} - \partial_{j} F_{i} + H_{ij},
\\
\end{array} \right| &&
\begin{array}{l}
\xi^{0} = \xi_{0} , \\
\xi^{i} = \xi_{i}^{T} + \partial_{i} \xi .
\end{array}
\end{eqnarray}
$\varphi$, $B$, $\Psi$, $E$, $\xi_{0}$ and $\xi$ are scalar fields under tree-dimensional rotations \footnote{The four scalar fields $\xi^\mu$ are scalars under the rotations of the three dimensional space. By rotations here, we mean the generalized rotations which include the rotations of the $\xi^i$ internal space along with the usual rotations of the three dimensional space. Such rotations leave Lagrangian (\ref{eq:FP_LIMG_action}) invariant.}. $S_{i}$, $F_{i}$ and $\xi^{T}_{i}$ are transverse vector fields satisfying
\begin{eqnarray*}
\partial_{i} S_{i} = \partial_{i} F_{i} =  \partial_{i} \xi_{i}^{T} = 0 .
\end{eqnarray*}
Summation over spatial indices is performed with the Euclidean metric. Because of the previous relations, each of these three vectors possess only two independent components. $H_{ij}$ is a transverse and traceless tensor field
\begin{eqnarray} \label{eq:FP_Hij}
\partial_{i} H_{ij} = H_{ii} = 0,
\end{eqnarray}
which possesses only two independent components. Therefore, in total there is one tensor perturbation $H_{ij}$
consisting of two components, tree vectors $S_i$, $F_i$, $\xi_i^{T}$, consisting of two components each, and 6 scalars $\varphi$,
$B$, $\Psi$, $E$, $\xi^0$, $\xi$. Since gauge-invariant quantities play important role in gravity, it will be useful to introduce gauge-invariant fields. One vector and two scalar perturbations are gauge degrees of freedom. As a consequence, there are only two gauge-invariant vector fields
\begin{eqnarray*}
\varpi_{i} = S_{i} + \dot{F}_{i} , & & \sigma_{i} = \xi_{i}^{T} - F_{i} ,
\end{eqnarray*}
and four scalar gauge-invariant fields
\begin{eqnarray*}
\Phi = \varphi - \ddot{E} - \dot{B} , & \Xi = \xi - E , & \Xi^{0} = \xi^0 - \dot{E} - B ,
\end{eqnarray*}
and $\Psi$. The tensor perturbation $H_{ij}$ is also gauge-invariant. Over-dot denotes the time derivative.

This decomposition is a powerful tool to explore any metric theory of gravity. Indeed, let $e_{i}^{\sm{(1)}}$ and $e_{i}^{\sm{(2)}}$ be two tree-dimensional, orthogonal, unit-vectors perpendicular to the momentum $k_i$ of the gravitational field. Then, any transverse and traceless tensor $H_{ij}$ can be decomposed as
\begin{eqnarray*}
H_{ij} = h_{\times} e_{ij}^{\sm{(1)}} + h_{+} e_{ij}^{\sm{(2)}} ,
\end{eqnarray*}
with the two following polarization tensors
\begin{eqnarray*}
e_{ij}^{\sm{(1)}} = \dfrac{1}{\sqrt{2}} \left( e_{i}^{\sm{(1)}} e_{j}^{\sm{(2)}} + e_{i}^{\sm{(2)}} e_{j}^{\sm{(1)}} \right) , & & e_{ij}^{\sm{(2)}} = \dfrac{1}{\sqrt{2}} \left( e_{i}^{\sm{(1)}} e_{j}^{\sm{(1)}} - e_{i}^{\sm{(2)}} e_{j}^{\sm{(2)}} \right).
\end{eqnarray*}
It is straightforward to see that this decomposition verifies the four relations (\ref{eq:FP_Hij}), and that these polarization tensors satisfy \mbox{$e_{ij}^{\sm{(a)}} e_{ij}^{\sm{(b)}} = \delta^{\sm{(ab)}}$}, where summation over $(a)$ indices is performed with the Euclidean, two-dimensional metric. Moreover, any transverse vector field $v_i$ could also be written as
\begin{eqnarray*}
\partial_i v_i = 0 \rightarrow & v_i = v_{\sm{(a)}} e_{i}^{\sm{(a)}} \rightarrow & v_i v_i = v_{\sm{(a)}} v_{\sm{(a)}}.
\end{eqnarray*}

The notations introduced here will be of constant use in this thesis to describe perturbations in Lorentz-invariant models of massive gravity as well as in Lorentz-violating theories. They will be generalized as to describe perturbations about a flat Friedmann-Lema\^{\i}tre-Robertson-Walker (FLRW) background in chapter \ref{ch:cosmo}.

\subsection{Getting around ghosts}

The previous notations enable one to simplify the analysis of the field content of the theory \refp{eq:FP_LIMG_action} because they preserve the rotational invariance of the Euclidean space. Therefore, if $\tilde{h}_{\mu\nu}$ and $\xi^\mu$ are decomposed according to the notations introduced previously (\ref{eq:Pert_Mk}), the Lagrangian of the massive gravitational field decomposes into scalar, vector and tensor parts
\begin{eqnarray*}
\lag = \lag_{\tx{tensor}} + \lag_{\tx{vector}} + \lag_{\tx{scalar}} \,.
\end{eqnarray*}
These tree sectors could then be consider separately.

Let us start with the tensor sector. This sector describes the usual transverse and traceless graviton modes already present in GR (although they are massless in GR). They correspond to degrees of freedom with helecities $\pm 2$, and have a mass equal to $m$
\begin{eqnarray*}
\lag_{\tx{tensor}} = - \dfrac{1}{4} H_{ij} \left( \Box + m^2 \right) H_{ij} ,
\end{eqnarray*}
where $\Box = \partial^\mu \partial_\mu$. This sector has healthy kinetic term and is free of tachyons provided that $m^2 > 0$. Indeed, by making use of the decomposition of $H_{ij}$ into its two polarization states, the tensor Lagrangian reads
\begin{eqnarray*}
\lag_{\tx{tensor}} = - \dfrac{1}{4} h_{\times} \left( \Box + m^2 \right) h_{\times} - \dfrac{1}{4} h_{+} \left( \Box + m^2 \right) h_{+}.
\end{eqnarray*}
Hence, we will assume that $m^2 > 0$ in what follows to avoid tachyon instabilities.

The vector part of the theory is described by the following Lagrangian
\begin{eqnarray*}
\lag_{\tx{vector}} = \dfrac{1}{2} \left[ - \varpi_{i} \partial_k^2 \varpi_{i} + m^2 \left( \varpi_{i}^2 - \left( \partial_{k } \sigma_{i} \right)^2 + 2 \varpi_{i} \dot{\sigma}_{i} + \dot{\sigma}_{i}^2 \right) \right] .
\end{eqnarray*}
As can be seen from this relation, $\varpi_{i}$ is a non dynamical field since it appears without time derivatives in the Lagrangian. This field can therefore be integrated out. Indeed, recalling that $\varpi_{i}^2 = \varpi_{\sm{(a)}}^2$ and that $\varpi_{i} \sigma_{i} = \varpi_{\sm{(a)}} \sigma_{\sm{(a)}}$, the variation of the action with respect to $\varpi_{\sm{(a)}}$ gives a vectorial equation\footnote{The variation of the quadratic Lagrangian with respect to $\varpi_{\sm{(a)}}$ gives $\left( \partial_k^2 - m^2 \right) \varpi_{\sm{(a)}}  - m^2 \dot{\sigma}_{\sm{(a)}} = 0$ which is a vectorial equation with indices $(a) = 1,2$. Since the vector fields are transverse, $\varpi_i = \varpi_{\sm{(a)}} e_{i}^{\sm{(a)}}$ and $\sigma_i = \sigma_{\sm{(a)}} e_{i}^{\sm{(a)}}$, this equation implies $\left( \partial_k^2 - m^2 \right) \varpi_{i}  - m^2 \dot{\sigma}_{i} = 0$ which is a vectorial equation with indices $i=1,2,3$.}
\begin{eqnarray*}
0 = \left( \partial_k^2 - m^2 \right) \varpi_{i}  - m^2 \dot{\sigma}_{i} ,
\end{eqnarray*}
which can be used to express $\varpi_{i}$ in terms of $\sigma_{i}$ in the Lagrangian. Note that in GR, this equation implies that $\varpi_{i} = 0$, and there are then no propagating modes of helicity $\pm 1$. For the massive case, the vector part of the theory becomes
\begin{eqnarray*}
\lag_{\tx{vector}} = - \dfrac{m^2}{2} \left[ \ddot{\sigma}_{i} \dfrac{\partial_k^2}{\partial_k^2 - m^2} \sigma_{i} - \sigma_{i} \partial_{k}^2 \sigma_{i} \right] .
\end{eqnarray*}
In the momentum space
\begin{eqnarray*}
\dfrac{\partial_k^2}{\partial_k^2 - m^2} \rightarrow \dfrac{k^2}{k^2 + m^2} > 0 ,
\end{eqnarray*}
where $k = \sqrt{k_i^2}$ is the three-dimensional momentum. As a consequence, the term with time derivatives has the correct sign in the previous Lagrangian provided that $m^2 > 0$. Hence this conditions does not only guarantee that the tensor degrees of freedom are not tachyons, but also that there are no ghosts (fields with wrong sign of the kinetic terms) in the vector part of the theory. The following redefinition field
\begin{eqnarray*}
\sigma_{i}^{c} = m \, \sqrt{\dfrac{2 \partial_k^2}{\partial_k^2 - m^2}} \, \sigma_{i},
\end{eqnarray*}
enables one to write the action of the vector sector in the canonical form
\begin{eqnarray} \label{eq:FP_LIMG_vect}
\lag_{\tx{vector}} = - \dfrac{1}{4} \sigma_{i}^{c} \left( \Box + m^2 \right) \sigma_{i}^{c} .
\end{eqnarray}
The vector sector becomes dynamical since the Lorentz - invariant mass terms introduced in the quadratic Lagrangian (\ref{eq:FP_LIMG_action}) contain kinetic term for the transverse vector fields. Hence, the vector Lagrangian describes two propagating modes of mass $m$ and helicities $\pm 1$.

Finally, the scalar sector describes the helicity $0$ degree of freedom. 
For a general value of $\alpha$, there are two dynamical modes in the scalar sector whose  Lagrangian is given by
\begin{eqnarray*}
\lag_{\tx{scalar}} &=& 6 \Psi \ddot{\Psi} + 2 \Psi \partial_{i}^2 \left( 2 \Phi - \Psi \right) - m^2 \left[ \left( \Phi - \dot{\Xi}^{0} \right)^2 - \dfrac{1}{2} \left( \partial_{i} \Xi^0 - \partial_{i} \dot{\Xi} \right)^2 + \left( \Psi \delta_{ij} + \partial_{i} \partial_{j} \Xi \right)^2 \right] \nonumber \\
& & - m^2 \alpha \left( \Phi - \dot{\Xi}^{0} - 3 \Psi - \partial_{k}^2 \Xi \right)^2 .
\end{eqnarray*}
In GR, $m = 0$ and $\Phi$ is a non-dynamical field. Still, this field appears linearly in the Einstein-Hilbert Lagrangian with consequence that it can not be just integrated out as it was the case for the vector field $\varpi_{i}$. Indeed, varying the action of GR with respect to $\Phi$ gives an equation which is actually a constraint on $\Psi$
\begin{eqnarray*}
\partial_{k}^2 \Psi = 0,
\end{eqnarray*}
and which could therefore not be used to integrate $\Phi$ out of the Lagrangian. This situation occurs in GR because $\Phi$ is in fact a Lagrange multiplier imposing a constraint on $\Psi$. Using this constraint, the equation obtained by varying the Lagrangian with respect to $\Psi$ becomes $\partial_{k}^2 \Phi = 0$. Hence, there is no propagating modes in the scalar sector of GR.

The situation is different for the massive gravitational field. The Lagrangian of the scalar sector can be written as follow
\begin{eqnarray} \label{eq:FP_LIMG_scalar}
\lag_{\tx{scalar}} &=& 6 \Psi \ddot{\Psi} + 2 \Psi \partial_{i}^2 \left( 2 \mathcal{A} + 2 \dot{\mathcal{B}} + 2 \ddot{\Xi} - \Psi \right) - m^2 \left[ \mathcal{A}^2 - \dfrac{1}{2} \left( \partial_{i} \mathcal{B} \right)^2 + \left( \Psi \delta_{ij} + \partial_{i} \partial_{j} \Xi \right)^2 \right] \nonumber \\
& & - m^2 \alpha \left( \mathcal{A} - 3 \Psi - \partial_{k}^2 \Xi \right)^2 ,
\end{eqnarray}
where $\mathcal{A} \equiv \Phi - \dot{\Xi}^{0}$ and $\mathcal{B} \equiv \Xi^0 - \dot{\Xi}$. In the massive case, the role of $\Phi$ is played by the field $\mathcal{A}$, which is not a dynamical fields since it enters the action without time derivatives. Let us first discuss the class of models for which $\alpha \neq - 1$. As can be seen from (\ref{eq:FP_LIMG_scalar}), $\mathcal{A}$ is not a Lagrange multiplier; it appears quadratically in the Lagrangian. Therefore, provided that $\alpha \neq - 1$, the equation obtained by varying the Lagrangian with respect to $\mathcal{A}$ gives a relation which can be used to integrate $\mathcal{A}$ out. Since there is no terms with $\mathcal{B}$ and two time derivatives, this field is not dynamical either implying that it can also be integrated out. The equations obtained by varying the action with respect to $\mathcal{A}$ and $\mathcal{B}$ are
\begin{eqnarray*}
0 = 2 \partial_{k}^2 \Psi - m^2 \left( \alpha + 1 \right) \mathcal{A} + m^2 \alpha \left( 3 \Psi + \partial_{k}^2 \Xi \right) , & & 0 = 4 \dot{\Psi} + m^2 \mathcal{B} .
\end{eqnarray*}
By making use of these two equations, the Lagrangian of the scalar sector becomes
\begin{eqnarray*}
\lag_{\tx{scalar}} &=& 6 \Psi \ddot{\Psi} + 2 \Psi \partial_{i}^2 \left( \dfrac{2}{m^2 \left( \alpha + 1 \right)} \partial_{k}^2 \Psi - \dfrac{4 \ddot{\Psi}}{m^2} + 2 \ddot{\Xi} - \Psi + \dfrac{6 \alpha}{\left( \alpha + 1 \right)} \left( 3 \Psi + \partial_{k}^2 \Xi \right) \right) \\
& & - m^2 \left( \Psi \delta_{ij} + \partial_{i} \partial_{j} \Xi \right)^2 - m^2 \dfrac{\alpha}{\alpha + 1} \left( 3 \Psi + \partial_{k}^2 \Xi \right)^2.
\end{eqnarray*}
This Lagrangian looks quite complicated. But at this point it is clear that both $\Psi$ and $\Xi$ are dynamical fields. One of them is a ghost, i.e. a field with the wrong sign of the kinetic terms. This can be seen from the terms with time derivatives. These terms can schematically be written as
\begin{eqnarray*}
\lag_{\tx{time}} = \dot{\Psi} \left( a \dot{\Psi} + 2 b \dot{\Xi} \right) = a \left[ \left( \dot{\Psi} + \dfrac{b}{a} \dot{\Xi} \right)^2 - \dfrac{b^2}{a^2} \dot{\Xi}^2 \right] ,
\end{eqnarray*}
where $a$ and $b$ are some constants. Depending of the sign of $a$, either $\Psi + \dfrac{b}{a} \Xi$ or $\Xi$ is a ghost, since these two fields have an opposite sign in front of their kinetic term. Therefore, Lorentz-invariant massive gravity models with $\alpha \neq - 1$ have pathological behavior at arbitrarily high spatial momentum.

The Lorentz-invariant massive gravity model characterized by $\alpha = - 1$ is known as the Fierz-Pauli (FP) theory \cite{Fierz:1939ix}. In this particular model, there is no quadratic terms in $\mathcal{A}$ in the Lagrangian \refp{eq:FP_LIMG_scalar}. This field is therefore a Lagrange multiplier as in GR, imposing the following relation between $\Psi$ and $\Xi$
\begin{eqnarray*}
0 = 2 \partial_{k}^2 \Psi - m^2 \left( 3 \Psi + \partial_{k}^2 \Xi \right) .
\end{eqnarray*}
This constraint kills one dynamical field out of two, leaving only one helicity-$0$ mode in the Lagrangian. Using this constraint, and after integrating out $\mathcal{B}$ as previously, the Lagrangian of the scalar sector becomes
\begin{eqnarray} \label{eq:FP_LIMG_psi}
\lag_{\tx{scalar}} = - 6 \Psi \left( \Box + m^2 \right) \Psi .
\end{eqnarray}
This Lagrangian is healthy and describes one scalar propagating mode corresponding to the helicity $0$ mode of the massive graviton. Therefore, the Lorentz-invariant theory \refp{eq:FP_LIMG_action} with $m^2 > 0$ and $\alpha = - 1$ describes five dynamical modes corresponding to the five helicity states of a massive spin-2 graviton.

\subsection{The Goldstone sector} \label{sc:FP_Goldst}

The previous conclusion concerning the vector and scalar sector could have been found directly from the study of the Goldstone sector of the theory. The four Goldstone fields have been introduced to restore the gauge invariance of the massive gravitational field. Yet, there is another motivation for the introduction of these fields: these scalar fields give a convenient way to single out and study the dangerous degrees of freedom of the theory \cite{Arkani-Hamed:2002sp,Dubovsky:2004sg}. Since the Goldstone fields only appear in the vector and scalar sectors, let us concentrate on these two sectors, described schematically by the following Lagrangian
\begin{eqnarray*}
\lag_{\tx{vector}} + \lag_{\tx{scalar}} &=& h_{\mu\nu} \left( \mathcal{D}^{\mu\nu\alpha\beta} + M_{h}^{\mu\nu\alpha\beta} \right) h_{\alpha\beta} + 2 M_{m}^{\mu\nu\alpha} h_{\mu\nu} \xi_{\alpha} + \xi_{\mu} D_{G}^{\mu\nu} \xi_{\nu},
\end{eqnarray*}
where $\mathcal{D}$ is the usual two-derivative graviton kinetic operator coming from the Einstein-Hilbert Lagrangian, $M_{h}$ accounts for the metric mass terms, $M_{m}$ is a one-derivative operator describing the mixing between the metric and Goldstone fields, and finally $D_{G}$ is a two-derivative operator describing the kinetic terms for the Goldstone fields. The field equations read
\begin{eqnarray*}
\left( \begin{array}{cc}
\mathcal{D}^{\mu\nu\alpha\beta} + M_{h}^{\mu\nu\alpha\beta} & M_{m}^{\mu\nu\alpha} \\
M_{m}^{\alpha\beta\mu} & D_{G}^{\mu\alpha}
\end{array} \right) \left( \begin{array}{c}
h_{\alpha\beta} \\
\xi_{\alpha}
\end{array} \right) = 0 .
\end{eqnarray*}
Here, one assume that the gauge symmetry is fixed in the gravity sector, by choosing for example the \emph{de Donder} gauge\footnote{In the \emph{de Donder} (or harmonic) gauge, the perturbations of the metric satisfy $\partial^\nu \left( 2 h_{\mu\nu} - \eta_{\mu\nu} h \right) = 0$.}. Then, the Einstein-Hilbert operator $\mathcal{D}$ is non-degenerate and the first set of equations may be used to express the metric perturbations in terms of the Goldstone fields
\begin{eqnarray*}
h_{\lambda\gamma} = - \dfrac{1}{\mathcal{D}^{\lambda\gamma\mu\nu} + M_{h}^{\lambda\gamma\mu\nu}} M_{m}^{\mu\nu\alpha} \xi_{\alpha}.
\end{eqnarray*}
This therefore gives a closed set of equations for $\xi_{\alpha}$
\begin{eqnarray*}
\left( D_{G}^{\alpha\beta} - M_{m}^{\lambda\gamma\alpha} \dfrac{1}{\mathcal{D}^{\lambda\gamma\mu\nu} + M_{h}^{\lambda\gamma\mu\nu}} M_{m}^{\mu\nu\beta} \right) \xi_{\beta} = 0 .
\end{eqnarray*}
For momenta larger than the mass, $k \gg m$, the mass matrix $M_{h}$ could be neglected with respect to $\mathcal{D}$. Then, the second term in the last equation is of order $m^4$ while the first term is of order $(\omega^2, k^2) \times m^2$. Consequently, when working at $k \gg m$, the field equations reduce to
\begin{eqnarray*}
D_{G}^{\alpha\beta} \xi_{\beta} = 0 ,
\end{eqnarray*}
equations which are given by the Goldstone sector of the theory
\begin{eqnarray*}
\lag_{\tx{Gold.}} = \xi_{\mu} D_{G}^{\mu\nu} \xi_{\nu} .
\end{eqnarray*}

This short calculation shows that neglecting the metric perturbations is legitimate as long as modes with $k^2 \gg m^2$ are considered (except for the tensor modes which are completely described by the metric perturbations). This result is compatible with the previous analysis of the vector and scalar sector of the Lorentz-invariant theory. Indeed, with the decomposition (\ref{eq:Pert_Mk}) into transverse and longitudinal fields, the Goldstone sector of the Lagrangian \refp{eq:FP_LIMG_action} reduces to
\begin{eqnarray*}
\lag_{\tx{Gold.}} = - \dfrac{m^2}{2} \left[ \xi_{i}^{T} \Box \xi_{i}^{T} + 2 \left( 1 + \alpha \right) \left( \dot{\xi}_{0}^2 + \left( \partial_{i}^2 \xi \right)^2 \right) - \left( \partial_{i} \xi_{0} \right)^2 - \left( \partial_i \dot{\xi} \right)^2 + 2 \left( 1 + 2 \alpha \right) \dot{\xi}_{0} \partial_{i}^2 \xi \right] .
\end{eqnarray*}
This relation makes it clear that the $\xi_{i}^{T}$-terms are perfectly healthy provided that $m^2 > 0$, and describe two propagating modes of helicity $\pm 1$. This vector sector is exactly the high-momentum $k^2 \gg m^2$ limit of the vector sector \refp{eq:FP_LIMG_vect} which has been discussed previously.

In order to discuss the scalar part of the Goldstone sector, let us focus first on the theories for which $\alpha \neq - 1$. In the four-momentum space, the scalar-Goldstone sector is given by
\begin{eqnarray*}
\lag_{\tx{Gold. sc.}} = \dfrac{m^2}{2} \left( \begin{array}{cc}
\xi^{*}_0 & k \xi^{*}
\end{array} \right) M \left( \begin{array}{c}
\xi_0 \\ k \xi
\end{array} \right) , & & M = \left( \begin{array}{cc}
- 2 \left( 1 + \alpha \right) \omega^2 + k^2 & i \left( 1 + 2 \alpha \right) \omega k \\
- i \left( 1 + 2 \alpha \right) \omega k & \omega^2 - 2 \left( 1 + \alpha \right) k^2
\end{array} \right) .
\end{eqnarray*}
The dispersion relations of these two scalar fields are obtained by requiring that $\det M = 0$
\begin{eqnarray*}
\det M = - 2 \left( 1 + \alpha \right) \left( \omega^2 - k^2 \right)^2 = 0 &\rightarrow& \omega^2 = k^2 .
\end{eqnarray*}
This relation clearly shows that there are two dynamical modes in the scalar-Goldstone sector, characterized by $w^2 = k^2$. Noting that $\det M = \lambda_{+} \lambda_{-}$, where $\lambda_{\pm}$ are the eigenvalues of $M$, the dispersion relations of the scalar-Goldstone fields could also be determined by finding the solutions to the following equations
\begin{eqnarray} \label{eq:Goldst_eig}
\lambda_{\pm} \left( \omega^2 \right) = 0 &\tx{with}& \lambda_{\pm} = \dfrac{\tx{Tr} M \pm \sqrt{(\tx{Tr} M)^2 - 4 \det M}}{2} .
\end{eqnarray}
Two conditions have to be met in order for the Goldstone scalar sector to be free of instability. The first one requires $\omega^2 > 0$ to guarantee the absence of classical instability. This condition is satisfied here. The second condition consist in requiring that the eigenvalues of $M$ be positive definite on their roots $\omega_{\pm}$
\begin{eqnarray} \label{eq:Goldst_ghost}
\dfrac{\partial \lambda_{\pm} }{\partial \omega^2} {\Big |}_{\omega^2 = \omega^2_{\pm}} > 0 .
\end{eqnarray}
If satisfied, this second condition implies the absence of ghosts. Therefore, to avoid ghosts it is necessary that each eigenvalue $\lambda_{\pm}$ has one zero, otherwise one of the eigenvalue will have two zeros and one of them will necessary violate the condition (\ref{eq:Goldst_ghost}). This last observation enable to understand whether there is a ghost in the Goldstone scalar sector or not. Indeed, let us relate the dispersion relations to the eigenvalues of $M$.
\begin{itemize}
\item If $\alpha > - 1 / 2$, the two dynamical modes are both solutions of the equations $\lambda_{+} = 0$, since the trace of $M$ is negative for these modes. Therefore, $\lambda_{-}$ has no zero while $\lambda_{+}$ has two, implying that one of them is necessarily a ghost.
\item If $\alpha < - 1 / 2$, the trace of $M$ is positive when $\omega^2 = k^2$ implying that the two dynamical modes are solutions of $\lambda_{-} = 0$. Hence, there is also a ghost for these value of $\alpha$.
\item If $\alpha = - 1 / 2$, $\lambda_{+}$ and $\lambda_{-}$ both have one zero on $\omega^2 = k^2$.
\end{itemize}
This discussion makes clear that, depending of the value of $\alpha$ (and recalling that we have excluded $\alpha = - 1$ from the beginning), there are two groups of solutions. First, if $\alpha = - 1 / 2$, the two propagating modes are associated to different eigenvalues, and it is straightforward to see from the matrix $M$ that in this case $\xi_{0}$ is a ghost-like field while $\xi$ has a healthy kinetic behavior. For the second group of solution characterized by $\alpha \neq - 1 / 2$, the two propagating modes are associated to the same eigenvalue implying that one of the modes is a ghost\footnote{This situation is similar to what happens in theories with four derivatives acting on a single scalar field. It is well known that such models posses ghost \cite{Pais:1950za,deUrries:1998bi}.}. One should therefore conclude that there is always a ghost like mode in Lorentz-invariant massive gravity. Yet, there is one special case for which this conclusion does not hold. Indeed, the only way to get rid of this pathological behavior is to choose $\alpha = - 1$. In this particular model corresponding to the FP theory, there is no dynamical scalar mode in the Goldstone Lagrangian.

The previous discussion shows that the study of the Goldstone sector is a powerful tool to analyze the field content of a theory. Yet, we have just shown that for $\alpha = - 1$ there are no helicity $0$ mode in the Goldstone sector while we have explicitly found such a mode in the previous section (\ref{eq:FP_LIMG_psi}). This apparent contradiction is a particularity of the FP theory, in which the kinetic terms for the scalar mode are originated in the mixing between the metric perturbations and the Goldstone fields (recall that $\mathcal{A}$ is a Lagrange multiplier which impose a constraint between these terms). Therefore, the only scalar terms present in the Goldstone sector of the FP theory are mass terms which are negligible with respect to the kinetic terms coming from the mixing, and there is no contradiction. This particularity of the FP theory is directly related to the vDVZ discontinuity of this theory, which will be discussed in the next section.

Before concluding the discussion of the Goldstone sector, it is important to remember that this sector is the high energy limit of the massive theory, corresponding to the $k^2 \gg m^2$ limit. Therefore, by studying the Goldstone sector we could miss some other modes with more slowly oscillating behavior. This remark should be kept in mind for the coming study of Lorentz-violating models.

\subsection{Interaction between two sources}

To agree with Solar System observations, the FP theory of gravity should reproduce GR predictions in the limit of vanishing graviton mass, since GR is quite successful in describing the whole Solar System. The predictions of the massive theory can be compared to those of GR by studying the interaction between two massive bodies, or between one massive body and light. For this purpose, let $\mathcal{T}$ and $\mathcal{T}'$ be two small, conserved energy-momentum tensors, corresponding for example to two massive bodies. These two energy-momentum tensors both source gravity through the linearized gravitational equation. For both GR and the FP theory in the unitary gauge (in which there is no scalar fields), the gravitational field produced by $\mathcal{T}$ is given by
\begin{eqnarray*}
h_{\mu\nu} = \mathcal{P}^{-1}_{\mu\nu\alpha\beta} \mathcal{T}^{\alpha\beta},
\end{eqnarray*}
where $\mathcal{P}^{-1}_{\mu\nu\alpha\beta}$ is the graviton propagator, obtained by inverting the equations of motion (EoM)\footnote{This could be done with the spin formalism of Barnes and Rivers \cite{Rivers:1964,VanNieuwenhuizen:1973fi}, but also with the transverse and longitudinal decomposition introduced previously. Indeed, the propagator could easily be calculated by adding sources in each of the three sectors of the theory, and then inverting the EoM.}. The interaction between this gravitational field and $\mathcal{T}'$ is given by the action of this gravitational source \cite{Boulware:1973my} which contains all interaction terms between the metric and $\mathcal{T}'$
\begin{eqnarray*}
\mathcal{S}_{\tx{int.}} \equiv \int \dif^4 x \, \mathcal{T}'^{\mu\nu} h_{\mu\nu} = \int \dif^4 x \, \mathcal{T}'^{\mu\nu} \mathcal{P}^{-1}_{\mu\nu\alpha\beta} \mathcal{T}^{\alpha\beta} .
\end{eqnarray*}
From this relation, it is straightforward to understand that it is sufficient to compare the graviton propagator in both theories to see if GR and the FP theory have similar predictions. After a four-dimensional Fourier transformation, the graviton propagator reads
\begin{center}
\begin{tabular}{lll}
GR : & & $\mathcal{P}^{-1}_{\mu\nu\alpha\beta} = \dfrac{1}{\Mpl k_{\lambda} k^{\lambda}} \left[ \eta_{\mu\alpha} \eta_{\nu\beta} + \eta_{\mu\beta} \eta_{\nu\alpha} - \eta_{\mu\nu} \eta_{\alpha\beta} + \tx{$k$-dependent terms} \right]$ , \\
FP : & & $\mathcal{P}^{-1}_{\mu\nu\alpha\beta} = \dfrac{1}{\Mpl \left(k_{\lambda} k^{\lambda} - m^2 \right)} \left[ \eta_{\mu\alpha} \eta_{\nu\beta} + \eta_{\mu\beta} \eta_{\nu\alpha} - \dfrac{2}{3} \eta_{\mu\nu} \eta_{\alpha\beta} + \tx{$k$-dependent terms} \right]$ .
\end{tabular}
\end{center}
The $k$-dependent terms are gauge-dependent terms. They give a null contribution to $\mathcal{S}_{\tx{int.}}$ since they are contracted with conserved energy-momentum tensors. Apart from these terms and from the presence of a non-zero mass, we see that the propagator in the FP theory differs from Einstein's one by a factor of $2/3$ in front of the third term. This difference, which is mass-independent and therefore remains when the limit $m \rightarrow 0$ is considered, is known as the \emph{van Dam-Veltman-Zakharov} discontinuity \cite{vanDam:1970vg,Zakharov:1970cc,PhysRevD.2.2255}, or vDVZ discontinuity. This discontinuity implies that the addition of FP mass terms in the action of the gravitational field changes drastically the interaction between two arbitrary sources, even for arbitrarily small masses.

The origin of this discontinuity can be traced back to the scalar sector of the FP model. Indeed, as discussed previously, the FP theory possesses five propagating modes corresponding to the five helicity states of the massive graviton: two tensor modes already present in GR, two vector modes and one scalar. It follows from the definition of the canonical vector field $\sigma_{i}^{c}$ that the vector Lagrangian \refp{eq:FP_LIMG_vect} reduces to zero in the limit of vanishing graviton mass; i.e. the vector modes get kinetic terms from the Goldstone sector and decouple in the limit $m \rightarrow 0$. Therefore only the standard contribution coming from GR survives in the limit of vanishing mass. The situation is completely different in the scalar sector. The kinetic terms for the scalar mode come from the mixing between $h_{\mu\nu}$ and the Goldstones. For this reason, the scalar mode does not decouple in the limit of vanishing mass, and the Lagrangian of the scalar sector does not reduce to zero when $m \rightarrow 0$. Therefore, the FP theory reduces to a scalar-tensor theory of the Jordan-Brans-Dicke type \cite{Brans:1961sx}. Indeed, in the presence of a conserved source $\mathcal{T}$, the Lagrangian (\ref{eq:FP_LIMG_psi}) of the scalar sector generalizes to
\begin{eqnarray*}
\lag_{\tx{scalar}} = - 6 \Psi \left( \Box + m^2 \right) \Psi + \dfrac{2 \Psi}{\Mpl} \left( \dfrac{3 \Box}{\partial_i^2} \mathcal{T}_{00} - \mathcal{T} \right) ,
\end{eqnarray*}
where $\mathcal{T} = \mathcal{T}_\mu^\mu$. We conclude from this Lagrangian that, in the limit $m \rightarrow 0$, the $\Psi$ field becomes
\begin{eqnarray} \label{eq:FP_Psi_matter}
\Psi = \dfrac{1}{\Mpl} \left( \dfrac{\Box}{\Box + m^2} \dfrac{1}{2 \partial_i^2} \mathcal{T}_{00} - \dfrac{1}{6 \left( \Box + m^2 \right)} \mathcal{T} \right) &\rightarrow& \Psi = \dfrac{1}{\Mpl} \left( \dfrac{1}{2 \partial_i^2} \mathcal{T}_{00} - \dfrac{1}{6 \Box} \mathcal{T} \right) ,
\end{eqnarray}
where $\Box^{-1}$ and $\left( \Box + m^2 \right)^{-1}$ have to be understood as the Green functions of the d'Alembert and Klein-Gordon equations, respectively. The first term in this relation is the usual GR contribution to $\Psi$, while the second, proportional to the trace of the energy-momentum tensor, is a new contribution. Along with the contribution coming from $\Phi$ and which is not independent, this new term is responsible for a $1/3 \eta_{\mu\nu} \eta_{\alpha\beta}$ contribution in the propagator, which combined to the contributions of the tensor modes gives the abnormal $2/3$ factor.

The third term in the propagator is responsible for the coupling between the traces of the two energy-momentum tensors. Therefore, this extra term will contribute in the interaction between two massive bodies, but not in the interaction of a massive body and a light beam since $\mathcal{T}_{\mu}^{\mu}$ is identically zero for light\footnote{The energy-momentum tensor of light is given by $\mathcal{T} = \tx{diag.} \left( \rho,- p,- p,- p \right)$. This tensor is trace-less. Indeed, $\rho$ and $p$ satisfy the following equation of state $3 p = - \rho$.}. To illustrate the prediction differences between the Einstein theory and the FP one, let us assume that $\mathcal{T}$ and $\mathcal{T}'$ correspond to two static massive bodies. For such bodies, the only non-zero component of their energy-momentum tensor is the zero-zero component. Therefore, the gravitational interaction between two static massive bodies is given by
\begin{center}
\begin{tabular}{lll}
GR : & & $\mathcal{T}^{\mu\nu} \mathcal{P}^{-1}_{\mu\nu\alpha\beta} \mathcal{T}^{\alpha\beta} = \dfrac{\mathcal{T}_{00} \mathcal{T}_{00}'}{\Mpl k_{\lambda} k^{\lambda}}$ , \\
FP : & & $\mathcal{T}^{\mu\nu} \mathcal{P}^{-1}_{\mu\nu\alpha\beta} \mathcal{T}^{\alpha\beta} = \dfrac{4}{3} \dfrac{\mathcal{T}_{00} \mathcal{T}_{00}'}{\Mpl \left( k_{\lambda} k^{\lambda} - m^2 \right)}$ .
\end{tabular}
\end{center}
This implies that, in the most favorable situation corresponding to the $m \rightarrow 0$ limit, the FP theory predict an interaction which is $4/3$ stronger than GR, unless the constant $\Mpl$ in the FP theory is $4/3$ bigger than its value in GR. This will re-conciliate the FP model with the Newtonian potential, and therefore with the observations of gravitational interaction between massive bodies, but will give a wrong angle deflection for light beam passing close to the Sun. Indeed, the energy-momentum tensor of light is trace-less. Therefore, the interaction between a massive static body and a beam of light is given by
\begin{center}
\begin{tabular}{lll}
GR : & & $\mathcal{T}^{\mu\nu} \mathcal{P}^{-1}_{\mu\nu\alpha\beta} \mathcal{T}^{\alpha\beta} = \dfrac{2 \mathcal{T}_{00} \mathcal{T}_{00}'}{\Mpl k_{\lambda} k^{\lambda}}$ , \\
FP : & & $\mathcal{T}^{\mu\nu} \mathcal{P}^{-1}_{\mu\nu\alpha\beta} \mathcal{T}^{\alpha\beta} = \dfrac{2 \mathcal{T}_{00} \mathcal{T}_{00}'}{\Mpl \left( k_{\lambda} k^{\lambda} - m^2 \right)}$ .
\end{tabular}
\end{center}
Then, if $\Mpl$ is $4/3$ bigger in the FP theory than its value in GR, these last relations implies that the interaction between a massive body and a light beam is $3/4$ the value predicted by GR and confirmed by observations.

\subsection{Breakdown of linearity} \label{sc:FP_vainst}

The vDVZ discontinuity seems to indicate that the FP theory can be ruled out by Solar System observations, since this theory predicts a deflection of light by the Sun which is $25\%$ smaller than the value predicted by GR and confirmed by observations. In fact, this conclusion is not trustworthy \cite{Vainshtein:1972sx}. Indeed, the appearance of the vDVZ discontinuity relies on a linear approximation of a non linear theory. In GR, this approximation is valid in the entire Solar-System, while it is possible to show that the linearized FP theory is not valid in the Solar System for arbitrarily small masses.

In both theories, GR and FP, the metric around the Sun can be linearized provided that there is a small parameter $\epsilon$ in the metric to support the Taylor expansion
\begin{eqnarray*}
\dif s^2 = \left[ g_{\mu\nu} + h_{\mu\nu} + \mathcal{O} \left( \epsilon^2 \right) \right] \dif x^\mu \dif x^\nu , & & | h_{\mu\nu} | \sim \epsilon .
\end{eqnarray*}
Then, if $\epsilon \ll 1$ the second order terms could be neglected as compared to the first order terms, and the gravitational interaction is correctly described by the linearized approximation. In GR, the metric around the Sun is given to a very good approximation by the Schwarzschild solution, and the linear approximation is valid at distances
\begin{eqnarray*}
r \gg 2 G M_{\odot} \sim 3 \, \tx{km} .
\end{eqnarray*}
Therefore, in GR the gravitational field around the Sun is correctly described by the linearized approximation for distances much larger than $3 \, \tx{km}$, which corresponds to the Schwarzschild radius of the Sun and which is actually much smaller than the radius of the Sun.

In the FP theory, the small parameter controlling the validity of the linearized approximation is $r_v / r$, where $r_v$ is the Vainshtein radius \cite{Vainshtein:1972sx}. As in GR, the linearized approximation is valid for distances much bigger than $r_v$
\begin{eqnarray*}
r \gg r_v , & & r_v^5 = \dfrac{2 G M_{\odot}}{m^4} .
\end{eqnarray*}
For a graviton mass as big as $m \sim \left( 10^{15} \tx{cm} \right)^{-1}$ (which corresponds more or less to the size of the Solar System), the Vainshtein radius $r_v \sim 10^{8}\, \tx{km}$ is of the same order as the Earth orbit. For a much smaller mass corresponding to the Hubble scale, $m \sim H_{0}^{-1}$, the Vainshtein radius is $r_v \sim 100\, \tx{kpc}$ which is bigger than the actual size of the Milky way.

The Vainshtein radius is singular in the graviton mass, and therefore the higher order corrections to the linearized theory are also singular in the graviton mass. This has for consequence that the linear theory breaks at a distance much larger than the Schwarzschild radius of GR. One has to conclude from this that the linearized FP theory can not describe gravity within the Solar System if the mass of the graviton is too small, since then the distance below which non-linearities become important is too big, and the light passing not far from the Sun will feel the non-linearity of the FP theory. This could solve the problem arising from the vDVZ discontinuity. Indeed, with the account of the full non-linear theory, one may hope that the non-linear completion of the model will modify the theory in such a way as to smooth the massless limit, and reconcile the model with observations. One can for example imagine that the full non-linear solution is continuous in the graviton mass, and still be the summation of singular terms. This is exactly what happens in the DGP model \cite{Dvali:2000hr}, in which the higher order corrections affect the scalar sector and decouple it from other modes in the small mass limit \cite{Deffayet:2001uk,Gruzinov:2001hp}. It is currently debated whether non-linear completion of the FP theory could solve the problems raised by the vDVZ discontinuity, since the full non-linear FP theory is not defined unambiguously. Recently, it has been argued by studying the Goldstone sector of FP theory that for some non-linear models there is a Vainshtein-like recovery of the solutions of GR \cite{Babichev:2009us}. Still, ghost-like mode seems unavoidable when accounting for higher order terms \cite{Creminelli:2005qk}. This ghost can be seen as responsible for the cancellation of the attraction exerted by the non-decoupling scalar mode \cite{Deffayet:2005ys}. But it also indicates that the theory is unstable inside the Vainshtein radius.

Non-linearities could also solve the vDVZ problem by selecting a non-asymptotically flat vacuum \cite{Damour:2002ws,Damour:2002gp}. Indeed, there is no mass discontinuity in de Sitter or Anti - de Sitter space-times \cite{Kogan:2000uy,Porrati:2000cp,Deser:2001wx,Porrati:2001db}.

\subsection{Strong coupling}

Both GR and the FP theory should be treated as low energy effective theories valid at energies below a certain energy scale $\Lambda$, which corresponds to the energy at which an ultra-violet completion of the theory is needed. Below $\Lambda$, all quantum effect are negligible. Above  this scale, terms with higher order derivatives become important and quantum effect could not been neglected. In Einstein theory, it is natural for dimensional reasons to identify this UV energy scale with the Planck mass $\textrm{M}_{\textrm{pl}}$ \cite{Burgess:2003jk}. One would therefore trust GR below this energy scale and for distances bigger than the Planck length $\textrm{M}_{\textrm{pl}}^{-1}$. Above this energy scale, it is supposed that GR has to be extended in a quantum theory of gravity.

In the FP theory, the energy scale at which quantum corrections become important is much lower. This is related to the non-linearity problems of the FP theory, which at the quantum level implies that the theory becomes strongly coupled at an energy scale \cite{Arkani-Hamed:2002sp}
\begin{eqnarray*}
\Lambda = \left( m^4 \textrm{M}_{\textrm{pl}} \right)^{1/5} .
\end{eqnarray*}
Therefore, the FP theory is trustable at distances larger than $\Lambda^{-1}$. For a mass \mbox{$m \sim \left( 10^{15} \tx{cm} \right)^{-1}$} of the order of the Solar System size, the effective theory is valid for distances larger than $4.5\, \tx{km}$. For a much smaller mass corresponding to the Hubble scale, quantum correction are important for distances smaller than $10^{16}\, \tx{cm}$, which encompass the entire Solar System. Since Newton's law has been tested to sub-millimeter scales \cite{Long:2003ta,Long:2003dx,Hoyle:2004cw,Smullin:2005iv}, this strong coupling effect is one of the biggest problems of the FP theory.

The energy scale $\Lambda$ is related to higher order operators (terms with more than two derivatives) which should be present in the UV completion of the theory. Yet, there is another source for such terms. Indeed, the Goldstone fields have been introduced to restore the gauge invariance of the theory. But actually the transformation law of the metric perturbations is not strictly linear\footnote{This assumes that the gauge invariance of the full non-linear theory is the same as in GR.}
\begin{eqnarray*} \label{eq:Gauge_quadr}
\tilde{h}_{\mu\nu} \rightarrow \tilde{h}_{\mu\nu} + \partial_{\mu} \varsigma_{\nu} + \partial_{\nu} \varsigma_{\mu} + \partial_{\mu} \varsigma_{\alpha} \partial_{\nu} \varsigma^{\alpha} + \ldots \,.
\end{eqnarray*}
Therefore, the definition of $\tilde{h}_{\mu\nu}$ should contains terms quadratic in the Goldstone fields, otherwise the theory would not be gauge invariant at the quadratic level. The previous definition of $\tilde{h}_{\mu\nu}$ should then be completed
\begin{eqnarray} \label{eq:FP_def_h_tilde}
h_{\mu\nu} = \tilde{h}_{\mu\nu} - \partial_{\mu} \xi_{\nu} - \partial_{\nu} \xi_{\mu} - \partial_{\mu} \xi_{\alpha} \partial_{\nu} \xi^{\alpha} + \ldots \,.
\end{eqnarray}
Consequently, the Lagrangian of the FP theory actually contains higher order terms which we have neglected up to now. These terms are responsible for the particularly high UV cutoff of the FP theory.

Let us introduce one vector $\hat{\xi}_{\mu}$ and one scalar $\phi$ fields through the following relations
\begin{eqnarray*} \label{eq:FP_U1decomp}
\xi_\mu = \hat{\xi}_{\mu} + \partial_\mu \phi.
\end{eqnarray*}
These notations replace four variable by five. Therefore, this introduces a fake $\mathcal{U}(1)$ symmetry
\begin{eqnarray*}
\hat{\xi}_\mu \rightarrow \hat{\xi}_\mu + \partial_\mu \lambda , & & \phi \rightarrow \phi - \lambda .
\end{eqnarray*}
Then, the kinetic terms of the scalar sector can be diagonalized by the following conformal transformation
\begin{eqnarray*}
\tilde{h}_{\mu\nu} = \hat{h}_{\mu\nu} + \eta_{\mu\nu} m^2 \phi,
\end{eqnarray*}
which eliminates the kinetic mixing between $\hat{h}_{\mu\nu}$ and $\phi$. If $\hat{h}_{\mu\nu}$ and $\hat{\xi}_{\mu}$ are decomposed into transverse and longitudinal fields according to (\ref{eq:Pert_Mk}), the scalar Lagrangian reads\footnote{$\phi$ is a gauge invariant field.}
\begin{eqnarray*}
\lag_{\tx{scalar}} = - 6 \left( \Psi - \dfrac{m^2}{2} \phi \right) \left( \Box + m^2 \right) \left( \Psi - \dfrac{m^2}{2} \phi \right)
\end{eqnarray*}
The $\mathcal{U}(1)$ symmetry enable to cancel either $\Psi$ or $\phi$, since these fields transform as follows:
\begin{eqnarray*}
\Psi \rightarrow \Psi - \dfrac{m^2}{2} \lambda , & & \phi \rightarrow \phi - \lambda.
\end{eqnarray*}
Let $\lambda$ be such that $\Psi = 0$. The higher-order interaction terms for the Goldstone fields are all $\mathcal{U}(1)$-invariant. Therefore, with the addition of these higher-order operators, and recalling that there is an overall $\Mpl$ factor in the definition of the Lagrangian, the scalar Lagrangian is
\begin{eqnarray} \label{eq:FP_Lag_phi}
\Mpl \mathcal{L}_{\tx{scalar}} = - \dfrac{3}{2} \phi^c \left( \Box + m^2 \right) \phi^c - \dfrac{1}{2 \Lambda^5} \Box \phi^c \left( \partial_\mu \partial_\nu \phi^c \partial^\mu \partial^\nu \phi^c - \left( \Box \phi^c \right)^2 \right) + \ldots \, ,
\end{eqnarray}
where the following rescaling $\phi^c = m^2 \tx{M}_{\tx{pl}} \phi$ has been made in order to get the canonically normalized scalar Lagrangian. It is obvious from this relation that the cubic interaction terms for $\phi$, coming from the definition of $\tilde{h}_{\mu\nu}$, are suppressed by a parameter which is precisely $\Lambda^5 = m^4 \tx{M}_{\tx{pl}}$. Since other higher order operators, coming from the non-linear and from the UV completions of the linearized theory, are less coupled than the cubic ones \cite{Arkani-Hamed:2002sp}, $\Lambda$ correspond to the strong coupling scale. This value of $\Lambda$ has been confirmed by direct calculations of the four-graviton scattering amplitude \cite{Aubert:2003je}.

One can imagine that the UV completion of the theory will bring terms which will exactly cancel these cubic operators. If such a UV completion exist, the strong coupling scale will be lowered to $\Lambda = \left( m^2 \tx{M}_{\tx{pl}} \right)^{1/3}$ which is the lowest that can be achieved
\cite{Arkani-Hamed:2002sp}. For a graviton mass corresponding to the Hubble scale, this new cutoff scale is still too low, $\Lambda = \left( 10^{3}\, \tx{km} \right)^{-1}$. In principle, this low cutoff scale could be raised in a non-flat background due to curvature terms. This occurs in the DGP model \cite{Nicolis:2004qq}, but apparently such effects are absent in the FP theory because of the presence of higher order, ghost-like propagating mode \cite{Creminelli:2005qk,Deffayet:2005ys}

It is worth noting that Lagrangian \refp{eq:FP_Lag_phi} enables one to understand the origin of the Vainshtein radius and its link to the strong coupling problem. Indeed, let us consider a static massive body for which $\mathcal{T}_{\mu\nu} = \delta_{\mu0} \delta_{\nu0} M \delta^{3} \left( x \right)$. Then, if we neglect the higher order operators, it is not difficult to convince ourselves from relation \refp{eq:FP_Psi_matter} that $\phi$ is given by
\begin{eqnarray} \label{eq:FP_phi_back}
\left( \Box + m^2 \right) \phi^c = \dfrac{4}{3 \tx{M}_\tx{pl}} \mathcal{T}_{00} &\rightarrow& \phi^c = \dfrac{M \ex{-m r}}{3 \pi \tx{M}_\tx{pl} r} ,
\end{eqnarray}
where the homogeneous solution has been neglected since it does not play any role in the present discussion. For $r \ll m^{-1}$, this solution can be approximated by $\phi^c \sim M r^{-1}$. Hence, the cubic operators could be neglected as compared to the second order terms if and only if
\begin{eqnarray*}
\phi^c \partial^2 \phi^c \gg \dfrac{\left( \partial^2 \phi \right)^3}{\Lambda^5} &\rightarrow& r^5 \gg \dfrac{M}{\pi \tx{M}_\tx{pl} \Lambda^5} \sim r_V^5.
\end{eqnarray*}
Apart from the coefficient of order one, this is exactly the definition of the Vainshtein radius.

\subsection{Instabilities in curved backgrounds} \label{sc:FP_BD_inst}

In curved background, the relation $\alpha = -1$ is violated by curvature terms. This relation, which defines the FP theory in Minkowski space-time, guarantees that the theory is free of pathological behavior. Its violation implies that ghost instabilities reappear in the spectrum. Such instabilities in curved background are known as \emph{Boulware-Deser} instabilities \cite{Boulware:1973my}. They exist for arbitrarily high spatial momenta and are therefore unacceptable.

The origin of the \emph{Boulware-Deser} instability is easy to understand from the study of the field $\phi$. Let us consider a nearly-flat background\footnote{In Ref.~\cite{Rubakov:2008nh}, these \emph{Boulware-Deser} instabilities are explicitely found and discussed for a Friedmann-Lema\^{\i}tre-Robertson-Walker (FLRW) background. In Ref.~\cite{Dubovsky:2005dw}, this instability is discussed in a fully non-linear generalization of the FP theory through the analyze of the Goldstone sector.}. If the curvature is small enough, the background could be described to a good approximation by the first order perturbation theory about Minkowski solution. Then, the $\phi$ perturbations about this nearly flat background will be given by expanding the Lagrangian for $\phi$ at the quadratic order in the field $\chi \equiv \phi^c - \phi_{b}$, where $\phi_{b}$ is the background. The Lagrangian for $\chi$ will contain four time derivatives. Indeed, cubic terms $(\partial^2 \phi)^3$ in Lagrangian (\ref{eq:FP_Lag_phi}) give $\partial^2 \phi_b (\partial^2 \chi)^2$. Independently of the way that the FP theory is generalized in curved space-time, the quadratic Lagrangian for $\chi$ reads schematically \cite{Creminelli:2005qk}
\begin{eqnarray*}
\lag_{\chi} \sim \left( \partial \chi \right)^2 - m^2 \chi^2 - \dfrac{\partial^2 \phi_b}{\Lambda^5} \left( \partial^2 \chi \right)^2 .
\end{eqnarray*}
This Lagrangian has four derivatives acting on a single scalar fields. It is well known that such model possesses a ghost \cite{Pais:1950za,deUrries:1998bi} \footnote{In general, the presence of ghost in theories with fourth-order operators acting on a single scalar fields can be demonstrated by reformulating the theory as a standard two-derivative Lagrangian for two independent scalar fields through the definition $\theta \equiv \ddot{\chi}$.}. This \emph{Boulware-Deser} ghost as a coordinate dependent mass given by
\begin{eqnarray*}
m^2_{\tx{ghost}} \sim \dfrac{\Lambda^5}{\partial^2 \phi_b} .
\end{eqnarray*}
Since the FP theory is only valid for energies below $\Lambda$, this ghost instability has no physical meaning as long as $m_{\tx{ghost}} > \Lambda$. Moreover, when approaching Minkowski background, the mass of the ghost increases and therefore the distance below which this instability shows up decreases. But this does not save the FP theory. Indeed, at distances $r \gg r_v$ from a static source, the background solution $\phi_b$ is given by \refp{eq:FP_phi_back}. Then, the ghost mass is of order
\begin{eqnarray*}
m^2_{\tx{ghost}} \sim \dfrac{r^3}{r_v^5},
\end{eqnarray*}
and the ghost instability occurs for distances
\begin{eqnarray*}
r_v < r < r_{\tx{ghost}} &\rightarrow& \left( \dfrac{M}{\tx{M}_\tx{pl}} \right)^{1/5} < \Lambda r < \left( \dfrac{M}{\tx{M}_\tx{pl}} \right)^{1/3},
\end{eqnarray*}
where $r_{\tx{ghost}}$ is the distance at which $m_{\tx{ghost}}$ drops below $\Lambda$. This interval is not empty with consequence that FP theory has ghost instability in curved background.

To summarize, the FP theory is the only Lorentz-invariant model describing massive gravitational waves which is stable in Minkowski background. However, its suffers from several problems. Some of them might be cured by higher-order terms, or by an UV completion of the theory. Recently, it has even been argued that the helicity $\pm 1$ modes could be superluminal in slightly curved background \cite{Osipov:2008dd}. Although the FP theory might be less pathological if the vacuum is different from Minkowski space-time \cite{Kogan:2000uy,Porrati:2000cp,Deser:2001wx,Porrati:2001db}, it seems that the FP theory has too many pathologies to be considered as a serious candidate for a theory of gravity. This justifies the study of Lorentz-violating massive gravity, since breaking this symmetry is a way to go around all the usual problems of massive gravity discussed in this section.

\section{Lorentz-violating massive gravity} \label{sc:LVMG}

Recently, it has been argue that the standard problems of massive gravity could be avoided just by renouncing the Lorentz-invariance of the background \cite{Dubovsky:2004sg,Rubakov:2004eb}. Those works were part of a more general tendency based on the idea that, within the current observational limits, the Lorentz symmetry could be spontaneously broken in the gravity sector\footnote{There are some hints that quantum gravity and string theory may both suggest that Lorentz-invariance is broken at high energies \cite{Kostelecky:1989jw,AmelinoCamelia:2003uc}.}. This would appear audacious, as the Lorentz-invariance is perhaps the most important symmetry in physics responsible for the giving up of absolute time and the emergence of the space-time continuum. Likewise, Lorentz-invariance could be the limit of a more general unknown symmetry, just as the Galilean group is the small velocity limit of the Lorentz group.

Up to now, Lorentz-invariance has been tested in many sectors of the Standard Model of Electroweak Interactions \cite{Coleman:1998ti,Jacobson:2001tu,Mattingly:2005re,Bailey:2006fd}. Yet, the idea that Lorentz-invariance could be broken by the vacuum expectation value of some field is not conceptually different from the way that any time-dependent cosmological fluid actually breaks this symmetry. However, those cosmological fluids are not the vacuum state of the theory since they carry energy and dilute away with the Universe's expansion. An example of such a cosmological fluid is given by the Cosmic Microwave Background (CMB). Indeed, any observer at rest with respect to the CMB would conclude that the CMB is highly homogenous and isotropic, while other observers working in other reference frames would observe a dipole because of theirs own motion with respect to the CMB's photons.

Different models have emerged with spontaneous breaking of the Lorentz-invariance \cite{Kostelecky:1989jw,Clayton:1998hv,Jacobson:2000xp,Mattingly:2001yd,Kostelecky:2003fs,Gripaios:2004ms,Bluhm:2004ep,Carroll:2004ai,Libanov:2005nv,Cheng:2006us}. One common feature of all these models is that they \textquotedblleft dynamically\textquotedblright break the Lorentz-invariance. Indeed, if the preferred frame were to be fixed by an external mechanism, it will violate the general diffeomorphisms invariance of the theory, and the divergence of the matter energy-momentum tensor will not vanish\footnote{In GR, the gauge invariance implies that the divergence of the Einstein tensor is identically zero with consequence that the matter energy-momentum tensor is also divergence-less.}. Therefore, the Lorentz symmetry has to be broken dynamically by fields, whose EoM will imply the conservation of the matter energy-momentum tensor.

The Ghost Condensate model \cite{Arkani-Hamed:2003uy} is an example of a theory in which the Lorentz-invariance is broken by the time-dependent vacuum expectation value of a scalar field. The condensation of this field leads to infrared modifications of the gravitational interaction, without involving any of the pathologies discussed previously. Although gravitons are massless in this model, the Ghost Condensate shares several properties with the Lorentz-violating massive gravity theory discussed in this thesis. Another interesting class of model is given by bigravity theories \cite{Blas:2007zz,Berezhiani:2007zf}. There are two dynamical metrics in those models, one being not Lorentz-invariant. Therefore, bigravity theories posses two graviton fields associated to each metrics, one being massive while the other massless.

In those two last examples, as in Lorentz-violating massive gravity, the Lorentz symmetry is broken by a mechanism which could be seen as an analog of the Brout-Englert-Higgs Mechanism for gravity. Lorentz-violating models of massive gravity will be introduced in the rest of this chapter. The purpose of this discussion is the motivate the particular class of models which is the subject of this thesis and which can be thought of as \textquotedblleft minimal models\textquotedblright of massive gravity.

\subsection{Brout-Englert-Higgs Mechanism for gravity}

Gauge interactions are mediated by massless particles. Photons are one example of such gauge particles corresponding to the unbroken $\mathcal{U}(1)$ gauge symmetry of electromagnetic interactions. Upon spontaneous breaking of a gauge symmetry, the gauge particle acquire a mass via the Brout-Englert-Higgs Mechanism. An example of this mechanism can be found in the Standard Model of Electroweak Interactions, in which some of the bosons associated to the electroweak $\mathcal{SU}(2)\times\mathcal{U}(1)$ gauge interaction become massive through their coupling to the Brout-Englert-Higgs field.

The gauge group of GR is the general reparametrization invariance associated to a massless spin 2 particle, the graviton. If the diffeomorphism invariance is spontaneously broken, the graviton is expected to acquire a mass. In four dimensional space-time, there are four reparametrization invariances associated to the four coordinates. Each of these symmetries could then be individually broken, for example by a scalar field with vacuum expectation value depending on one particular coordinate. Such scenario could be thought at as the low energy limit of some kind of Brout-Englert-Higgs Mechanism for gravity \footnote{The analog of the Brout-Englert-Higgs Mechanism for gravity will include the UV completion of the theory.}. As an example, let us consider the following Lagrangian which depends on four scalar fields $\phi^0$, $\phi^1$, $\phi^2$ and $\phi^3$ through a function $\mathcal{P}(Q)$
\begin{eqnarray*}
\lag = \sqrt{- g} \left[ - \mathcal{R} + \dfrac{\Lambda^4}{\Mpl} \mathcal{P} \left( Q \right) \right] , & & Q = g^{\mu\nu} \partial_\mu \phi^\alpha \partial_\nu \phi^\beta q_{\alpha\beta} ,
\end{eqnarray*}
where $q_{\alpha\beta}$ is a constant metric for the scalar fields\footnote{For related works, see \cite{'tHooft:2007bf,Kakushadze:2000zn}.}. The parameter $\Lambda$ is a UV cutoff. The gravitational field equations for this theory read
\begin{eqnarray*}
\mathcal{R}_{\mu\nu} - \dfrac{1}{2} g_{\mu\nu} \mathcal{R} = \dfrac{\Lambda^4}{\Mpl} \left( - \dfrac{1}{2} g_{\mu\nu} \mathcal{P} + \mathcal{P}' \partial_\mu \phi^\alpha \partial_\nu \phi^\beta q_{\alpha\beta} \right) .
\end{eqnarray*}
The Minkowski vacuum is then solution to the EoM if the scalar fields are given by
\begin{eqnarray*}
\begin{array}{cccc}
\phi^{0} = a t , & \phi^1 = b x^{1} , & \phi^2 = c x^{2} , & \phi^3 = d x^{3} ,
\end{array}
\end{eqnarray*}
where $a$, $b$, $c$ and $d$ are four constants which values are determined by the requirement that the energy-momentum tensor associated with these four scalar fields vanishes in the Minkowski background
\begin{eqnarray*}
\begin{array}{cccc}
a^2 = \dfrac{\mathcal{P}}{2 q_{00} \mathcal{P}'} , & b^2 = - \dfrac{\mathcal{P}}{2 q_{11} \mathcal{P}'} , & c^2 = - \dfrac{\mathcal{P}}{2 q_{22} \mathcal{P}'} , & d^2 = - \dfrac{\mathcal{P}}{2 q_{33} \mathcal{P}'} .
\end{array}
\end{eqnarray*}
These four equations are satisfied for example if $\mathcal{P}(Q) = \mathcal{P}'(Q) = 0$. For arbitrary value of the metric $q_{\alpha\beta}$, each of these four Goldstone scalar fields breaks one of the four coordinate invariances of the Lagrangian by selecting one coordinate parametrization among all possibilities. Therefore, these scalars are selecting one reference frame.

Rather than breaking the whole diffeomorphism invariance of the theory, this mechanism could be used to break some specific symmetries while keeping the residual reparametrization invariance. For example, the background is invariant under the group $\mathcal{SO}(3)$ of three-dimensional rotations if $q_{0i} = 0$ and $q_{ij} \propto \delta_{ij}$, because then $b = c = d$ and the Lagrangian is invariant under a $\mathcal{SO}(3)$ rotation of the fields $\phi^i \rightarrow O^{i}_{\phantom{i}j} \phi^j$. Likewise, the invariance under boosts will not be broken if $q_{\mu\nu} \propto \eta_{\mu\nu}$. In these two last examples, the invariance under three-dimensional rotations is not broken and the transverse and traceless graviton modes with helicities $\pm 2$ satisfy the following equation
\begin{eqnarray*}
0 = \left( \Box - m^2 \mathcal{P} \right) H_{ij} ,
\end{eqnarray*}
which shows that the gravitons are massive provided that $\mathcal{P}(Q) < 0$ in Minkowski background. Yet, this example does not represent the most general case. For arbitrary function $\mathcal{P}$, the four constants $a$, $b$, $c$ and $d$ have to be such that $\mathcal{P}(Q) = 0$ in order for Minkowski space-time to be solution of the EoM. In general, the mass of the gravitational waves has to include the second derivative of $\mathcal{P}$ in order to be non-zero.

Before introducing the action for massive gravity, let us remark that bigravity theories with spontaneous breaking of the Lorentz-invariance are based on such symmetry breaking as well. Indeed, these models includes two Einstein-Hilbert terms along with a coupling function $\mathcal{P}(g^{\mu\alpha} f_{\alpha\nu})$ \cite{Blas:2007zz,Berezhiani:2007zf}. Then, Lorentz-invariance is broken by assuming light propagates at different speed in the two space-times. Those models are somehow related to the mechanism presented here, since $\partial_\mu \phi^\alpha \partial_\nu \phi^\beta q_{\alpha\beta}$ could be seen as a second space-time metric \footnote{Of course, bigravity theories are completely different from the present example because of their field content.}. Then, the speed of light in this second space-time will be given by $a^2 q_{00} / (b^2 q_{11})$ while being equal to one in the first metric. In the Ghost Condensate model \cite{Arkani-Hamed:2003uy}, the metric for the scalar fields satisfy $q_{\alpha\beta} = \delta_{\alpha0} \delta_{\beta0}$. Therefore, the invariance under boosts is broken in this model. $\mathcal{P}$ and $\mathcal{P}'$ have to vanish in Minkowski space-time in order to satisfy the EoM for this background implying that the gravitons remain massless. Yet, this model has a very interesting phenomenology. 

\subsection{The action for massive gravity} \label{sc:LVMG_Action}

This mechanism for the spontaneous breaking of the reparametrization invariance of the gravitational field can be used to construct a full non-linear theory of massive gravity, with a dynamical breaking of Lorentz-invariance by the space-time dependence of four Goldstone scalar fields. Indeed, the idea is to add terms to the action of GR in order to describe massive graviton without Lorentz symmetry. This could be done by adding an arbitrary non-Lorentz-invariant function of the metric to the usual Einstein-Hilbert action
\begin{eqnarray*}
\act = \act_{\tx{GR}} + \act_{\phi} &\tx{with}&
\act_{\phi} = \Lambda^4 \int \dif^4 x \, \sqrt{- g} \mathcal{F} \left( g^{00}, g^{0i}, g^{ij} \right) .
\end{eqnarray*}
Since this function involves only the metric without derivatives, one expects that such terms will give a mass to the gravitons. But in this action, the Lorentz symmetry is not dynamically broken with consequences that the action is not gauge invariant. Yet, it can be seen as the unitary gauge description of a gauge invariant action through the St\"{u}ckelberg trick. Indeed, the previous model can be generalized in a gauge invariant formulation by adding four scalar fields $\phi^0$, $\phi^1$, $\phi^2$ and $\phi^3$ to the action, which then reads \cite{Dubovsky:2004sg}
\begin{eqnarray} \label{eq:MG_action_V1}
\act_{\phi} = \Lambda^4 \int \dif^4 x \, \sqrt{- g} \mathcal{F} \left( X, V^{i}, W^{ij} \right) ,
\end{eqnarray}
with
\begin{eqnarray} \label{eq:MG_XVY}
\begin{array}{ccc}
X = g^{\mu\nu} \partial_\mu \phi^0 \partial_\nu \phi^0 , & V^{i} = g^{\mu\nu} \partial_\mu \phi^0 \partial_\nu \phi^i , & W^{ij} = g^{\mu\nu} \partial_\mu \phi^i \partial_\nu \phi^j - \dfrac{V^i V^j}{X} .
\end{array}
\end{eqnarray}
Note that the fields $\phi^\mu$ have dimensions of length so that $X$, $V^i$ and $W^{ij}$ are dimensionless. This new action defines a class of models which are gauge and Lorentz-invariant. As previously, $\Lambda$ is a UV cutoff with dimension of mass. Therefore, these theories have to be understood as low energy effective theories valid for energies below $\Lambda$. This low energy effective action will be our starting point in the study of Lorentz-violating massive gravity. But before studying these models in details, two remarks are in order:
\begin{itemize}
\item For general functions $\func$ of $X$, $V^i$ and $W^{ij}$, we could replace $W^{ij}$ in the action by a simpler variable $Y^{ij} = \partial_\mu \phi^i \partial^\mu \phi^j$ without loss of generality, $\mathcal{F} \left( X, V^{i}, W^{ij} \right) = \tilde{\mathcal{F}} \left( X, V^{i}, Y^{ij} \right)$. The reasons why $W^{ij}$ has been chosen instead of $Y^{ij}$ will appear in the next sections.
\item Instead of breaking the whole Lorentz group, one requires that the background breaks only the invariance under boosts. Then, the function $\mathcal{F}$ has to be understood as a function of the following scalar variables
\begin{eqnarray*}
\begin{array}{cccc}
X , & V^T V , & \tx{Tr} \left( W^n \right) , & \ldots ,
\end{array}
\end{eqnarray*}
where the dots stand for scalar terms which mix $V^i$ and $W^{ij}$ (with all indices contracted) while $n = 1,2,3$. This last requirement guarantees the invariance of the action under the rotations $\phi^i \rightarrow O^{i}_{\phantom{i}j} \phi^j$ of the $\phi^i$ internal space, which is necessary for the background to be invariant under the $\mathcal{SO}(3)$ group of three-dimensional rotations.
\end{itemize}

Now, let us assume that the following field configuration is the vacuum solution of these theories
\begin{eqnarray} \label{eq:MG_vacuum_V1}
g_{\mu\nu} = \eta_{\mu\nu} , & \phi^0 = a t , & \phi^i = b x^i ,
\end{eqnarray}
where $a$ and $b$ are two constants. This background configuration obviously breaks the Lorentz symmetry. For this solution, one has that $X = a^2$, $V^i = 0$ and $W^{ij} = - \delta^{ij} W / 3$ with $W = b^2$. Consequently, in this vacuum the first derivative of $\mathcal{F}$ with respect to $V^{i}$ is identically zero while the first derivative of $\mathcal{F}$ with respect to $W^{ij}$ is proportional to $\delta_{ij}$
\begin{eqnarray*}
\dfrac{\partial \mathcal{F}}{\partial W^{ij}} \equiv \delta_{ij} \mathcal{F}_{W} .
\end{eqnarray*}
The requirement that (\ref{eq:MG_vacuum_V1}) is a solution of the EoM implies that the energy-momentum tensor of the Goldstone fields has to vanish in Minkowski space-time. If $\mathcal{F}_{X}$ is the first derivative of $\mathcal{F}$ with respect to $X$, this condition translates into two algebraic equations
\begin{eqnarray} \label{eq:MG_T_vacuum}
0 = \mathcal{F} - 2 X \mathcal{F}_{X} , & & 0 = \mathcal{F} + 2 W \mathcal{F}_{W} ,
\end{eqnarray}
which determine the value of the two constants $a$ and $b$. Therefore, the vacuum configuration (\ref{eq:MG_vacuum_V1}) is a good ansatz and there is a solution for any generic function $\mathcal{F}$. Actually, once a function $\mathcal{F}$ is chosen, the two constants $a$ and $b$ could be set to one by a redefinition of the scalar fields. Therefore, we could assume that this redefinition has been done and only consider functions for which $a = b = 1$. This will be assumed in what follows, and therefore the vacuum solution will be
\begin{eqnarray} \label{eq:MG_vacuum}
g_{\mu\nu} = \eta_{\mu\nu} , & \phi^0 = t , & \phi^i = x^i .
\end{eqnarray}

Before discussing these models in more details, it is worth noting that the action (\ref{eq:MG_action_V1}) is invariant under the usual translations of the coordinates and under shifts of the scalar fields $\phi^\alpha \rightarrow \phi^\alpha + \lambda^\alpha$ by a constant vector $\lambda^\alpha$. The vacuum configuration (\ref{eq:MG_vacuum}) breaks both invariances. However, for the class of models based on (\ref{eq:MG_action_V1}), the usual invariance of the background under four dimensional translations is replaced by translations of the coordinates along with a shift of the scalar fields
\begin{eqnarray} \label{eq:MG_bkg_trans}
x^\alpha \rightarrow x^\alpha - \lambda^\alpha , && \phi^\alpha \rightarrow \phi^\alpha + \lambda^\alpha .
\end{eqnarray}
As a result, there are different possible definitions of the energy for the Goldstone scalar fields \cite{Arkani-Hamed:2005gu}. For usual matter fields, the gravitational energy is the one which couples to gravity. This energy is given by the energy-momentum tensor, and the conservation of this tensor emerges naturally from the requirement that the matter fields action be invariant under four-dimensional diffeomorphisms \cite{Wald:1984rg}. The particle physics energy, or inertial energy, is the Noether charge of the time translation symmetry of the vacuum. For usual matter fields, these two definitions are equivalent in Minkowski space-time. But things are different for the Goldstone fields. Indeed, their vacuum expectation value breaks the time translational symmetry down to symmetry (\ref{eq:MG_bkg_trans}). Consequently, their gravitational energy is not equal to their inertial energy. Indeed, the four currents associated to the four symmetries (\ref{eq:MG_bkg_trans}) read
\begin{eqnarray*}
\mathcal{J}^\mu_\alpha = - 2 m^2 t^\mu_{\alpha} + J^\mu_{\alpha} ,
\end{eqnarray*}
where $t^\mu_{\alpha}$ is the energy-momentum tensor deduced from $\act_\phi$, while $J^\mu_{0}$, $J^\mu_{1}$, $J^\mu_{2}$ and $J^\mu_{3}$ are the currents associated to the shift symmetries of the four scalar fields.

\subsection{The Goldstone sector} \label{sc:MG_Goldst}

It is not expected that all models described by an action such as (\ref{eq:MG_action_V1}) will be well-defined theories free of pathologies. The study of the Goldstone sector of this action enables to restrain the class of viable models by excluding those possessing instabilities \cite{Dubovsky:2004sg}. Indeed, as discussed in section \ref{sc:FP_Goldst} for Lorentz-invariant massive gravity, this sector describes the $k^2 > m^2$ limit of the vector and scalar sectors of the perturbation theory about Minkowski space-time. Therefore, any dangerous instability present in these two sectors will show up in the Goldstone sector.

Let us consider perturbations of the vacuum solution (\ref{eq:MG_vacuum}),
\begin{eqnarray} \label{eq:MG_pert_Mk}
\begin{array}{ccccc}
g_{\mu\nu} = \eta_{\mu\nu} + h_{\mu\nu} , & \phi^0 = t + \xi^0 , & \phi^i = x^i + \xi^i .
\end{array}
\end{eqnarray}
Then, the Goldstone sector is given by all the terms of the Lagrangian quadratic in the Goldstone perturbations. After decomposing $\xi^i$ into transverse and longitudinal fields through (\ref{eq:Pert_Mk}), and for an arbitrary function $\mathcal{F}$, the Goldstone sector of the massive gravity models described by (\ref{eq:MG_action_V1}) reads\footnote{The Goldstone sector of Lorentz-violating massive gravity was first studied in Ref.~\cite{Dubovsky:2004sg}. In this section, we resume the arguments presented in that paper.}
\begin{eqnarray*}
\lag_{\tx{Golds.}} &=& - \dfrac{1}{2} \xi_i^T \left( m_1^2 \partial_0^2 - m_2^2 \partial_i^2 \right) \xi_j^T \\
& & + m_0^2 \dot{\xi}_0^{2} + \dfrac{m_1^2}{2} \left[ \left( \partial_i \dot{\xi} \right)^2 + \left( \partial_i \xi_0 \right)^2 \right] + \left( m_3^2 - m_2^2 \right) \left( \partial_i^2 \xi \right)^2 + \left( 2 m_4^2 - m_1^2 \right) \dot{\xi_0} \partial_i^2 \xi \nonumber ,
\end{eqnarray*}
where $m_n^2$ ($n = 0, \ldots, 4$) are five Lorentz-violating graviton mass parameters. They are combinations of the function $\mathcal{F}$ and its first and second derivatives with an overall factor $m^2$ (see appendix \ref{app:sc:mass_param} for a detailed expression of these mass parameters). It is worth noting that the FP theory is obtained when $m_0^2 = 0$ and all other masses are equal
\begin{eqnarray*}
FP: & m_0^2 = 0 , & m_1^2 = m_2^2 = m_3^2 = m_4^2 .
\end{eqnarray*}
Here and below, $m$ denotes a parameter of dimension of mass which gives the scale to the Lorentz-violating masses
\begin{eqnarray*}
m_i^2 \propto m^2 &\tx{with}& m^2 \equiv \dfrac{\Lambda^4}{\Mpl} .
\end{eqnarray*}
The Goldstone sector does not tell anything about the tensor modes, which will be explicitely discussed in the next chapter. However, before going further in the study of the Goldstone sector, one should stress that the mass parameter $m_2^2$ is special in the sense that it corresponds to the mass of the helicity $\pm 2$ modes. Therefore, one has a first constraint on the function $\mathcal{F}$ which is that $m_2^2$ has to be strictly positive. Hence, only functions satisfying $m_2^2 > 0$ will be considered in the rest of this thesis.

The vector part of the Goldstone sector is given by the quadratic terms in $\xi_i^T$. For $m_1^2 > 0$, there are two healthy propagating degrees of freedom with helicities $\pm 1$ in the Goldstone sector. If $m_1^2$ is negative, the vector sector possesses two ghosts while for $m_1^2 = 0$ there are no helicity $\pm 1$ modes. Therefore, we will assume that the two following conditions are satisfied
\begin{eqnarray*}
m_2^2 > 0 , && m_1^2 \geq 0 \, .
\end{eqnarray*}

As in the case of Lorentz-invariant massive gravity, the scalar sector is much more tricky. Indeed, in the four-momentum space, the Goldstone scalar Lagrangian reads
\begin{eqnarray*}
\lag_{\tx{Gold. sc.}} = \dfrac{m_1^2}{2} \left( \begin{array}{cc}
\xi^{*}_0 & k \xi^{*}
\end{array} \right) M \left( \begin{array}{c}
\xi_0 \\ k \xi
\end{array} \right) , & & M = \left( \begin{array}{cc}
\mu_1 \omega^2 + k^2 & - i \mu_3 \omega k \\
i \mu_3 \omega k & \omega^2 + \mu_2 k^2
\end{array} \right) ,
\end{eqnarray*}
where
\begin{eqnarray*}
\mu_1 = \dfrac{2 m_0^2}{m_1^2} , & \mu_2 = 2 \dfrac{m_3^2 - m_2^2}{m_1^2} , & \mu_3 = \dfrac{2 m_4^2 - m_1^2}{m_1^2} .
\end{eqnarray*}
Therefore, the question we would like to answer here is whether it is possible to find values of the three parameters $\mu_i$ such that the Goldstone scalar Lagrangian is free of classical instabilities and ghosts. Let us first assume that $m_1^2 \neq 0$. The determinant and the trace of the matrix $M$ are given by
\begin{eqnarray*}
\det M = \mu_1 \omega^4 + \mu_2 k^4 + \omega^2 k^2 \left( 1 + \mu_1 \mu_2 - \mu_3^2 \right) , && \tr M = \omega^2 \left( \mu_1 + 1 \right) + k^2 \left( \mu_2 + 1 \right) .
\end{eqnarray*}
The dispersion relations of this sector are given by the equation $\det M = 0$ which has two solutions, namely $\omega^2_{+}$ and $\omega^2_{-}$. The product of these two solutions reads
\begin{eqnarray*}
\omega^2_{+} \omega^2_{-} = \dfrac{\mu_2}{\mu_1} k^4 .
\end{eqnarray*}
There are therefore three different classes of models:
\begin{itemize}
\item For $\mu_1 \neq 0$ and $\mu_2 \neq 0$, there are two propagating modes in the Goldstone scalar sector.
\item For $\mu_1 = 0$, there is at least one non-propagating mode in this sector characterized by $k^2 = 0$.
\item For $\mu_2 = 0$, there is at least one non-propagating mode in this sector characterized by $\omega^2 = 0$.
\end{itemize}
For the sake of example, let us discuss the two first class of models. A complete discussion of the third class can be found in Ref.~\cite{Dubovsky:2004sg}.

\subsubsection{Phase $\mu_1 \neq 0$ and $\mu_2 \neq 0$}

In this class of models, there are two propagating scalar modes. The absence of classical instabilities implies among others that
\begin{eqnarray*}
\dfrac{\mu_2}{\mu_1} > 0 .
\end{eqnarray*}
As discussed in section \ref{sc:FP_Goldst}, the requirement that the scalar sector is free of ghost implies that each eigenvalue of $M$ (given by eq.~(\ref{eq:Goldst_eig})) has to have one zero. This condition is not satisfied if $\mu_1, \mu_2 > 0$, since then the trace of $M$ is positive-definite. This implies that the equation $\lambda_+ = 0$ has no roots while the equation $\lambda_- = 0$ has two roots, one corresponding to a ghost.

If $\mu_1, \mu_2 < 0$, the Goldstone sector also possesses a ghost. Indeed, if we assume that $\lambda_-$ has only one zero, this mode correspond to a ghost since $\det M < 0$ for large value of $\omega^2$ implying that $\lambda_-$ is negative at large $\omega^2$. Therefore, the condition (\ref{eq:Goldst_ghost}) cannot be satisfied.

\subsubsection{Phase $\mu_1 = 0$}

For $\mu_1 = 0$, the solutions to the equation $\det M = 0$ reads
\begin{eqnarray*}
k^2 = 0 , && \omega^2 = \dfrac{\mu_2}{\mu_3^2 - 1} k^2 .
\end{eqnarray*}
This class of solutions reduce to three subgroups.
\begin{itemize}
\item For $\mu_2 \neq 0$ and $\mu_3^2 \neq 1$, there is one propagating mode in the Goldstone scalar sector:
\begin{itemize}
\item If $\mu_2$ and $\mu_3^2 - 1$ are both positive, the trace of $M$ is positive definite, implying that this propagating mode is a zero of $\lambda_-$. Since $\det M < 0$ and therefore $\lambda_- < 0$ for large $\omega^2$, one has to conclude that this propagating mode is a ghost; it does not verify the condition (\ref{eq:Goldst_ghost}).
\item If $\mu_2$ and $\mu_3^2 - 1$ are both negative, the trace and determinant of $M$ are both positive at large $\omega^2$, implying that both eigenvalues are also positives in this limit. Therefore, the derivative of the eigenvalue which has a zero is positive in its zero, and the propagating mode is perfectly healthy.
\end{itemize}
\item For $\mu_2 = 0$ and $\mu_3^2 \neq 1$, there are two non-propagating modes in the Goldstone scalar Lagrangian, with dispersion relation $k^2 = 0$ and $\omega^2 = 0$.
\item For $\mu_2 \neq 0$ and $\mu_3^2 = 1$, there are two non-propagating modes in the Goldstone scalar Lagrangian with dispersion relation $k^2 = 0$.
\end{itemize}
Finally, there is a possibility that both $\mu_2$ and $\mu_3^2 - 1$ vanish. This will be equivalent to the situation of the FP theory, where $\det M$ is identically zero and nothing could be said from the analysis of the Goldstone scalar sector.

\subsubsection{Phase $m_1^2 = 0$}

The three previous class of models assume that $m_1^2 \neq 0$. Therefore, there is a fourth class of models characterized by $m_1^2 = 0$. If this last relation holds, the determinant of the matrix $M$ reduces to
\begin{eqnarray*}
\det M = 4 \left[ m_0^2 \left( m_3^2 - m_2^2 \right) - m_4^4 \right] k^2 \omega^2 .
\end{eqnarray*}
Consequently, if $m_0^2 \left( m_3^2 - m_2^2 \right) - m_4^4 \neq 0$, there are two non-propagating modes with dispersion relations $k^2 = 0$ and $\omega^2 = 0$. If this relation between the parameters $\mu_i$ is not satisfied, then $\det M$ is identically zero and the Goldstone sector does not tell anything about the scalar sector.

To conclude this discussion of the Goldstone sector of Lorentz-violating massive gravity theories, let us stress that models with $m_1^2 = 0$ have only two propagating massive tensor modes, and can be thought of as minimal models of massive gravity. The exact linearized approximation of these models will be discussed in the next chapter.

\subsection{Symmetry as stabilizer}

Imposing certain relations between the five mass parameters enables to get rid of ghosts scalar degrees of freedom. But this is not sufficient to make the theory healthy. Indeed, as discussed previously for Lorentz-invariant massive gravity, the absence of ghost is guaranteed by the fine-tuning relation $\alpha = - 1$. However, instabilities may reappear in non-flat background since curvature terms could be responsible for a violation of this fine-tuning relation. Likewise, the stability of Lorentz-violating massive gravity in Minkowski space-time is also protected by fine-tuning relations, implying that \emph{Boulware-Deser} instabilities may also be present in those models. Moreover, instabilities may appear in the spectrum upon the addition of higher order operators.

To illustrate this problem, let us once again consider the Goldstone scalar sector. In a nearly flat arbitrary background, the Lagrangian of this sector reads
\begin{eqnarray*}
\lag_{\tx{Golds. sc.}} &=& \left( m_0^2 + \ldots \right) \dot{\xi}_0^{2} + \dfrac{1}{2} \left( m_1^2 + \ldots \right) \left( \partial_i \dot{\xi} \right)^2 + \dfrac{1}{2} \left( m_1^2 + \ldots \right) \left( \partial_i \xi_0 \right)^2 \\
& & + \left( m_3^2 - m_2^2 + \ldots \right) \left( \partial_i^2 \xi \right)^2 + \left( 2 m_4^2 - m_1^2 + \ldots \right) \dot{\xi_0} \partial_i^2 \xi + \ldots \nonumber ,
\end{eqnarray*}
where the dots stand for curvature terms. If the five mass parameters are non zero, it has been demonstrated in section \ref{sc:MG_Goldst} that such Lagrangian describes two propagating degrees of freedom, one of them being a ghost like mode. Yet, it is possible to get rid of this dangerous scalar mode by imposing for example that there are no quadratic terms for $\xi_0$ with two time-derivatives. This last requirement reads
\begin{eqnarray*}
m_0^2 + \ldots = 0 .
\end{eqnarray*}
This condition is easily satisfied in Minkowski space-time if the function $\mathcal{F}$ is chosen such that $m_0^2 = 0$. Still, this condition seems impossible to satisfy for arbitrary curved space-time. Indeed, let us take for example the following function
\begin{eqnarray*}
\mathcal{F} = X^2 + 2 X W^{ii} - W^{ij} W^{ij} + V^i V^i .
\end{eqnarray*}
Then, the vacuum configuration (\ref{eq:MG_vacuum}) with $a = b = 1$ is a solution of the EoM with \mbox{$m_0^2 = m_4^2 = 0$}. According to the previous discussion of the Goldstone Lagrangian in flat space-time, this particular model possesses four massive modes with helicities $\pm2$ and $\pm1$, respectively. The two scalar modes are non-dynamical with a dispersion relation $k^2 = 0$ implying that the vacuum is perfectly stable. However, this theory is unstable for the following background configuration\footnote{The Friedmann-Lema\^{\i}tre-Robertson-Walker (FLRW) background will correspond to this configuration of the scalar field in the limit of small curvature.}
\begin{eqnarray*}
\phi^0 = t + \xi^0_{\tx{bg}}(t), & \phi^i = x^i + \xi^i_{\tx{bg}} ,
\end{eqnarray*}
where $\xi^0_{\tx{bg}}(t)$ and $\xi^i_{\tx{bg}}$ are solutions of the linearized EoM above Minkowski space-time. Indeed, for such a background one has the following terms in the Goldstone scalar sector
\begin{eqnarray*}
\left( m_0^2 - 12 m^2 \partial_0 \xi^0_{\tx{bg}}(t) g^{00} + \ldots \right) \dot{\xi}_0^{2} \neq 0 .
\end{eqnarray*}
Therefore, there are quadratic terms for $\xi_0$ with two time-derivatives in the Goldstone Lagrangian and this model possesses a ghost-like mode in curved space-time. Fortunately, there is an elegant solution to such problems.

In several cases, requiring that these fine-tuning relations are a consequence of unbroken gauge symmetry is sufficient to protect Lorentz-violating models of massive gravity from becoming pathological upon addition of curvature terms and/or higher order operators \cite{Dubovsky:2004sg}. Indeed, if a symmetry forbids one particular term, this term will reappear neither in curvature terms nor in the UV completion of the theory. Therefore, if the absence of this term is sufficient to guarantee the stability of the model under consideration, then this model will be UV stable and free of instabilities of the \emph{Boulware-Deser} type. An example of theory whose stability is insensitive to its UV completion is given by the Ghost Condensate. This model is required to be invariant under space-time dependent spatial reparametrization
\begin{eqnarray} \label{eq:MG_symm_ghost}
x^i \rightarrow x^i + \varsigma^i \left( t, x^i \right) ,
\end{eqnarray}
implying that all masses but $m_0$ have to vanish. There are other symmetries one can think of. For example, the symmetry
\begin{eqnarray} \label{eq:MG_symm_Y}
t \rightarrow t + \varsigma^0 \left( t, x^i \right) ,
\end{eqnarray}
guarantees that $m_0 = m_1 = m_4 = 0$ while the symmetry
\begin{eqnarray} \label{eq:MG_symm_MG}
x^i \rightarrow x^i + \varsigma^i \left( t \right) ,
\end{eqnarray}
implies that $m_1$ has to be zero.

It is worth stressing that precautions have to be taken when considering models where some of the fields are non-dynamical; they may become dynamically unstable after taking into account higher-order terms and/or curved background. An example of such behavior is given by models invariant under the symmetry (\ref{eq:MG_symm_Y}), which become dynamically unstable after accounting for higher order operators \cite{Rubakov:2008nh}. Let us discussed the two other symmetries separately.

\subsection[The Ghost Condensate]{The symmetry $x^i \rightarrow x^i + \varsigma^i \left( t, x^i \right)$ and the Ghost Condensate}

The class of models invariant under the symmetry (\ref{eq:MG_symm_ghost}) are known as Ghost Condensate models \cite{Arkani-Hamed:2003uy}. Because of the relation between the four scalar vacuum expectation values and the four coordinates, the requirement that the theory is invariant under this symmetry independently of the gauge is equivalent to requiring that the full action is invariant under the following transformation of the scalar fields
\begin{eqnarray} \label{eq:sym_ghost_cond}
\phi^i \rightarrow \phi^i + \chi^i \left( \phi^0, \phi^k \right),
\end{eqnarray}
where $\chi^i$ are arbitrary functions of $\phi^0$ and $\phi^k$. Therefore, the action of Ghost Condensate models depend only on the field $\phi^0$ and reads
\begin{eqnarray} \label{eq:act_ghost_cond}
\act_{\phi} = \Lambda^4 \int \dif^4 x \, \sqrt{- g} \mathcal{F} \left( X \right) .
\end{eqnarray}
The Minkowski vacuum is a solution of the EoM for $\phi^0 = t$ if the function $\mathcal{F}$ is chosen such that the following two relations hold
\begin{eqnarray*}
\mathcal{F} |_{X=1} = 0 &\textrm{and}& \mathcal{F}_X |_{X=1} = 0 .
\end{eqnarray*}

At the quadratic level, the Lagrangian of the Ghost Condensate simply reads
\begin{eqnarray*}
\lag = \lag_{EH} + \dfrac{m_0^2}{4} \left( h_{00} - 2 \dot{\xi}_0 \right)^2 .
\end{eqnarray*}
Therefore, the new term will not affect the tensor and vector sectors of the gravitational field; i.e., the tensor modes are massless and there is no vector degrees of freedom. It is then straightforward to see that there is no scalar propagating mode either, as the dispersion relation deduced from the Goldstone sector is simply $\omega^2 = 0$. Indeed, the field equations for the scalar sector are
\begin{eqnarray*}
0 = 2 \partial_i^2 \Psi + m_0^2 \left( \Phi - \dot{\Xi}_0 \right) , & 0 = \ddot{\Psi} + \partial_i^2 \left( \Phi - \Psi \right) , & 0 = m_0^2 \left( \dot{\Phi} - \ddot{\Xi}_0 \right) ,
\end{eqnarray*}
whose solutions read
\begin{eqnarray*}
\Psi = \Psi_0 \left( x^i \right) , & \Phi = \Psi_0 \left( x^i \right) , & \dot{\Xi}_0 = \left( 1 + 2 \dfrac{\partial_i^2}{m_0^2} \right) \Psi_0 \left( x^i \right) ,
\end{eqnarray*}
where $\Psi_0 \left( x^i \right)$ is a time-independent integration constant. Note that if $\Psi_0 = 0$, this solution reduces to the solution of GR. Ghost Condensate models could therefore be seen as Einstein gravity plus an initial condition on the potentials.

The UV stability and the absence of dangerous \emph{Boulware-Deser} instabilities of Ghost Condensate models is easily understood by studying the Goldstone sector. The absence of gradient terms in this sector signals that the non-propagating modes will become dynamical upon addition of curvature terms or higher order operators. Let us first concentrate on the UV completion, which brings terms with more than one derivative acting on $\phi^0$. Because of the symmetry (\ref{eq:sym_ghost_cond}), the most general quadratic Goldstone Lagrangian for the perturbation $\xi_0$ is given, in the four-momentum space and up to fourth order operators, by
\begin{eqnarray*}
\lag = m_0^2 \left( \omega^2 - c_0 \dfrac{\omega^4}{\Lambda^2} - c_{1} \dfrac{k^2 \omega^2}{\Lambda^2} - c_{2} \dfrac{k^4}{\Lambda^2} + \ldots \right) \xi_0^* \xi_0
\end{eqnarray*}
where $c_i$ are constants of order one. These higher order terms are suppressed by the cutoff scale $\Lambda^2 = m \, \tx{M}_\tx{pl}$ and are therefore small corrections to the low energy effective action (\ref{eq:act_ghost_cond}). The dispersion relations deduced from this Lagrangian are
\begin{eqnarray} \label{eq:modes_Ghost_Cond}
\omega^2 = \dfrac{\Lambda^2}{c_0} - \dfrac{c_2}{c_1} k^2 - c_2 \dfrac{k^4}{\Lambda^2} + \mathcal{O} \left( \dfrac{1}{\Lambda^4} \right) , & & \omega^2 = c_2 \dfrac{k^4}{\Lambda^2} + \mathcal{O} \left( \dfrac{1}{\Lambda^4} \right) .
\end{eqnarray}
For the low energy effective theory with $\omega^2 \ll \Lambda^2$, the first relation is irrelevant. The second dispersion relation is the one of interest. It shows how the non-propagating mode with $\omega^2 = 0$ becomes a slowly varying dynamical mode upon addition of higher order terms. One could therefore conclude that imposing $m_0^2 > 0$ is sufficient to guarantee that the Ghost Condensate will be free of ghosts after considering higher order terms. Note that $c_2$ has to be positive otherwise this modes will be a slowly varying classical instability.

Therefore, Ghost Condensate models are truly healthy in the absence of gravity in contrast with the analogous sector of the DGP model which only becomes healthy trough its coupling to gravity \cite{Nicolis:2004qq}. When coupled to gravity, the scalar sector is classically unstable at low momenta. Indeed, after taking into account the higher order terms which modify the relation $\omega^2 = 0$, the full scalar Lagrangian reads in four-dimensional Fourier space
\begin{eqnarray*}
\lag = \left( \begin{array}{ccc} \Psi^* & \Phi^* & \Xi^{0*} \end{array} \right) \left( \begin{array}{ccc} - 6 \omega^2 + 2 k^2 & - 2 k^2 & 0 \\
- 2 k^2 & m_0^2 & i m_0^2 \omega \\
0 & - i m_0^2 \omega & m_0^2 \left( \omega^2 - c_{2} \dfrac{k^4}{\Lambda^2} \right) \end{array} \right) \left( \begin{array}{c} \Psi \\ \Phi \\ \Xi^0 \end{array} \right) + \ldots .
\end{eqnarray*}
The determinant of this matrix shows that two of the three scalar fields are non dynamical with the dispersion relation $k^2 = 0$ while the third field satisfies the relation
\begin{eqnarray} \label{eq:ghost_cond_mode}
\omega^2 = c_{2} \dfrac{k^2}{\Lambda^2} \left( k^2 - \dfrac{m_0^2}{2} \right) + \mathcal{O} \left( \dfrac{1}{\Lambda^4} \right) .
\end{eqnarray}
This last relation shows that when coupled to gravity, the ghost condensate is classically unstable for small momenta $2 k^2 \ll m_0^2$. Such low-momentum instabilities are nothing else than the analogs of the Jeans instability for ordinary matter fluid.

Similarly, Ghost Condensate models do not exhibit dangerous \emph{Boulware-Deser} instabilities, contrary to Lorentz-invariant massive gravity. Indeed, the propagating mode will never be a ghost since curvature terms cannot change the sign of the leading term proportional to $\omega^2$. This is completely different from the situation encountered in the FP theory discussed in section \ref{sc:FP_BD_inst}. In that model it is a non-propagating ghost which becomes dynamical in slightly curved space-time. Still, in slightly curved background, the dispersion relation (\ref{eq:ghost_cond_mode}) could acquire additional terms with small coefficient controlled by the curvature. Those terms could for example be gradient terms with a negative sign $- k^2$ implying a Jeans-like instability at low spatial momenta. However, such low momenta classical instabilities are not dangerous and are even interesting from a phenomenological point of view. Indeed, in GR such instabilities are believed to be responsible for the formation of the structures of the Universe.

Ghost Condensate models are interesting as toy models which enable one to understand the features of theories with spontaneous breaking of Lorentz-invariance. Moreover, some properties of those models leads to an interesting phenomenology. For example, the Newtonian potential is modified at distances larger than $m_0^{-1}$. However, this modification only shows up after a time $t_c = \Lambda / m_0^2$ \cite{Arkani-Hamed:2003uy}, since it is a consequence of the $k^4$ term which is suppressed by the UV cutoff $\Lambda$ (recall that, without the UV completion there is no modification of the scalar potentials). This modification of the Newtonian potential could be interpreted as resulting from an effective mass term produced by lumps of the scalar field excitation $\xi_0$. Such lumps could have a positive or negative effective mass leading to an attractive or repulsive contribution to the Newtonian potential.

Since the retardation effects are very strong in the Goldstone sector, the time delay between the passage of an object at a given point an the appearance of its track in the potential at this point is of order $t_c$. Consequently, the Ghost Condensate could be thought of as a bubble chamber in which all moving massive objects leave tracks \cite{Peloso:2004ut,Dubovsky:2004qe}, and observer in motion will not have time to see these new effects in the potential while an observer at rest will see them after a time $t_c$.

Due to the breaking of Lorentz-invariance, the non-linear dynamics of Ghost Condensate models is quite rich. For example, Ghost Condensate tends to form caustics \cite{Arkani-Hamed:2005gu}. The Ghost Condensate seems also to be non-perturbatively unstable. This leads to the formation of microscopic bubbles with negative energy density in their centers \cite{Krotov:2004if}. However, those \textquotedblleft holes\textquotedblright of negative energy do not expand and have little effect on the outer part of the bubbles.

From a cosmological point of view, Ghost Condensate could be a dark matter candidate \cite{Arkani-Hamed:2005gu,Kiselev:2004bh}. Moreover, there is a term proportional to $\mathcal{F}$ in the EoM which characterizes a fluid with equation of state $\rho = - p$. This fluid could give rise to a de Sitter expansion phase of the Universe without need for a cosmological constant $\Lambda_{c}$ \cite{Arkani-Hamed:2003uy}. It has also been shown that Ghost Condensate could drive an inflationary de Sitter phase of expansion if the shift symmetry is broken by a potential $V(\phi^0)$ \cite{ArkaniHamed:2003uz}.

To conclude this discussion of the Ghost Condensate, let us mention another very interesting property of Ghost Condensate models concerning black holes physics. Due to the higher order terms, there is an accretion of the Ghost Condensate onto black holes with a modification of the usual Schwarzschild black hole solution \cite{Frolov:2004vm,Mukohyama:2005rw}. Moreover, black holes may have hair \cite{Dubovsky:2007zi} as informations may escape frome black holes through a possible violation of the usual laws of thermodynamics \cite{Dubovsky:2006vk}.

\subsection[A minimal model of massive gravity]{The symmetry $x^i \rightarrow x^i + \varsigma^i \left( t \right)$ and a minimal model of massive gravity} \label{sc:Sym_Massive_Gravity}

Similarly to the situation of the Ghost Condensate model, massive gravity theories invariant under the transformation $x^i \rightarrow x^i + \varsigma^i \left( t \right)$ are perfectly healthy \cite{Dubovsky:2004sg,Dubovsky:2005dw}. In the Goldstone picture, this symmetry reads
\begin{eqnarray} \label{eq:MG_Sym_phi}
\phi^i \rightarrow \phi^i + \chi^i \left( \phi^0 \right),
\end{eqnarray}
where $\chi^i$ are arbitrary functions of $\phi^0$. This transformation is a symmetry of the action if it has the following form
\begin{eqnarray} \label{eq:MG_action}
\act = \int \dif^4 x \, \sqrt{- g} \left[ -\Mpl \ricci + \Lambda^4 \mathcal{F} \left( X, W^{ij} \right) \right].
\end{eqnarray}
Indeed, both $X$ and $W^{ij}$ are invariant under this transformation while $V^i$ is not. This last observation motivates the fact that up to now $W^{ij}$ has been used as a variable of $\mathcal{F}$ while it would have seem simpler to use $Y^{ij}$. At the level of the quadratic action, the requirement that $\mathcal{F}$ does not depend on $V^i$ implies that $m_1^2 = 0$. Therefore, according to the discussions of section \ref{sc:MG_Goldst}, in Minkowski space-time there are only two modes with helicities $\pm2$. But contrary to Ghost Condensate models, these modes are now massive with consequences that (\ref{eq:MG_action}) leads to true modifications of the gravitational interaction.

To address the issue of the UV stability of these models, let us once again concentrate on the Goldstone sector. Under the assumption that higher order operators are also invariant under the residual gauge symmetry \mbox{$x^i \rightarrow x^i + \varsigma^i \left( t \right)$}, the UV completion cannot contain $\left( \partial_0^2 \xi^i \right)^2$ and is therefore given up to fourth order derivatives by
\begin{eqnarray*}
\lag_{\tx{UV.}} &=& - \dfrac{m^2_0}{\Lambda^2} \left( c_0 \ddot{\xi}_0^2 + c_1 \left( \partial_i \dot{\xi}_0 \right)^2 + c_2 \left( \partial_i \partial_j \xi_0 \right)^2 \right) + \dfrac{m_4^2}{\Lambda^2} c_6 \partial_i \partial_0 \xi^0 \partial_i \partial_j \xi^j \\
& & + \dfrac{m_2^2}{\Lambda^2} \left( c_3\left( \partial_i \dot{\xi}^j \right)^2 - c_4 \left( \partial_i \partial_j \xi^k \right)^2 + c_5 \left( \partial_j \partial_i \xi^i \right)^2 \right) + \ldots ,
\end{eqnarray*}
where the $c_i$ with $i = 0,\ldots,6$ are constants of order one. Let us first have a look at the vector sector. The Goldstone vector sector in the Fourier space reads
\begin{eqnarray*}
\lag_{\tx{Golds. vect.}} = m_2^2 c_3 \dfrac{k^2}{\Lambda^2} \left( \omega^2 - \dfrac{\Lambda^2}{2 c_3} - \dfrac{c_4}{c_3} k^2 \right) |\xi^T_i|^2 .
\end{eqnarray*}
One conclude from this Lagrangian that upon addition of higher order derivatives, the non-propagating vector modes already present at the classical level remain non-dynamical. Indeed, the dispersion relation $k^2 = 0$ is not affected by the UV completion. Moreover, there is a second vector mode which falls out of the validity region of the low energy effective theory since it describes a mode with frequency of the order of $\Lambda$. Therefore, the vector sector is not sensitive to its UV completion.

The Goldstone scalar sector up to fourth order derivatives is much more tricky. For simplicity, let us neglect the contributions proportional to $c_0$ and $c_1$ \footnote{Recall that we have seen in the discussion of the Ghost Condensate that these terms give rise to a mode (\ref{eq:modes_Ghost_Cond}) which is unimportant for the low energy effective theory.}. Then, the field equations for the scalar sector are given in the four momentum space by
\begin{eqnarray*}
\left( \begin{array}{cc}
m_0^2 \left( \omega^2 - c_2 \dfrac{k^4}{\Lambda^2} \right) & - i \omega k m_4^2 \left( 1 + c_6 \dfrac{k^2}{\Lambda^2} \right) \\
i \omega k m_4^2 \left( 1 + c_6 \dfrac{k^2}{\Lambda^2} \right) & k^2 \left( m_3^2 - m_2^2 + \dfrac{m_2^2}{\Lambda^2} \left( c_3 \omega^2 + \left( c_5 - c_4 \right) k^2 \right) \right)
\end{array} \right) \left( \begin{array}{c}
\xi_0 \\ k \xi
\end{array} \right) = 0 .
\end{eqnarray*}
As already discussed, the dispersion relation are simply given by the equation $\det M = 0$, where $M$ is the matrix of the EoM. It is straightforward to see that these dispersion relations are
\begin{eqnarray*}
\left( 1 + \dfrac{m_4^4}{m_0^2 \left( m_2^2 - m_3^2 \right)} \right) \omega^2 = c_2 \dfrac{k^4}{\Lambda^2} + \ldots , & & 0 = k^2 \left( \omega^2 - \tx{const.} \, \Lambda^2 - \tx{const.} \, k^2 + \ldots \right) .
\end{eqnarray*}
The first of these two relations is the generalization of the relation (\ref{eq:modes_Ghost_Cond}) already met in the study of the Ghost Condensate. Indeed, this first relation precisely describes the Ghost Condensate mode. Therefore, the Ghost Condensate is somehow included in these massive gravity models. The condition $m_0^2 > 0$ that guarantees that this mode is not a ghost-like degree of freedom in the Ghost Condensate models is replaced here by
\begin{eqnarray} \label{eq:MG_ghost_UV}
m_0^2 + \dfrac{m_4^4}{m_2^2 - m_3^2} > 0 .
\end{eqnarray}
It is sufficient to impose this condition for the $\omega^2$ term present in the classical two-derivative theory to have the correct sign. As for the Ghost Condensate models, this mode is a slowly varying classical instability if $c_2 < 0$.

Similarly to the situation in the vector sector, the second scalar dispersion relation $k^2 = 0$ is in-sensitive to the UV completion. This mode is still a solution of $\det M = 0$ after taking into account higher-derivatives terms. This is a direct consequence of the residual symmetry which forbids terms quadratic in $\xi$ without at least two spatial derivatives. Similarly to the vector sector, there is a new mode $\omega^2 = \tx{const.} \, \Lambda^2 + \tx{const.} \, k^2 + \ldots$ which falls out of the low energy effective theory. Consequently, we have just demonstrated that massive gravity theories based on (\ref{eq:MG_action}) are perfectly healthy upon addition of higher order operators.

Those models do not exhibit dangerous \emph{Boulware-Deser} instabilities either, since curvature terms cannot change the sign of the term proportional to $\omega^2$ or introduce such terms for the $\xi$ fields. Hence, there is no ghost propagating mode in curved space-time contrarily to the situation of the FP theory. Still, in slightly curved background, the dispersion relation $\omega^2 = 0$ could acquire additional terms with small coefficient controlled by the curvature. Those terms could for example be gradient terms with a negative sign $- k^2$ implying a Jeans-like instability at low spatial momenta.

At this stage, it is clear that these models are candidates for consistent massive gravity theories. Consequently, from now on we will only focus on these models which are very interesting from a phenomenological point of view even without taking into account their UV completions.

\chapter{Weak gravitational interaction} \label{ch:weak_field}

GR is a full non-linear theory, with consequence that only a few exact solution of the Einstein's field equations are known. Fortunately, the gravitational interaction is so weak that for most practical situations it is sufficient to study the solution of the linearized EoM. For instance, in GR the gravitational field of the Sun is static to a very good approximation implying that the metric around the Sun is given by the Schwarzschild solution
\begin{eqnarray*}
\tx{d}s^2 = \left( 1 + 2 \Phi \right) \tx{d}t^2 - \left( 1 + 2 \Phi \right)^{-1} \tx{d}r^2 - r^2 \left( \tx{d}\theta^2 + \sin^2 \theta \tx{d}\varphi^2 \right) , && \Phi = - \dfrac{r_s}{2 r} .
\end{eqnarray*}
Thus, for distances much larger than the Schwarzschild radius $r_s \equiv 2 G M_{\odot} \sim 3 \tx{km}$, the potential $\Phi$ is a small quantity. Consequently, the Schwarzschild metric can be approximated by its Taylor expansion around $r_s = 0$, and all terms of order $(r_s/r)^n$ with $n \geq 2$ are completely negligible as compared to the linear terms
\begin{eqnarray*}
\tx{d}s^2 = \left( 1 - \dfrac{r_s}{r} \right) \tx{d}t^2 - \left( 1 + \dfrac{r_s}{r} \right) \tx{d}r^2 - r^2 \left( \tx{d}\theta^2 + \sin^2 \theta \tx{d}\varphi^2 \right) + \ldots \, .
\end{eqnarray*}
Therefore, the gravitational field of the Sun is correctly described by the linearized approximation for distances $r \gg 3 \tx{km}$.

In this chapter, we discuss the perturbations above Minkowski space-time induced by a small-amplitude gravitational source $\delta \mathcal{T}_{\mu\nu}$ in Lorentz-violating models of massive gravity described by action (\ref{eq:MG_action}). The linearized theory which results from this corresponds to the weak field limit of the gravitational interaction produced by $\delta \mathcal{T}_{\mu\nu}$ as long as these perturbations remain small as compared to the metric of flat space-time. This chapter has two aims. The first one is to introduce a sub-class of minimal models for which there is no modifications of Newton's potential. The second one is to provides the equations which will be the starting point of chapter \ref{ch:inst_int}.

\section{Energy-momentum tensor} \label{sc:Energy-momentum}

Before studying the linearized approximation of the massive gravitational field, it is worth noting that there is a convenient way to parameterize the energy-momentum tensor of any gravitational source. Let $v^\mu$ with $v_\mu v^\mu = 1$ be the velocity of an observer comoving with a fluid described by the energy-momentum tensor $\mathcal{T}_{\mu\nu}$. The two following tensors are projection tensors along and perpendicular to $v^\mu$ \cite{Uzan:2005}, respectively,
\begin{eqnarray*}
U^{\mu\nu} \equiv v^\mu v^\nu , & j^{\mu\nu} \equiv v^\mu v^\nu - g^{\mu\nu} ,
\end{eqnarray*}
satisfying
\begin{eqnarray*}
U^{\mu\alpha} U_{\alpha\nu} = U^\mu_\nu , & U^{\mu}_\mu = 1 , & U^{\mu\nu} u_\nu = u^\mu , \\
j^{\mu\nu} u_\nu = 0 , & j^\mu_{\mu} = 3 , & j^{\mu\alpha} j_{\alpha\nu} = - j^\mu_{\nu} .
\end{eqnarray*}
These two projection operators enable one to introduce natural notions of time and space for the observer since the space-time interval can be written
\begin{eqnarray*}
\tx{d}s^2 = \left( u_\mu \tx{d}x^\mu \right)^2 - j_{\mu\nu} \tx{d}x^\mu \tx{d}x^\nu .
\end{eqnarray*}
With these projectors, any energy-momentum tensor can be decomposed as \cite{Weinberg:1972,Uzan:2005}
\begin{eqnarray} \label{eq:En_Mo_tens}
\mathcal{T}_{\mu\nu} = \rho v_{\mu} v_{\nu} + q_\mu v_\nu + q_\nu v_\mu + p j_{\mu\nu} + \pi_{\mu\nu} ,
\end{eqnarray}
where $\rho = \mathcal{T}_{\mu\nu} v^\mu v^\nu$ and $p = \mathcal{T}_{\mu\nu} j^{\mu\nu} / 3$ are the energy and pressure densities measured by the comoving observer respectively, $q^\mu = - \mathcal{T}_{\gamma\lambda} v^\lambda j^{\gamma\mu}$ is the energy flux perpendicular to $v^\mu$ and $\pi^{\mu\nu} = \mathcal{T}_{\lambda\gamma} j^{\lambda\mu} j^{\gamma\nu} - j^{\mu\nu} \mathcal{T}_{\lambda\gamma} j^{\lambda\gamma} / 3$ is the anisotropic pressure tensor. The following relations are consequences of the previous definitions
\begin{eqnarray*}
q_\mu v^\mu = 0 ,& \pi_{\mu\nu} v^\nu = 0 , & \pi_\mu^\mu = 0 .
\end{eqnarray*}

\section{Weak field} \label{sc:Weak_Field}

In order to determine the weak field approximation in Lorentz-violating massive gravity models, one has to consider small perturbations about the flat vacuum solution (\ref{eq:MG_vacuum}) for which the energy-momentum tensor of the usual matter is zero, $\mathcal{T}_{\mu\nu} = 0$. A small perturbation $|\delta \mathcal{T}_{\mu\nu}| \ll 1$ of the matter energy-momentum tensor will produce metric $\delta g_{\mu\nu} \equiv h_{\mu\nu}$ and Goldstone perturbations $\delta \phi^\mu \equiv \xi^\mu$ about the vacuum solution.

With the notations introduced previously, the energy-momentum tensor of a small-amplitude source of gravitational field above Minkowski space-time is parameterized as follows,
\begin{eqnarray*}
\delta \mathcal{T}_{\mu\nu} = \left( \delta \rho + \delta p \right) v_{\mu} v_{\nu} - \eta_{\mu\nu} \delta p + \left( v_{\mu} \delta q_{\nu} + v_{\nu} \delta q_{\mu} \right) + \delta \pi_{\mu\nu} , \label{eq:energy-tensor-pert}
\end{eqnarray*}
where $\delta \rho$ and $\delta p$ are the matter density and pressure measured by the comoving observer, $\delta q_\mu$ is the energy flux perpendicular to $v_\mu$ ($v^\mu \delta q_\mu = 0$) and $\delta \pi_{\mu\nu}$ is the anisotropic pressure tensor ($v^\mu \delta \pi_{\mu\nu} = \delta \pi_{\mu}^\mu = 0$). The velocity $v_\mu$ of the observer obeys the geodesic equation $v^\nu \partial_\nu v^\mu = 0$, with $v^\mu v_\mu = 1$. This implies that the affine parameter of the observer can be chosen such that $v_\mu = \left( 1, 0, 0, 0 \right)$. As a consequence, $\delta q_0 = \delta \pi_{0\nu} = 0$ and
\begin{eqnarray*}
\delta \mathcal{T}_{00} = \delta \rho , & \delta \mathcal{T}_{0i} = \delta q_i , & \delta \mathcal{T}_{ij} = \delta_{ij} \delta p + \delta \pi_{ij} .
\end{eqnarray*}
The energy flux $\delta q_i$ and the anisotropic stress $\delta \pi_{ij}$ can be parameterized in the following way,
\begin{eqnarray*}
\delta q_i &=& \zeta_i + \partial_i \zeta , \\
\delta \pi_{ij} &=& \left( 3 \partial_i \partial_j - \delta_{ij} \partial_k^2 \right) \pi + \partial_i \pi_j + \partial_j \pi_i + \pi_{ij} ,
\end{eqnarray*}
where the vector perturbations $\zeta_i$ and $\pi_i$ are transverse while the tensor perturbation $\pi_{ij}$ is transverse and traceless \footnote{For illustration, an imperfect fluid with shear viscosity $\eta$ will have a transverse and traceless anisotropic stress given by $\pi_{ij} = - \eta \dot{H}_{ij}$ \cite{Weinberg:1972}.}. With these notations, the energy-momentum conservation reads
\begin{eqnarray} \label{eq:En_Cons_Mink}
\dot{\delta \rho} = \partial_i^2 \zeta , & \dot{\zeta} = \delta p + 2 \partial_i^2 \pi , & \dot{\zeta_i} = \partial_j^2 \pi_i .
\end{eqnarray}
Hence, the ten independent components of the linearized energy-momentum tensor are expressed through four gauge-invariant scalars $\delta\rho$, $\delta p$, $\zeta$ and $\pi$, four vector degrees of freedom in the form of two transverse gauge-invariant vectors $\zeta_i$ and $\pi_i$, and two tensor degrees of freedom through the transverse and traceless gauge-invariant tensor $\pi_{ij}$.

As mention earlier, the presence of the source $\delta \mathcal{T}_{\mu\nu}$ implies a modification of the solution (\ref{eq:MG_vacuum}) for the metric and Goldstone fields. The metric and Goldstone perturbations are described through the decomposition (\ref{eq:Pert_Mk}) into transverse and longitudinal fields. With the account of all these notations, and because of the invariance of the vacuum solution under three-dimensional rotations, the linearized Einstein equations split into scalar, vector and tensor equations (appendix \ref{app:sc:line_equa}).
\[
\xy
(-37,0)*{\mathcal{G}_{\mu\nu} = \dfrac{1}{\Mpl} \left( \mathcal{T}_{\mu\nu} + t_{\mu\nu} \right)}; (46,15)*{\tx{tensor equations}}; (46,0)*{\tx{vector equations}}; (46,-15)*{\tx{scalar equations}}; (-37,20)*{g_{\mu\nu} = \eta_{\mu\nu} + h_{\mu\nu}}; (-37,-20)*{\phi^{\mu} = x^{\mu} + \xi^{\mu}}; (7,0)*{\begin{tabular}{c}
\cellcolor{gray} \\
\white{\tx{\textbf{linearization}}} \cellcolor{gray} \\
\cellcolor{gray}
\end{tabular}};
{\ar (22,2)*{}; (32,15)*{}};
{\ar (22,0)*{}; (32,0)*{}};
{\ar (22,-2)*{}; (32,-15)*{}};
{\ar (-37,15)*{}; (-37,6)*{}};
{\ar (-37,-15)*{}; (-37,-6)*{}};
{\ar@{|->} (-18,0)*{}; (-7,0)*{}};
\endxy
\]
Where $\mathcal{G_{\mu\nu}}$ is the Einstein tensor and where $t_{\mu\nu}$ is the energy-momentum tensor of the four Goldstone fields deduced from $\mathcal{S}_\phi$ (see appendix \ref{app:ch:equations} for a complete expression of this tensor). This splitting has already been observed in the previous chapter, where the quadratic Lagrangian of the gravitational field decomposes into scalar, vector and tensor sectors. The equations of these three sectors could then be consider separately. The discussion presented here was first made in Ref.~\cite{Dubovsky:2004ud}.

\subsection{The tensor and vector sectors}

The tensor linear equations states that any transverse and traceless anisotropic pressure tensor $\pi_{ij}$ will generate massive gravitational waves
\begin{eqnarray} \label{eq:Gravitons}
0 = \left( \Box + m_2^2 \right) H_{ij} + \dfrac{2 \pi_{ij}}{\Mpl} .
\end{eqnarray}
The solution to this relation can be formulated in the following way
\begin{eqnarray} \label{eq:Gravitons_Sol}
H_{ij} = - \dfrac{2}{\Mpl} \dfrac{\pi_{ij}}{\Box + m_2^2} ,
\end{eqnarray}
where $\left( \Box + m_2^2 \right)^{-1}$ has to be understood as the Green function of the Klein-Gordon equation with mass $m_2$. In this equation, we neglect the tensor modes which are solutions to the homogeneous equation $\left( \Box + m_2^2 \right) H_{ij} = 0$. These modes are present in the vacuum and do not play any role in the interaction between gravitational sources.

There are two vector equations deduced from the linearized EoM
\begin{eqnarray} \label{eq:MG_vector_eq}
0 = \partial_j^2 \varpi_{i} - \dfrac{2 \zeta_{i}}{M_{pl}^{2}} , & 0 = \dot{\varpi}_i - m_2^2 \sigma_i - \dfrac{2}{\Mpl} \pi_i .
\end{eqnarray}
One concludes from these two equations that there is no difference in the vector sector between GR and the model of massive gravity given by (\ref{eq:MG_action}). Indeed, the first equation is exactly the equation deduced in GR while the second equation reduces to $m_2^2 \sigma_i = 0$. Then, using the energy-momentum conservation, the equation for $\varpi_{i}$ can be written as $\Box \varpi_{i} = 2 \tx{M}_{\tx{pl}}^{-2} \left( \dot{\pi}_{i} - \zeta_{i} \right)$ implying that the solutions to the vector equations are given by
\begin{eqnarray} \label{eq:Vector_Sol}
\varpi_{i} = 2 \dfrac{\dot{\pi}_{i} - \zeta_{i}}{\Mpl \Box} , && \sigma_i = 0 ,
\end{eqnarray}
since we are interested in models with massive gravitons, i.e. $m_2^2 > 0$ \footnote{For models characterized by $m_2^2 = 0$, gravitational waves are massless and there are no constraints on the value of $\sigma_i$.}. To conclude the discussion of the tensor and vector sectors, it is worth stressing that there is no vDVZ discontinuity in these two sectors; the solutions to the tensorial and vectorial equations both correspond to the solutions of GR in the limit of vanishing mass $m^2 \rightarrow 0$.

\subsection{The scalar sector}

As usual, the scalar sector is more complicated. There are four scalar equations which read
\begin{eqnarray}
0 &=& - 2 \partial_i^2 \Psi + \tx{M}_{\tx{pl}}^{-2} \delta \rho + m_0^2 \left( \dot{\Xi}^0 - \Phi \right) + m_4^2 \left( \partial_i^2 \Xi + 3 \Psi \right) , \label{eq:MG_equat_Sc_01} \\
0 &=& - 2 \dot{\Psi} + \tx{M}_{\tx{pl}}^{-2} \delta_\zeta , \label{eq:MG_equat_Sc_02} \\
0 &=& \Phi - \Psi + m_2^2 \Xi + 3 \tx{M}_{\tx{pl}}^{-2} \pi , \label{eq:MG_equat_Sc_03} \\
0 &=& 2 \ddot{\Psi} + \partial_i^2 \left( \Phi - \Psi \right) - \tx{M}_{\tx{pl}}^{-2} \left( \delta p - \partial_i^2 \pi\right) + m_3^2 \left( 3 \Psi + \partial_i^2 \Xi \right) - m_2^2 \Psi \nonumber \\
& & - m_4^2 \left( \Phi - \dot{\Xi}^{0} \right) \label{eq:MG_equat_Sc_04} .
\end{eqnarray}
By making use of the conservation relations (\ref{eq:En_Cons_Mink}), eq.~(\ref{eq:MG_equat_Sc_02}) implies that
\begin{eqnarray} \label{eq:MG_Psi}
\Psi = \dfrac{\delta \rho}{2 \Mpl \partial_i^2} + \Psi_0 \left( x^i \right) .
\end{eqnarray}
The first term in this relation is the usual GR contribution to the potential $\Psi$. $\Psi_0$ is a function of the space coordinates only which corresponds to a new contribution to this potential. Combining the three other equations one finds that Newton's potential $\Phi$ is given by
\begin{eqnarray} \label{eq:MG_Phi}
\Phi = \dfrac{1}{\Mpl} \left( \dfrac{\delta \rho}{2 \partial_i^2} - 3 \pi \right) + \dfrac{3 \mu - 2}{2 \mu} \dfrac{m_2^2 \delta \rho}{\Mpl \partial_i^4} + \left( 1 - \dfrac{2 m_4^2}{\mu m_0^2} + \dfrac{3 \mu - 2}{\mu} \dfrac{m_2^2}{\partial_i^2} \right) \Psi_0 \left( x^i \right) ,
\end{eqnarray}
with
\begin{eqnarray*}
\mu = \dfrac{m_0^2 \left( m_2^2 - m_3^2 \right) + m_4^4}{m_2^2 m_0^2} .
\end{eqnarray*}
The first contribution to the potential is the usual GR term while the two other contributions are new terms absent in GR. Since $\Phi$ is gauge-invariant, the previous relation is valid provided that $\Phi \ll 1$. Therefore, the second contribution proportional to $\partial_i^{-4}$ is responsible for a breakdown of perturbation theory at large distances. Indeed, the gravitational potential of a static source $\delta \mathcal{T}_{\mu\nu} = \delta_{0\mu} \delta_{0\nu} M \delta^3 \left( x \right)$ has the form
\begin{eqnarray*}
\Phi = G M \left( - \dfrac{1}{r} + \dfrac{2 - 3 \mu}{2 \mu} m_2^2 r \right) + \left( 1 - \dfrac{2 m_4^2}{\mu m_0^2} + \dfrac{3 \mu - 2}{\mu} \dfrac{m_2^2}{\partial_i^2} \right) \Psi_0 \left( x^i \right),
\end{eqnarray*}
where $G = \left( 8 \pi \Mpl \right)^{-1}$ is Newton's constant. Since the second contribution to this potential grows linearly, the perturbation theory breaks down when this term reaches one, that is, at distances
\begin{eqnarray*}
\dfrac{1}{G M m_2^2} \dfrac{2 \mu}{2 - 3 \mu} \lesssim r .
\end{eqnarray*}
As an example, considering the Sun and assuming a graviton mass of the order $m \sim \left( 10^{15} \tx{cm} \right)^{-1}$ (which correspond more or less to the size of the Solar System), this last expression tell us that the linearized approximation breaks at distances $r \gtrsim \left( G M m^2 \right)^{-1} \sim 10^{24} \tx{cm}$ which is beyond the actual tests of the Solar gravitational field. For a smaller mass of the order of the Hubble scale $m \sim H_0^{-1}$, the linearized approximation breaks at distances much larger than the actual Hubble scale. Therefore, this breakdown of perturbation theory is not problematic from a phenomenological point of view contrarily to the breakdown of the FP theory who occurs at relative small distances (section \ref{sc:FP_vainst}). Still, this growing term puts strong constrains on the graviton mass $\left( 2 - 3 \mu \right) m_2^2 / 2 \mu \ll r^{-2}$ since Newton's potential is verified to a very good approximation in the Solar System \cite{Will:2005va}.

With the previous solutions for the two gravitational potentials $\Psi$ and $\Phi$, one can invert eqs.~(\ref{eq:MG_equat_Sc_03}) and (\ref{eq:MG_equat_Sc_01}) to obtain the following expressions for $\Xi$ and $\Xi^0$,
\begin{eqnarray}
\Xi &=& \dfrac{2 - 3 \mu}{2 \mu} \dfrac{\delta \rho}{\Mpl \partial_i^4} + \left( \dfrac{2 m_4^2}{\mu m_0^2 m_2^2} - \dfrac{3 \mu - 2}{\mu \partial_i^2} \right) \Psi_0 , \label{eq:MG_Xi} \\
\dot{\Xi}^{0} &=& \dfrac{1}{\Mpl} \left[ \left( 1 - \dfrac{2 m_4^2}{\mu m_0^2} \right) \dfrac{\delta \rho}{2 \partial_i^2} - 3 \pi + \dfrac{3 \mu - 2}{2 \mu} \dfrac{m_2^2 \delta \rho}{\partial_i^4} \right] \nonumber \\
& & + \left[ 2 \left( \dfrac{\mu m_0^2 m_2^2 - m_4^4}{\mu m_0^4 m_2^2} \right) \partial_i^2 + 1 - \dfrac{4 m_4^2}{\mu m_0^2} + \dfrac{3 \mu - 2}{\mu} \dfrac{m_2^2}{\partial_i^2} \right] \Psi_0 \label{eq:MG_Xi0} .
\end{eqnarray}
These last two equations complete the analysis of the scalar sector of the massive gravitational field in the presence of a sources $\delta \mathcal{T}_{\mu\nu}$. Note that all these relations imply that while $\Psi$, $\Phi$ and $\Xi$ are static for a static source, $\Xi^0$ is growing linearly with time. Therefore, for a static source, the solution found here is the approximation of a stationary solution for which there is an accretion of the Ghost Condensate fluid $\Xi^0$ by $\delta T_{\mu\nu}$.

The previous solution for the scalar sector contains the time-independant function $\Psi_0$. This function corresponds to a mode with dispersion relation $\omega^2 = 0$ (section \ref{sc:MG_Goldst}) analogous to the Ghost Condensate mode which should become dynamical upon addition of higher-derivative terms. Yet, this function is fixed at the two-derivative level by the initial conditions on the gravitational system. It follows from eqs.~(\ref{eq:MG_Xi}) and (\ref{eq:MG_Xi0}) that a non-zero $\Psi_0$ would mean the presence of a non-trivial condensation of the scalar fields $\phi^\mu$. This function does not depend on $\delta \mathcal{T}_{\mu\nu}$ and is therefore present in the solutions of the scalar sector even without matter fields. Hence, this function does not play any role in the interaction between sources of gravitational field and we will disregard its contribution when studying the interaction between gravitational sources in the rest of this thesis.

Contrarily to the situation of FP theory, there is no vDVZ discontinuity in the scalar sector of massive gravity models described by (\ref{eq:MG_action}). Indeed, $m_i^2 \propto m^2$ with the consequence that the parameters $\mu$ is of order one even in the limit $m \rightarrow 0$ and relation (\ref{eq:MG_Xi}) implies then that $\Psi_0 \propto m^2 \Xi \rightarrow 0$ in the limit of vanishing mass. Consequently, both $\Psi$ and $\Phi$ tend to their values in GR.

\section{Dilatation symmetry}

For particular models satisfying $3 \mu = 2$, the term responsible for the breakdown of the perturbation theory disappears and Newton's potential reduces to its value in GR (plus the contribution of $\Psi_0$). Therefore, those models are very interesting from a phenomenological point of view. They describe massive gravitons without any modifications in the vector and scalar sectors. A way to achieve this is to impose the following dilatation symmetry as a symmetry of the Goldstone action \cite{Dubovsky:2004ud}
\begin{eqnarray} \label{eq:dilatation_Z}
\phi^0 \rightarrow \lambda \phi^0 , & \phi^i \rightarrow \lambda^{-\gamma} \phi^i .
\end{eqnarray}
This dilatation is a symmetry of massive gravity theories if their action depends on the variables $X$ and $W^{ij}$ through a new variable $Z^{ij} = X^\gamma W^{ij}$
\begin{eqnarray} \label{eq:MG_action_Z}
\act = \int \dif^4 x \, \sqrt{- g} \left[ -\Mpl \ricci + \Lambda^4 \mathcal{F} \left( Z^{ij} \right) \right] .
\end{eqnarray}
For the vacuum configuration (\ref{eq:MG_vacuum}) one has $Z^{ij} = - \delta^{ij}$. For models based on a function $\mathcal{F}$ of a single variable $Z^{ij}$, there are two relations between the mass parameters (see appendix \ref{app:sc:mass_param})
\begin{eqnarray*}
m_0^2 = 3 \gamma m_4^2 , & m_4^2 = \gamma \left( 3 m_3^2 - m_2^2 \right) ,
\end{eqnarray*}
which imply that $3 \mu = 2$. Consequently, for those models the scalar potentials read
\begin{eqnarray} \label{eq:Scalar_F(Z)_Pot}
\Psi = \dfrac{\delta \rho}{2 M_{pl}^{2} \partial_i^2} + \Psi_0 \left( x^i \right) , & & \Phi = \dfrac{1}{M_{pl}^{2}} \left( \dfrac{\delta \rho}{2 \partial_i^2} - 3 \pi \right) + \left( 1 - \dfrac{1}{\gamma} \right) \Psi_0 \left( x^i \right) ,
\end{eqnarray}
while the Goldstone perturbations are given by
\begin{eqnarray} \label{eq:Scalar_F(Z)_Gold}
\Xi = \dfrac{1}{\gamma m_2^2} \Psi_0 , & \dot{\Xi}^{0} = \dfrac{1}{\Mpl} \left[ \left( 1 - \dfrac{1}{\gamma} \right) \dfrac{\delta \rho}{2 \partial_i^2} - 3 \pi \right] + \left[ \dfrac{m_2^2 - m_3^2}{\gamma m_4^2 m_2^2} \partial_i^2 + 1 - \dfrac{2}{\gamma} \right] \Psi_0 .
\end{eqnarray}
Note however that in the absence of anisotropic pressure $\pi = 0$, GR predicts $\Phi = \Psi$. This relation is not satisfied in the massive gravity model because of the appearance of the function $\Psi_0$. Hence, if $\Psi_0 \neq 0$ the gravitational potential $\Psi$ created by a massive source is not the potential $\Phi$ responsible for the geodesic motion of an observer around this source\footnote{The Cassini spacecraft has put a bound on the differences between those two potentials $\Psi / \Phi - 1 < 10^{-5}$ \cite{Bertotti:2003rm}.}. If we neglect this arbitrary function there is no difference in the scalar sector between GR and the models given by (\ref{eq:MG_action_Z}) at the linear level.

For such models, there are only a few constrains on the graviton mass coming from direct or indirect observations of gravitational waves. For instance, the secular decrease of orbital period of binary pulsar puts bounds on the graviton mass $m_2$; the slow-down of orbital period of binary pulsars has been shown to be compatible with the emission of gravitational waves as predicted by GR \cite{Taylor:1982zz,Hulse:1974eb}\footnote{For a recent discussion on bounding the graviton mass with binary pulsars, see \cite{Sutton:2001yj}.}. This implies that the mass of the gravitational waves cannot be larger than the frequency of the waves emitted by these system \cite{1979Natur.277..437T}
\begin{eqnarray*}
\dfrac{m_2}{2 \pi} < \omega \sim 3.6 \, 10^{-5} \, \tx{Hz} \sim \left( 10^{15} \, \tx{cm} \right)^{-1} \sim 10^{-28} \, \tx{GeV}.
\end{eqnarray*}
This frequency corresponds to the period of the orbital motion which is of order of 8 hours. Apart from the relatively high graviton mass allowed in such models, there is another motivation for studying the sub-class of models (\ref{eq:MG_action_Z}). Indeed, we will see in chapter \ref{ch:cosmo} that those models are attractors of the cosmological expansion \cite{Dubovsky:2005dw}.

To conclude, let us stress that the only differences between GR and massive gravity models described by (\ref{eq:MG_action_Z}) lei in the tensor sector which possesses two massive modes. This explain why these models have been studied intensively, and why most of the original contributions of this thesis will concern them. Before closing this chapter, let us note that the solutions to the linearized field equations which have been discussed here are the starting point of the coming study of physical instantaneous interactions. Moreover, these equations and their solutions will be generalized as to describe perturbations about a flat Friedmann-Lema\^{\i }tre-Robertson-Walker space-time in chapter \ref{ch:cosmo}, in order to study the formation of structures in massive gravity theories.

\part[Aspects of massive gravity]{\usefont{OT1}{pzc}{m}{n}\selectfont Aspects of massive gravity physics}

\chapter{Instantaneous interaction} \label{ch:inst_int}

One peculiarity of massive gravity models based on the action (\ref{eq:MG_action}) consist in the presence of a physical instantaneous interaction \cite{Bebronne:2008tr}. In the conventional GR, the gravitational potentials are all instantaneous potentials. Still, there is no physical instantaneous interaction since the instantaneous contributions cancel in the graviton propagator. An analogous situation exists in classical electrodynamics, where the instantaneous contributions to the potentials $A_0$ and $A_i$ cancel each other in the photon propagator, leaving the theory free of instantaneous interactions \cite{brill:832}. In models where these subtle cancellations are spoiled by the presence of fields that break Lorentz-invariance, physical instantaneous interactions are possible. An example of such models is given by Lorentz-violating electrodynamics \cite{Dvali:2005nt,Gabadadze:2004iv}.

Before considering the gravitational field, we review the Lorentz-violating electrodynamic model as a warming up. This model shared all the relevant features of massive gravity theories. The emergence of the electrodynamic instantaneous interaction in that model is analogous to the emergence of the instantaneous interaction in our class of massive gravity theories. Then, we will discuss the physical instantaneous interaction of massive gravity and give a concrete example consisting in a instantaneous frequency shift of light by a distant gravitational source. This chapter constitute the first original part of this thesis.

\section{Lorentz-violating massive electrodynamics} \label{sc:electro}

As a toy model for physical instantaneous interaction, consider the Lorentz-violating electrodynamics of Ref.~\cite{Dvali:2005nt,Gabadadze:2004iv}, described by the following Lagrangian
\begin{eqnarray*}
\mathcal{L} = - \dfrac{1}{4} F_{\mu\nu} F^{\mu\nu} + \dfrac{1}{2} m_0^2 A_0^2 - \dfrac{1}{2} m_1^2 A_i^2 + A_{\mu} J^{\mu} ,
\end{eqnarray*}
where $J^\mu$ is a conserved current $\partial_\mu J^\mu = 0$. The Lorentz-invariant Proca theory of massive electrodynamics is recovered by imposing $m_0 = m_1$. Note that the gauge invariance can be restored by making use of the St\"{u}ckelberg trick. Hence, this action may be thought of as the unitary gauge description of Lorentz-violating massive electrodynamics.

This Lagrangian is still invariant under spatial rotations implying that the following parametrization of the spatial vectors $A_i$ and $J_i$ will simplify the equations of motion:
\begin{eqnarray*}
A_i = A_i^T + \partial_i A^L , && J_i = J_i^T + \partial_i J^L ,
\end{eqnarray*}
where $A_i^T$ and $J_i^T$ are transverse. It is important to notice that the transverse and longitudinal currents $J_i^T$ and $\partial_i J^L$ extend over the whole space, even if $J_i$ is localized \cite{1975clel.book.....J}. Indeed, for all localized currents $J_i$, the transverse current $J_i^T$ possesses an delocalized contribution which is the opposite of the delocalized contribution of $\partial_i J^L$. As a concrete example one can think of a dipole which appears at $t = 0$ on $x^i = 0$
\begin{eqnarray*}
J_0 = - 4 \pi \mu_i \Theta \left( t \right) \partial_i \delta^3 \left( x \right) , & J_i = - 4 \pi \mu_i \delta^4 \left( x \right) .
\end{eqnarray*}
Despite the fact that this current is conserved and perfectly localized, its longitudinal and transverse components extend over the whole space
\begin{eqnarray*}
J^{T}_i = \delta \left( t \right) \mu_j \left[ \dfrac{\delta_{ij}}{r^3} - \dfrac{3 x^i x^j}{r^5} - \dfrac{8 \pi \delta_{ij}}{3} \delta^3 \left( x \right) \right] , & \partial_i J^{L} = \mu_j \delta \left( t \right) \left[ \dfrac{3 x^i x^j}{r^5} - \dfrac{\delta_{ij}}{r^3} - \dfrac{4 \pi}{3} \delta_{ij} \delta^3 \left( x \right) \right] .
\end{eqnarray*}
The transverse and longitudinal currents $J_i^T$ and $\partial_i J^L$ associated to the localized current $J_i$ can be determined using the transverse and longitudinal projection operator introduced in \cite{brill:832} (see appendix \ref{app:sc:Trans_Long_Op}). Therefore, for any localized current appearing in a small region of space there are one transverse and one longitudinal currents which extend to the entire space.

For an arbitrary current $J_\mu$, the EoM deduced from the previous Lagrangian are
\begin{eqnarray*}
\partial_i^2 \left( A_0 - \dot{A}^L \right) - m_0^2 A_0 - J_0 &=& 0 , \\
\partial_i \left[ \partial_0 \left( A_0 - \dot{A}^L \right) - m_1^2 A^L - J^L \right] &=& 0 , \\
\left( \Box + m_1^2 \right) A_i^T + J_i^T &=& 0 .
\end{eqnarray*}
The potential $A_0 - \dot{A}^L$ is invariant under the $\mathcal{U}(1)$ gauge transformations, and for $m_0 = m_1 = 0$, these equations are precisely those of classical electrodynamics. The first two equations imply the generalized Proca constraint $m_0^2 \dot{A}_0 = m_1^2 \partial_i^2 A^L$. Let us concentrate on the model defined by $m_1 > 0$ and $m_1 \neq m_0$. Then, $m_0$ has to be zero to guarantee the absence of ghosts and rapid instabilities. Since we are interested in solutions which do not grow at infinity, the Proca constraint implies then that $A^L = 0$. Hence, the field equations are equivalent to
\begin{eqnarray*}
A_0 = \dfrac{J_0}{\partial_i^2} , && \left( \Box + m_1^2 \right) A_i^T = - J_i^T .
\end{eqnarray*}
If $J_0 \neq 0$, it is obvious from these equations that $A_0$ is an instantaneous potential, since $J_0$ and $A_0$ have the same time dependence. In fact, $A_i^T$ is also an instantaneous potential. Indeed, the transverse current $J_i^T$ extends over all space and implies a non-vanishing $A_i^T$ in all space. As a consequence, for all localized currents $J_i$, $A_i^T$ is an instantaneous potential \footnote{There is only one possibility for $A_i^T$ and $A_0$ not to be instantaneous potentials : $A_i^T$ will not be an instantaneous potential if the localized source is transverse $J_i = J_i^T$ while $A_0$ will not be an instantaneous potential if the localized source is longitudinal $J_i = \partial_i J^L$. }.

Using the current conservation, the equation for $A_0$ can be written $\Box A_0 = \dot{J}^L - J_0$. Hence, the solutions to the field equations are
\begin{eqnarray*}
A_0 &=& \int \dif^4 x^\prime G^{+} \left( t, x , t^\prime, x^\prime \right) \left[ \dot{J}^L \left( t^\prime, x^\prime \right) - J_0 \left( t^\prime, x^\prime \right) \right] , \\
A_i^T &=& - \int \dif^4 x^\prime G_m^{+} \left( t, x , t^\prime, x^\prime \right) J_i^T \left( t^\prime, x^\prime \right) ,
\end{eqnarray*}
where $G^{+} \left( t, x , t^\prime, x^\prime \right)$ and $G_m^{+} \left( t, x , t^\prime, x^\prime \right)$ are the retarded Green functions of the d'Alembert and Klein-Gordon equations
\begin{eqnarray*}
\Box G^{+} \left( t, x , t^\prime, x^\prime \right) &=& \delta^4 \left( x - x^\prime \right) , \\
\left( \Box + m^2 \right) G_m^{+} \left( t, x , t^\prime, x^\prime \right) &=& \delta^4 \left( x - x^\prime \right) ,
\end{eqnarray*}
respectively.

In classical electrodynamics, the instantaneous contributions to the potentials $A_0$ and $A_i^T$ cancel each other in observables such as the electric and magnetic fields \cite{brill:832}. There is then no instantaneous interaction. This is no more true in the Lorentz-violating model, because the mass $m_1$ modifies the dispersion relation of the transverse modes $A^T_i$ without affecting the dispersion relation of $A_0$. Therefore, the instantaneous contributions to the potentials do not cancel each other anymore, giving rise to a physical instantaneous interaction. This can be illustrated by looking at the electric field $E_i \equiv F_{0i}$, which reads
\begin{eqnarray} \label{eq:ElectricField}
E_i &=& \int \dif^4 x^\prime G^{+} \left( t, x , t^\prime, x^\prime \right) \left[ \dfrac{\partial J_0 \left( t^\prime, x^\prime \right)}{\partial x^{i \prime}} - \dfrac{\partial J_i \left( t^\prime, x^\prime \right)}{\partial t^\prime} \right] \nonumber \\
& & - \int \dif^4 x^\prime \Delta G^{+} \left( t, x , t^\prime, x^\prime \right) \dfrac{\partial J_i^T \left( t^\prime, x^\prime \right)}{\partial t^\prime} ,
\end{eqnarray}
where $\Delta G^{+} \left( t, x , t^\prime, x^\prime \right) \equiv G_m^{+} \left( t, x , t^\prime, x^\prime \right) - G^{+} \left( t, x , t^\prime, x^\prime \right)$ is proportional to $m_1$. The first integral in (\ref{eq:ElectricField}) is the retarded electric field of classical electrodynamics. This term vanishes outside of the light-cone of the source: it involves only localized current $J_\mu$, whose effects are retarded because of the retarded Green function $G^{+}$. The second integral is a contribution specific to the massive case, which does not vanish outside of the light-cone of the source since it involves the unlocalized transverse current $J_i^T$. Therefore, this second integral extends over the whole space and represents an instantaneous contribution to the electric field.

It is worth noting that the presence of this physical instantaneous interaction is related to the breaking of Lorentz-invariance. Without breaking of this symmetry, the modifications to the dispersion relation of $A_0$ and $A^T_i$ would be proportional and the instantaneous contributions of these potentials would cancel each other in the expression (\ref{eq:ElectricField}). The discussion of physical consequences of instantaneous interactions in massive electrodynamics can be found in \cite{Dvali:2005nt}.

\section{Instantaneous gravitational potentials}

The emergence of a physical instantaneous interaction in the massive gravity model described by (\ref{eq:MG_action}) is also a consequence of the non-cancellation of unlocalized contributions \cite{Bebronne:2007qh}, whereas those contributions cancel in GR. Therefore, the same approach as the one used in the previous section will be useful to understand how a physical instantaneous interaction appears in this model. For simplicity and because of the particular interest of those models, we will focus in the current discussion on massive gravity theories based on a function of $Z^{ij}$ (action \ref{eq:MG_action_Z}).

As already mentioned, a small perturbation $\delta \mathcal{T}_{\mu\nu}$ of the matter energy-momentum tensor will produce metric and Goldstone perturbations $\delta g_{\mu\nu}$ and $\delta \phi^\mu$ of the vacuum solution, just as a non-vanishing current $J_\mu$ is responsible for the electromagnetic potentials $A_\mu$. It follows from the discussion of section \ref{sc:Weak_Field} that the tensor potential $H_{ij}$ is instantaneous for the same reason that the potential $A^T_i$ of electrodynamics. Indeed, in all physical situations $\delta \mathcal{T}_{ij}$ is localized in space. But its transverse and traceless part $\pi_{ij}$ extends over the whole space, implying a non-vanishing tensor potential $H_{ij}$ everywhere in space. Relation (\ref{eq:MG_vector_eq}) makes it clear that the vector potential $\varpi_{i}$ is also an instantaneous potential. By making use of the the Green functions of the d'Alembert and Klein-Gordon equations equation, the tensor and vector potentials (\ref{eq:Gravitons_Sol}) and (\ref{eq:Vector_Sol}) are given by
\begin{eqnarray}
H_{ij} &=& - 2 \int \dfrac{\dif^4 x^\prime}{\Mpl} G_m^{+} \left( t, x, t^\prime, x^\prime \right) \pi_{ij} \left( t^\prime, x^\prime \right) , \label{eq:II_Tensor_Gr} \\
\varpi_{i} &=& 2 \int \dfrac{\textrm{d}^4 x^\prime}{\Mpl} G^{\pm} \left( t , x , t^\prime , x^\prime \right) \left[ \dot{\pi}_{i} \left( t^\prime , x^\prime \right) - \zeta_{i} \left( t^\prime , x^\prime \right) \right] . \label{eq:II_Vector_Gr}
\end{eqnarray}
It is also obvious from eqs.~(\ref{eq:Scalar_F(Z)_Pot}) that the two gravitational potentials $\Psi$ and $\Phi$ are two instantaneous potentials, such as the Coulomb potential $A_0 - \dot{A}^L$ of electrodynamics. With the account of the energy-momentum conservation, and by making use of the Green's function of the d'Alembert equation, these scalar potentials read
\begin{eqnarray}
\Psi &=& \int \dfrac{\dif^4 x^\prime}{2 \Mpl} G^{+} \left( t, x, t^\prime, x^\prime \right) \left[ \dot{\zeta} \left( t^\prime, x^\prime \right) - \delta \rho \left( t^\prime, x^\prime \right) \right] \label{eq:II_Psi_Gr} , \\
\Phi &=& \int \dfrac{\dif^4 x^\prime}{\Mpl} G^{+} \left( t, x, t^\prime, x^\prime \right) \left[ 2 \dot{\zeta} \left( t^\prime, x^\prime \right) - 3 \ddot{\pi} \left( t^\prime, x^\prime \right) - \dfrac{\delta \rho \left( t^\prime, x^\prime \right) + 3 \delta p \left( t^\prime, x^\prime \right)}{2} \right] \label{eq:II_Phi_Gr} .
\end{eqnarray}
$\Psi_0$ has been neglected since this arbitrary function does not play any role in the interaction between gravitational sources.

The modification of the dispersion relation of $H_{ij}$ induced by $m_2$ is analogous to the modifications of the dispersion relation of the vector field $A_i^{T}$ of electrodynamics induced by $m_1$. Since there is no modifications of dispersion relations as compared to GR in the other sectors of the theory due to the breaking of the Lorentz-invariance, the instantaneous contributions to the potentials will not cancel themselves in the graviton propagator. Therefore, it is necessary to search for an observable involving $H_{ij}$ in order to see a physical instantaneous interaction.

\section{Frequency shift} \label{sc:ii_freq_shift}

To illustrate the physical instantaneous interaction present in our class of massive gravity models, consider the measured energy of a light beam. Let $v^\mu$ be the four-velocity of an observer, with $v^\mu v_\mu = 1$, and $u^\mu$ the vector tangent to the light's geodesic, with $u^\mu u_\mu = 0$. The light's momentum is defined by $p^\mu = \omega_0 u^\mu$, where $\omega_0$ is a constant, and the frequency measured by the observer is given by
\begin{eqnarray} \label{eq:frequency_mesured}
\omega = v_\mu p^\mu .
\end{eqnarray}
In the vacuum $\mathcal{T}_{\mu\nu} = 0$ and the solution (\ref{eq:MG_vacuum}) holds. Then, the affine parameter $\tau_{ob}$ of the observer can be chosen such that $v^\mu = \left( 1, 0, 0, 0 \right)$ while the affine parameters $\tau_{ph}$ of the light's geodesic can be chosen such that $u^\mu = \left( 1, n^i \right)$ with $n_i^2 = 1$. Therefore, the frequency measured by the observer is $\omega = \omega_0$.

If there is a small extra source for gravity $\mathcal{T}_{\mu\nu} = \delta \mathcal{T}_{\mu\nu}$, the vacuum solution is slightly modified. Because of the geodesic equations, the perturbations of the metric imply perturbations $\delta v^\mu$ of the four-velocity of the observer and perturbations $\delta u^\mu$ of the vector tangent to the light's geodesic. Therefore, $\delta v^\mu$ and $\delta u^\mu$ are given in terms of the metric perturbations and there is a shift of the measured frequency, $\delta \omega = \omega \left( \delta v_\mu u^\mu + v_\mu \delta u^\mu \right)$. At the linearized level, eq.~(\ref{eq:frequency_mesured}) gives
\begin{eqnarray} \label{eq:shift}
\dfrac{\delta \omega}{\omega} &=& \int \dif \tau_{ph} \left[ \dot{\Psi} - n^i \partial_i \Phi + n^i n^j \left( \dfrac{1}{2} \dot{H}_{ij} - \partial_{i} \varpi_j \right) \right] + \int \dif \tau_{ob} \, n^i \partial_i \Phi .
\end{eqnarray}
Here, we only present the results; the details of the calculations can be found in appendix \ref{app:ch:ii}. In a FLRW background, this relation is known as the \emph{Sachs-Wolfe} effect \cite{Sachs:1967er}. From (\ref{eq:shift}) it is obvious that any difference in the tensor sector of the theory as compared to GR will lead to a modification of the measured spectral shift.

\subsection{Frequency shift in GR}

In GR, the vector and scalar potentials are given by relations (\ref{eq:II_Vector_Gr}), (\ref{eq:II_Psi_Gr}) and (\ref{eq:II_Phi_Gr}). The tensor potential $H_{ij}$ is a solution to eq.~(\ref{eq:Gravitons}) with $m_2^2 = 0$. Therefore, this potential is given by eq.~(\ref{eq:II_Tensor_Gr}) where the Green function $G^{+}_m$ is replaced by the Green function $G^{+}$, since $G^{+}_m \rightarrow G^{+}$ when $m_2 \rightarrow 0$. With the account of these four relations, the frequency shift measured by the observer in GR reads
\begin{eqnarray} \label{eq:ShiftGR}
\dfrac{\delta \omega^{\tx{GR}} \left( t, x \right)}{\omega} &=& \int \dfrac{\textrm{d}^4 x^\prime}{\Mpl} G^{+} \left( t, x, t^\prime, x^\prime \right) \left[ \dfrac{\delta \mathcal{T}_{kk} \left( t^\prime , x^\prime \right) - \delta \mathcal{T}_{00} \left( t^\prime , x^\prime \right)}{2} - n^i n^j \delta \mathcal{T}_{ij} \left( t^\prime , x^\prime \right) \right] \nonumber \\
& & - \int \dfrac{\dif \tau_{ob}}{2 \Mpl} \int \textrm{d}^4 x^\prime G^{+} \left( t, x, t^\prime, x^\prime \right) n^i \dfrac{\partial}{\partial x^{\prime i}} \left[ \delta \mathcal{T}_{00} \left( t^\prime , x^\prime \right) + \delta \mathcal{T}_{kk} \left( t^\prime , x^\prime \right) \right] \nonumber \\
& & + n^i n^j \int \dfrac{\dif \tau_{ph}}{\Mpl} \int \textrm{d}^4 x^\prime G^{+} \left( t, x, t^\prime, x^\prime \right) \dfrac{\partial}{\partial x^{\prime i}} \left[ n^k \left( \delta \mathcal{T}_{jk} \left( t^\prime , x^\prime \right) + \delta_{jk} \delta \mathcal{T}_{00} \left( t^\prime , x^\prime \right) \right) \right. \nonumber \\
& & \left. + 2 \delta \mathcal{T}_{0j} \left( t^\prime , x^\prime \right) \right] .
\end{eqnarray}
The instantaneous contributions to the potentials cancel each others in this relation. Indeed, this expression involves only localized sources, i.e., the components of the localized energy-momentum tensor $\delta \mathcal{T}_{\mu\nu}$. The effects of these sources are retarded because of the retarded Green function $G^{+}$. As a consequence, in GR the frequency shift cancels outside of the light cone of the localized source of gravitational perturbations, and there is no instantaneous interaction.

\subsection{Frequency shift in massive gravity}

The situation is different in massive gravity models described by the action (\ref{eq:MG_action_Z}) for which $m_2^2 \neq 0$. Indeed, the vector and scalar potentials are still given by relations (\ref{eq:II_Vector_Gr}), (\ref{eq:II_Psi_Gr}) and (\ref{eq:II_Phi_Gr}). But the tensor potential is given by eq.~(\ref{eq:II_Tensor_Gr}) with $m_2^2 \neq 0$. Therefore, the spectral shift measured by an observer of velocity $v^\mu$ is expressed trough
\begin{eqnarray} \label{eq:ShiftMG}
\dfrac{\delta \omega}{\omega} = \dfrac{\delta \omega^\tx{GR}}{\omega} + \int \dif \tau_{ph} \dfrac{n^i n^j}{2} \dot{H}^\Delta_{ij} ,
\end{eqnarray}
where $H^\Delta_{ij} = H_{ij} - H_{ij} |_{\tx{GR}}$. Hence, there is only one extra term in relation (\ref{eq:ShiftMG}) as compared to GR. This term extends over the whole space and is responsible for an instantaneous shift of frequency. Therefore, the shift is observable outside of the light cone of the source. At this point, let us stress that this extra term looks like the extra electric field (\ref{eq:ElectricField}) obtained in the Lorentz-violating electrodynamics model discussed before. Indeed, these two extra terms involve the same function $\Delta G^{+}$ with sources which extend over the whole space
\begin{eqnarray} \label{eq:DeltaTensor}
H^\Delta_{ij} \left( t, x \right) = - 2 \int \dfrac{\textrm{d}^4 x^\prime}{\Mpl} \Delta G^{+} \left( t, x, t^\prime, x^\prime \right) \pi_{ij} \left( t^\prime , x^\prime \right) .
\end{eqnarray}

To demonstrate that this term is responsible for an instantaneous shift of frequency, we will focus for simplicity reasons on the derivative of the spectral shift with respect to the affine parameters $\tau_{ph}$ of the light geodesic
\begin{eqnarray} \label{eq:derivatives_shift}
\textbf{a}_\omega \equiv \dfrac{\dif}{\dif \tau_{ph}} \left( \dfrac{\delta \omega}{\omega} - \dfrac{\delta \omega^{\tx{GR}}}{\omega} \right) .
\end{eqnarray}
This quantity will now be determined for a concrete example of gravitational source.

\subsection{Example of instantaneous interaction} \label{sc:II_ex}

As an explicit example of physical instantaneous interaction, suppose we have a source which appears at $t = 0$, described by the following energy-momentum tensor
\begin{eqnarray}
\delta \mathcal{T}_{00} &=& 2 \mu_{ij} t^2 \Theta \left( t \right) \partial_i \partial_j \delta^3 \left( x - x_s \right) , \nonumber \\
\delta \mathcal{T}_{0i} &=& 4 \mu_{ij} t \Theta \left( t \right) \partial_j \delta^3 \left( x - x_s \right) , \nonumber \\
\delta \mathcal{T}_{ij} &=& 4 \mu_{ij} \Theta \left( t \right) \delta^3 \left( x - x_s \right) , \label{eq:Sources}
\end{eqnarray}
where $x_s = \left( 0, 0, d \right)$. The presence of the Heaviside function in these expressions guarantees that $\delta \mathcal{T}_{\mu\nu} = 0$ before $t = 0$. Let the observer be located at $x^i = 0$ and the light's geodesic be along the $x$ direction, with $n^i = - \delta^i_1$.
\[
\xy
{\ar@{->} (-8,0)*{}; (-8,25)*{}};
{\ar@{-} (-8,0)*{}; (-2,0)*{}};
{\ar@{->} (2,0)*{}; (17,0)*{}};
{\ar@[]@{<-} (-18,-18)*{}; (-8,0)*{}};
(-19,-15)*{x};(15,-3)*{y};(-10,22)*{z};
(0,0)*{\color{bl}{{\bullet}}};
(9,3)*{\color{bl}{\tx{Observer}}};
(0,20)*{\red{\bullet}};
(22,20)*{\red{\tx{Gravitational source}\, \delta \mathcal{T}_{\mu\nu}}};
{\ar@[grey]@{<.>} (-3,1)*{}; (-3,19)*{}};
(-5,10)*{d};
{\ar@[]@{~>} (-12,-20)*{}; (-2,-2)*{}};
{\ar@[]@{~>} (-14,-20)*{}; (-4,-2)*{}};
{\ar@[]@{~>} (-10,-20)*{}; (0,-2)*{}};
(4,-16)*{\color{orange}{\tx{Light waves}}};
\endxy
\]
Since the distance between the observer and the source is $d$, and because the speed of light is one in our units $c = 1$, in GR the observer will not measure any spectral shift before $t = d$.

In the model (\ref{eq:MG_action_Z}), there is an instantaneous interaction and (\ref{eq:derivatives_shift}) is not zero for $t < d$. In order to determine the frequency shift, the integral (\ref{eq:DeltaTensor}) has to be evaluated for the source (\ref{eq:Sources}). If the source is chosen such that $\mu_{ij} = \delta_{2i} \delta_{2j}$, the only relevant component of the transverse and traceless tensor $\pi_{ij}$ is given by\footnote{For an arbitrary energy-momentum tensor $\delta \mathcal{T}_{\mu\nu}$, the transverse and traceless anisotropic stress tensor $\pi_{ij}$ can be determined using transverse and longitudinal operators. See appendix \ref{app:sc:ii_source_tens}.}
\begin{eqnarray*}
\pi_{11} = 2 \Theta \left( t \right) \left( \dfrac{\partial_x^2 \partial_y^2}{\partial_i^2 \partial_j^2} - \dfrac{\partial_z^2}{\partial_i^2} \right) \delta^3 \left( x - x_s \right) .
\end{eqnarray*}
Hence, using relations (\ref{eq:ShiftMG}) and (\ref{eq:DeltaTensor}), the derivative of the spectral shift with respect to the affine parameters of the light geodesic is expressed through
\begin{eqnarray*}
\textbf{a}_\omega \left( t, x \right) = - \dfrac{2}{\Mpl} \left( \dfrac{\partial_x^{2} \partial_y^{2}}{\partial_i^{2} \partial_j^{2}} - \dfrac{\partial_z^{2}}{\partial_i^{2}} \right) \Delta G^{+} \left( t, x, 0, x_s \right) .
\end{eqnarray*}
Since $\Delta G^{+}$ involves the Bessel function $\mathcal{J}_{1} \left( m_2 x \right)$ divided by its argument and since $m_2 x \ll 1$, the following Taylor expansion can be used to find an approximation of this last expression
\begin{eqnarray*}
\dfrac{\mathcal{J}_{1} \left( m_2 x \right)}{m_2 x} = \dfrac{1}{2} - \dfrac{m_2^2 x^2}{16} + \mathcal{O} \left( m_2^3 x^3 \right) .
\end{eqnarray*}
With the account of all these relations, the measured variation of the frequency, caused by the source located at distance $d$ of the observer, is simply given for $t < d$ (outside of the light cone of the source) by
\begin{eqnarray*}
\textbf{a}_\omega \left( t , 0 \right) = - \dfrac{m_2^4 \, d}{64 \pi \Mpl} t + \mathcal{O} \left( m_2^5 \right) , & 0 \le t < d .
\end{eqnarray*}
This last expression clearly shows a measurable instantaneous interaction, proportional to the mass of the graviton to the fourth power.

\section{Origin of the instantaneous interaction} \label{sc:ii_origin}

The origin of this physical instantaneous interaction can be traced back to the three scalar fields $\phi^i$ which couple to the metric through $W^{ij}$ in $\func \left( X, W^{ij} \right)$. For the sake of argument, let us redefine $W^{ij}$ as
\begin{eqnarray} \label{eq:W_epsilon}
W^{ij} = \left( g^{\mu\nu} - \epsilon \dfrac{\partial^\mu \phi^0 \partial^\nu \phi^0}{X} \right) \partial_\mu \phi^i \partial_\nu \phi^j ,
\end{eqnarray}
where $\epsilon \leq 1$ is a constant. The standard definition of $W^{ij}$ correspond to $\epsilon = 1$. Only for this value of $\epsilon$ are massive gravity models invariant under the symmetry (\ref{eq:MG_Sym_phi}) which guarantees the absence of instabilities. It follows from (\ref{eq:W_epsilon}) that the three fields $\phi^i$ live, and therefore propagate, in the effective metric
\begin{eqnarray} \label{eq:effectve_metric}
G^{\mu\nu} \equiv g^{\mu\nu} - \epsilon \dfrac{\partial^\mu \phi^0 \partial^\nu \phi^0}{X} .
\end{eqnarray}
Consequently, in the vacuum (\ref{eq:MG_vacuum}) these three fields propagate at a velocity given by
\begin{eqnarray*}
\omega^2 = \dfrac{1}{1 - \epsilon} k^2 &\Rightarrow& v_{\tx{group}} = \dfrac{1}{\sqrt{1 - \epsilon}} .
\end{eqnarray*}
Hence, for $\epsilon < 0$ these fields have a propagation velocity smaller than the speed of light while for $0 < \epsilon < 1$ they propagate faster than light. For the models of interest, $\epsilon = 1$ and these fields have an infinite velocity.

The presence of fields with an infinite propagation velocity is not enough to account for an instantaneous interaction. For instance, there is no modifications of the scalar and vector sectors of massive gravity models possessing the dilatation symmetry (\ref{eq:dilatation_Z}), with consequence that only the tensor part of the metric couples to these instantaneous fields. Hence, it is not surprising that one has to look for an observable involving tensor modes in order to see an instantaneous interaction. Moreover, if one gives up the requirement of the invariance under the dilatation symmetry, Newton's potential is modified because of a direct coupling between the scalar part of the metric and these instantaneous fields. Hence, an instantaneous interaction is also expected in the scalar sector for those models.

Before going to the conclusions of this chapter, let us consider the physical instantaneous interaction from a \textquotedblleft fluid mechanics\textquotedblright point of view for arbitrary function $\func \left( X , W^{ij} \right)$. The generic notations for the energy-momentum tensor introduced in section \ref{sc:Weak_Field} can also be used for the energy-momentum tensor of the Goldstone fields $\phi^{\mu}$,
\begin{eqnarray*}
t_{\mu\nu} = \left( \rho_{\phi} + p_{\phi} \right) v_\mu v_\nu - g_{\mu\nu} p_{\phi} + \pi_{\mu\nu}^{\phi} , & & v_\mu = \dfrac{\partial_\mu \phi^0}{\sqrt{X}} .
\end{eqnarray*}
The energy, pressure and anisotropic stress densities for the Goldstone fluid are given by
\begin{eqnarray} \label{eq:rho_p_Golds}
\rho_{\phi} &=& \Lambda^{4} \left( - \dfrac{1}{2} \mathcal{F} + X \dfrac{\partial \mathcal{F}}{\partial X} \right), \\
p_{\phi} &=& \Lambda^{4} \left( \dfrac{1}{2} \mathcal{F} - \dfrac{W^{ij}}{3} \dfrac{\partial \mathcal{F}}{\partial W^{ij}} \right), \nonumber \\
\pi_{\alpha\beta}^{\phi} &=& \Lambda^{4} \left[ \left( v_\mu v_\nu v_{\alpha} v_{\beta} - v_\nu v_{\beta} g_{\alpha\mu} - v_\mu v_{\alpha} g_{\beta\nu} + g_{\alpha\mu} g_{\beta\nu} \right) \partial^\mu \phi^i \partial^\nu \phi^j + j_{\alpha\beta} \dfrac{W^{ij}}{3} \right] \dfrac{\partial \mathcal{F}}{\partial W^{ij}} \nonumber ,
\end{eqnarray}
respectively. We know from GR that the tensor modes only couple to matter fields through the anisotropic stress tensor. Hence, these relations make clear that it is necessary to have a function $\func$ depending on $W^{ij}$ in order for the scalar fields to give a mass to gravitons.

The perturbations of the Goldstone fluid belong to two categories: isotropic sound waves which are perturbations of the energy and pressure densities and anisotropic sound waves originated in small modifications of $\pi_{\mu\nu}^{\phi}$. In the vacuum, the energy and pressure perturbations $\delta \rho_{\phi} = \delta t_{\mu\nu} v^\mu v^\nu$ and $\delta p_{\phi} = \delta t_{\mu\nu} j^{\mu\nu} / 3$ are given by
\begin{eqnarray*}
\delta \rho_{\phi} &=& \Mpl \left[ m_0^2 \left( \dot{\Xi}_{0} - \Phi \right) + m_4^2 \left( \partial_j^2 \Xi + 3 \Psi \right) \right] , \\
\delta p_{\phi} &=& \Mpl \left[ m_4^2 \left( \Phi - \dot{\Xi}_{0} \right) + \left( m_2^2 - 3 m_3^2 \right) \left( \Psi + \dfrac{1}{3} \partial_j^2 \Xi \right) \right] .
\end{eqnarray*}
After substituting in these relations the solutions (\ref{eq:MG_Psi}), (\ref{eq:MG_Phi}), (\ref{eq:MG_Xi}) and (\ref{eq:MG_Xi0}) of the EoM, one finds that the energy and pressure density perturbations of the Goldstone fluid are
\begin{eqnarray*}
\delta \rho_{\phi} = 0 , & \delta p_{\phi} = \dfrac{3 \mu - 2}{3 \mu} m_2^2 \dfrac{\delta \rho}{\partial_i^2} .
\end{eqnarray*}
Therefore, a perturbation $\delta \rho$ of the matter energy density produces an isotropic sound wave in the Goldstone fluid which propagates at an infinite velocity
\begin{eqnarray*}
c_s^2 \equiv \dfrac{\delta p_{\phi}}{\delta \rho_{\phi}} = \infty .
\end{eqnarray*}
Hence, there is an instantaneous interaction in the scalar sector. Note that massive gravity models possessing the dilatation symmetry (\ref{eq:dilatation_Z}) are an exception. For these models, $\delta p_{\phi} = 0$ with consequences that there is no isotropic sound wave at all. 

If the matter energy-momentum tensor has a non-trivial anisotropic stress $\delta \pi_{\mu\nu} = \delta \mathcal{T}_{\lambda\gamma}     j^{\lambda\mu} j^{\gamma\nu} - j^{\mu\nu} \delta \mathcal{T}_{\lambda\gamma} j^{\lambda\gamma} / 3$, it will generate an anisotropic sound wave in the Goldstone fluid ($\delta \pi_{0\mu}^{\phi} = 0$)
\begin{eqnarray*}
\delta \pi_{ij}^{\phi} = \dfrac{\Mpl m_2^2}{2} \left[ \partial_j \sigma_i + \partial_i \sigma_j + H_{ij} + \dfrac{2}{3} \left( 3 \partial_i \partial_j - \delta_{ij} \partial_k^2 \right) \Xi \right],
\end{eqnarray*}
which for the class of models considered here becomes
\begin{eqnarray*}
\delta \pi_{ij}^{\phi} = - m_2^2 \int \dif^4 x^\prime G_m^{+} \left( t, x, t^\prime, x^\prime \right) \pi_{ij} \left( t^\prime, x^\prime \right) + \dfrac{2 - 3 \mu}{6 \mu} m_2^2 \dfrac{3 \partial_i \partial_j - \delta_{ij} \partial_k^2}{\partial_k^4} \delta \rho.
\end{eqnarray*}
There are two terms in this relation, both responsible for an instantaneous anisotropic sound wave. While the second term vanishes for massive gravity models possessing the dilatation symmetry (\ref{eq:dilatation_Z}), the first term is the origin of the physical instantaneous interaction observed in the previous sections. Indeed, this anisotropic wave propagates at an infinite velocity because of the unlocalized transverse and traceless tensor $\pi_{ij}$ which implies that $\delta \pi_{ij}^{\phi}$ has a non-vanishing value in the entire space.

\section{Summary and prospects}

Let us conclude this chapter by summarizing the previous discussion. It has then been argued that physical instantaneous interactions are present in massive gravity models described by the function $\func \left( Z^{ij} \right)$. The origin of this interaction lies in an anisotropic sound wave which propagates at an infinite velocity in the Goldstone fluid. The existence of such interactions is related to the spontaneous breaking of the Lorentz-symmetry induced by space-time dependent vacuum expectation values of the four Goldstone scalar fields $\phi^\mu$. Indeed, this symmetry breaking allows for modifications of the dispersion relation of the tensor modes as compared to GR without affecting the dispersion relations of other sectors. As a consequence, the instantaneous contributions to the potentials do not cancel in the graviton propagator, unlike in GR. This is analogous to the situation in Lorentz-violating electrodynamics, where the instantaneous contributions to the potentials do not cancel in the photon propagator.

It has been demonstrated here that a gravitational source localized in space is responsible for an instantaneous frequency shift of light beams seen by an observer. It is worth noting that, at the linearized level, the amplitude of this instantaneous spectral shift is proportional to the graviton mass to the fourth power, and therefore is very small as compared to the usual retarded interaction between gravitational sources. Regardless of the strength, there is no causal paradox associated with superluminal propagation in this model as the breaking of the Lorentz symmetry implies a preferred frame.

The action (\ref{eq:MG_action_Z}) is a low-energy effective action. One should then expect corrections containing higher-derivative terms to be present in massive gravity models. These corrections could not be responsible for modifications of the dispersion relations of the scalar and vector fields proportional to $m_2^2$. Therefore, the instantaneous contributions coming from the mass of the graviton cannot be canceled by higher-derivative terms, and the inclusion of such terms will not affect the conclusion of this work.

Finally, the presence of the physical instantaneous interaction is supposed to be responsible for the violation of the black hole \textquotedblleft no-hair\textquotedblright theorem present in these models \cite{Dubovsky:2007zi} (the physics of black holes will be discussed in chapter \ref{ch:bh}). Indeed, a crucial feature of black holes in GR is the absence of \textquotedblleft hair\textquotedblright, which simply means that a black hole is entirely characterized by its mass, charge and angular momentum. This \textquotedblleft no-hair\textquotedblright theorem is a consequence of the existence of an horizon which divides the space-time in two regions: the outside and the inside of the black hole. The causal structure of the metric implies then that no signal with velocity smaller or equal to the velocity of light can escape from the inside of the black hole. It seems reasonable to expect that an instantaneous interaction such as the one discussed here could carry information through the black hole horizon while interactions which propagate at finite velocities $v \le c$ cannot. Hence, one does not expect the \textquotedblleft no-hair\textquotedblright theorem to hold in Lorentz-violating massive gravity. It is worth noting that this instantaneous interaction should also allow to look behind the cosmological horizon, since the cosmological horizon is somehow similar to the black hole horizon. It would be interesting to study this last issue in more detail.

\chapter{Black Holes} \label{ch:bh}

In GR, a crucial role is played by the spherically symmetric vacuum solution to the Einstein equations -- the Schwarzschild solution. This role is twofold. First, this solution describes the metric outside of spherical non-rotating bodies and gives rise, in the weak field limit, to Newtonian gravity. It provides therefore a useful approximation in many astrophysical situations. Second, the Schwarzschild solution describes the result of the gravitational collapse of massive stars, black holes \cite{1983mtbh.book.....C,Townsend:1997ku,Frolov:1998wf}. Although the existence of black holes has not yet been directly confirmed, there exists indirect evidences that some of binary stellar systems contain black holes as one of the companion \cite{1986ApJ...308..110M,McClintock:2003gx}, and that many galaxies, including the Milky Way, harbor super-massive black holes in their centers \cite{Genzel:1997im,Gillessen:2008qv}. It is conceivable that black holes will be directly observed in the near future, and that their properties, including the metric configuration near the horizon, will be quantitatively tested \cite{Poisson:1996tc,Rees:1997eb}, thus providing a probe of GR in a fully non-linear regime.

\section{Hairy black holes} \label{sc:hair}

Because of the horizon enclosing each black holes that acts as a \textquotedblleft one-way membrane\textquotedblright, the properties of black holes are extremely resistant to any modifications. Particles and radiation can enter the horizon from outside, but escaping to the exterior is impossible. Consequently, information cannot get away from a black hole and most of the physical properties of an object (or field) falling into a black hole become unobservable once the object (or field) crosses the horizon. For instance, the baryon number of a molecule falling into a black hole is not transferred to the black hole once the molecule enters the horizon: it simply becomes unobservable \cite{Bekenstein:1971hc,Bekenstein:1972ky}.

The uniqueness theorems for the Reissner-Nordstr\"{o}m \cite{Israel:1967wq} and Kerr \cite{Carter:1971zc,Wald:1971iw} black holes follow from this particularity of the horizon, as the statement by Wheeler that \textquotedblleft\emph{Black holes have no hair}\textquotedblright \cite{1971PhT....24a..30R}. These theorems claim that any collapsing object will result in a stationary black hole completely characterized by its mass, electric charge and angular momentum, regardless of the details of the collapse or the properties of the collapsing object \cite{Price:1971fb,Teitelboim:1972qx,Adler:1978dp} \footnote{For a recent discussion of stationary black holes with global charges, see \cite{Heusler:1998ua} and references therein.}. It is natural that these quantities remain observable since they are all conserved quantities subject to a Gauss law, implying that they can be measured by a distant observer. Hence, if one defines hair as those free parameters of the black hole which are not subject to a Gauss law \cite{Bekenstein:1996pn}, one indeed concludes that black holes have no hair in GR.

The Schwarzschild metric, together with properly arranged scalar fields, is a solution to Einstein equations in massive gravity as well \cite{Dubovsky:2007zi}. The following solution for the scalar fields
\begin{eqnarray*}
\phi^0 &=& t + 2 \sqrt{r r_s} + r_s \ln \dfrac{\sqrt{r} - \sqrt{r_s}}{\sqrt{r} + \sqrt{r_s}} , \\
\phi^i &=& x^i ,
\end{eqnarray*}
where $r_s$ is the Schwarzschild radius of the black hole, together with the Schwarzschild metric imply that the energy-momentum tensor for the Goldstone fields vanishes outside of the black hole\footnote{For this Schwarzschild solution of massive gravity, it is straightforward to show that $X = 1$ while $\tx{Tr} \left( W^n \right) = (-)^{n} 3$ as in Minkowski space-time. Hence, the energy-momentum tensor for the Goldstone fields reduces to its value in the vacuum which is zero.}. Hence, this field configuration is a solution to the massive gravitational field equations. Still, the properties of black holes are expected to be different in massive gravity with spontaneous breaking of Lorentz-invariance. For instance, rotating black holes carry a long-range tensor component (quadrupole moment) which is certainly modified when tensor modes acquire a mass \cite{Dubovsky:2007zi}.

More generally, black holes are expected to have hair \cite{Dubovsky:2007zi} in massive gravity theories because of the physical instantaneous interaction present in these models. As already discussed, this instantaneous interaction is supported by the three fields $\phi^i$ which propagate in the effective metric (\ref{eq:effectve_metric}) with $\epsilon = 1$. Hence, these fields have an infinite propagation velocity in the vacuum. In order to understand the dynamics of these fields in the Schwarzschild space-time, let us write the Schwarzschild solution of massive gravity in a coordinate system regular at the horizon :
\begin{eqnarray*}
\dif s^2 &=& \dif \tau^2 - \dfrac{r_s}{r \left( \tau, R , r_s \right)} \dif R^2 - r^2 \left( \tau, R , r_s \right) \dif \Omega^2 , \\
\phi^0 &=& \tau , \\
\phi^i &=& r \left( \tau, R , r_s \right) \times \left(\cos \varphi \sin \theta , \sin \varphi \sin \theta, \cos \theta \right) ,
\end{eqnarray*}
where
\begin{eqnarray*}
r \left( \tau, R , r_s \right) = \left[ \dfrac{3}{2} \left( R - \tau \right) \sqrt{r_s} \right]^{2/3} .
\end{eqnarray*}
This coordinate system, known as the Lema\^{\i}tre reference frame, corresponds to the frame of free falling observers \cite{Frolov:1998wf}. In this frame, the effective metric $G^{\mu\nu}$ in which the fields $\phi^i$ propagate is simply given by
\begin{eqnarray*}
\dif s_G^2 = \left( 1 - \epsilon \right) \dif \tau^2 - \dfrac{r_s}{r \left( \tau, R , r_s \right)} \dif R^2 - r^2 \left( \tau, R , r_s \right) \dif \Omega^2 .
\end{eqnarray*}
Although we are interested in the case $\epsilon = 1$, it is instructive to keep $\epsilon$ general for the sake of argument. The coefficient in front of $\dif \tau^2$ can be absorbed by the following rescaling
\begin{eqnarray*}
\tau \rightarrow \dfrac{\tau}{\sqrt{1 - \epsilon}} , & R \rightarrow \dfrac{R}{\sqrt{1 - \epsilon}},
\end{eqnarray*}
with consequence that the effective metric reduces to the Lema\^{\i}tre one with a different Schwarzschild radius $\tilde{r}_s$
\begin{eqnarray*}
\dif s_G^2 = \dif \tau^2 - \dfrac{\tilde{r}_s}{r \left( \tau, R , \tilde{r}_s \right)} \dif R^2 - r^2 \left( \tau, R , \tilde{r}_s \right) \dif \Omega^2 , & \tilde{r}_s = \left( 1 - \epsilon \right) r_s.
\end{eqnarray*}
This effective metric has an horizon at $r \left( \tau, R , \tilde{r}_s \right) = \tilde{r}_s$. Consequently, black holes appear larger for sub-luminal fields ($\epsilon < 0$) and smaller for superluminal ($0 < \epsilon < 1$) ones. Although the previous coordinates redefinition becomes singular in the limit $\epsilon \rightarrow 1$, one may wonder what is the value of $\tilde{r}_s$ in this limit which corresponds to the instantaneous fields of interest: the effective Schwarzschild radius $\tilde{r}_s$ goes to zero when $\epsilon \rightarrow 1$. Hence, one may conclude that there is no horizon for these fields, and therefore that they can escape from black holes.

As expected, this short discussion implies that instantaneous interaction can carry information outside of black holes, which should therefore have a large number of hair. This conclusion, based perhaps on naive arguments, is reinforced by direct calculations made in the context of GR with the model of Lorentz-violating electrodynamics introduced in section \ref{sc:electro} for which there is an instantaneous electromagnetic interaction. Indeed, by considering neutral non-spherically symmetric black holes, the authors of Ref.~\cite{Dubovsky:2007zi} have demonstrated the existence of hair which can be interpreted as the electric dipole moment of the black holes.

There are also thermodynamical arguments in favor of black hole hair in massive gravity models with spontaneous breaking of Lorentz symmetry. The fact that the black hole horizon appears larger for sub-luminal particles and smaller for superluminal ones implies that the temperature of the Hawking radiation emitted by the black hole depends on the propagation velocity of the fields emitted \cite{Dubovsky:2006vk}. Hence, the temperature of this radiation is not universal any longer, and one may think of processes such that the black hole mass and angular momentum remain constant while the entropy outside decreases \cite{Dubovsky:2006vk,Eling:2007qd}, invalidating the second law of thermodynamics. Still, on top of reducing the entropy outside of the black hole, such processes also produce change inside the black hole. Consequently, the validity of the second law could be restore if these changes are observable from outside, i.e., if black holes have hair on top of their mass and angular momentum \cite{Dubovsky:2007zi}.

The possible existence of hairy black hole in massive gravity models suggests that there might exist spherically symmetric solutions other than the Schwarzschild one. This supposition has been confirmed by finding of exact spherically symmetric vacuum solutions in massive gravity theories \cite{Bebronne:2009mz} corresponding to modified black holes. This chapter is devoted to these solutions.

\section{Static spherically symmetric ansatz and equations} \label{sc:BH_Ansatz}

The first step in the determination of exact static spherically symmetric solutions consist in finding a good ansatz to describe them. The ansatz for the metric is the same as in GR. Hence, the only subtlety consist in getting the correct configuration for the four scalar fields. With the knowledge of the vacuum solution (\ref{eq:MG_vacuum}), it is possible to convince oneself that any static spherically symmetric configuration in massive gravity models can be written in the following form,
\begin{eqnarray*}
\dif s^2 &=& \alpha \left( r \right) \dif t^2 + 2 \delta \left( r \right) \dif t \dif r - \beta \left( r \right) \dif r^2 - \kappa \left( r \right) \dif \Omega^2 , \nonumber \\
\phi^0 &=& t + h \left( r \right) , \nonumber \\
\phi^i &=& \phi \left( r \right) \dfrac{x^i}{r} .
\end{eqnarray*}
This field configuration is invariant under two residual coordinate transformations. The first one is an arbitrary change of the radial coordinate $r \rightarrow r^\prime = r^\prime \left( r \right)$ which allows to set either $\kappa = r^2$ or $\phi = r$. The second one
consist in redefining the time variable $t \rightarrow t^\prime = t + \tau \left( r \right)$. This last transformation allows one to cancel
either $\delta \left( r \right)$ or $h \left( r \right)$. We  choose the conditions $\kappa = r^2$ and $\delta = 0$. Thus, we get the following ansatz,
\begin{eqnarray} \label{eq:BH_ansatz}
\dif s^2 &=& \alpha \left( r \right) \dif t^2 - \beta\left( r \right) \dif r^2 - r^2 \left( \dif \theta^2 + \sin^2 \theta \dif \varphi^2 \right) , \nonumber \\
\phi^0 &=& t + h \left( r \right) , \nonumber \\
\phi^i &=& \phi \left( r \right) \dfrac{x^i}{r} .
\end{eqnarray}
As compared to GR, this configuration contains two additional radial functions, $h \left( r \right)$ and $\phi \left( r \right)$.

As has been pointed out in section \ref{sc:LVMG_Action}, the rotational invariance of the vacuum (and likewise, of the ansatz (\ref{eq:BH_ansatz})) requires that the function $\func \left( X , W^{ij} \right)$ depends on $W^{ij}$ through three scalar combinations $w_n = \tx{Tr}\left(W^n\right)$. Hence, the function $\func$ depends on four scalar variables which are expressed in terms of the radial functions and their derivatives as follows,
\begin{eqnarray*} \label{eq:w_n}
\begin{array}{lllllll}
X = \dfrac{\beta - \alpha h'^2}{\alpha\beta} , && w_{1} = - \left( f_{1} + 2 f_{2} \right) , && w_{2} = f_{1}^2 + 2 f_{2}^2 , && w_{3} = - \left( f_{1}^3 + 2 f_{2}^3 \right) ,
\end{array}
\end{eqnarray*}
where the two functions $f_{1}$ and $f_{2}$ are given by
\begin{eqnarray*}
f_{1} = \dfrac{\phi^{\prime 2}}{\alpha\beta X} , & f_{2} = \dfrac{\phi^{2}}{r^2} .
\end{eqnarray*}
In these expressions and in what follows, the prime denotes the derivative with respect to the radial coordinate $r$.

With the ansatz (\ref{eq:BH_ansatz}), six of the ten Einstein equations are identically satisfied. Consequently, there is only four equations left to be solved for four unknown radial functions,
\begin{eqnarray*}
\mathcal{G}_{0}^{0} = \dfrac{1}{\Mpl} t_{0}^{0} , & \mathcal{G}_{r}^{r} = \dfrac{1}{\Mpl} t_{r}^{r} , \\
\mathcal{G}_{\theta}^{\theta} =  \dfrac{1}{\Mpl}t_{\theta}^{\theta}, & t_{0}^{r} = 0 .
\end{eqnarray*}
The explicit expressions for the components of $\mathcal{G}_\mu^\nu$ and $t_\mu^\nu$ are given in the appendix \ref{app:sc:bh}. Consider first the equation $t_{0}^{r} = 0$. Assuming $h^\prime \neq 0$, this equation gives 
\begin{eqnarray} \label{eq:BH:T_0r=0}
0 = X \mathcal{F}_X + f_{1} \left( \mathcal{F}_1 - 2 f_1 \mathcal{F}_2 + 3 f_{1}^2 \mathcal{F}_3 \right),
\end{eqnarray}
where $\mathcal{F}_X \equiv \partial \mathcal{F}/ \partial X$ as previously while $\mathcal{F}_i \equiv \partial \mathcal{F}/ \partial w_i$. Furthermore, the time and radial components of the energy-momentum tensor differ by the quantity proportional to eq.~(\ref{eq:BH:T_0r=0}). Therefore, when this equation holds one has $t_{0}^{0} = t_{r}^{r}$. This implies that $\mathcal{G}_{0}^{0} = \mathcal{G}_{r}^{r}$ or, equivalently,
\begin{eqnarray*}
\alpha \left( r \right) \beta \left( r \right) = 1 ,
\end{eqnarray*}
in full analogy with the Schwarzschild solution in GR. Hence, the ten Einstein equations reduce to the following four equations,
\begin{eqnarray}
1 &=& \alpha \beta , \label{eq:BH:equat1} \\
0 &=& \dfrac{\alpha^{\prime}}{r} + \dfrac{\alpha - 1}{r^2} - \dfrac{m^2}{2} \left( \mathcal{F} - 2 X \mathcal{F}_X \right) , \label{eq:BH:equat2} \\
0 &=& \dfrac{\alpha^{\prime}}{r} + \dfrac{\alpha^{\prime\prime}}{2} - \dfrac{m^2}{2} \left( \mathcal{F} + X \mathcal{F}_X - w_{1} \mathcal{F}_{1}- 2 w_{2} \mathcal{F}_{2} - 3 w_{3} \mathcal{F}_{3} \right) , \label{eq:BH:equat3} \\
0 &=& X \mathcal{F}_X + f_{1} \left( \mathcal{F}_1 - 2 f_1 \mathcal{F}_2 + 3 f_{1}^2 \mathcal{F}_3 \right) , \label{eq:BH:equat4}
\end{eqnarray}
For a generic function $\mathcal{F}$, this system of equations is well defined. Indeed, since the function $h(r)$
enters eqs.~(\ref{eq:BH:equat1} - \ref{eq:BH:equat4}) only through the variable $X$, one may consider $X$ as an independent variable
instead of $h(r)$. Then the fourth equation allows to find $\phi$ in terms of $X$, while the first equation gives $\beta$ in terms of
$\alpha$. The second equation then gives $X$ in terms of $\alpha$ and the third equation allows to determine $\alpha$ as a function of $r$.

\section{Analytical example} \label{sc:BH_XW}

Finding analytical solutions of the non-linear system of equations like (\ref{eq:BH:equat1}--\ref{eq:BH:equat4}) is impossible for a generic
function ${\cal F}$. So, in order to get some insight into the behavior of the solutions, let us choose the function $\mathcal{F}$ in such a way that the resulting equations are solvable analytically.

Consider the function $\mathcal{F}$ of the following form,
\begin{eqnarray} \label{eq:BH:fct-XW}
\mathcal{F} = c_0 \left( \dfrac{1}{X} + w_1 \right) + c_1 \left( w_{1}^3 - 3 w_{1} w_{2} - 6 w_1 + 2 w_{3} - 12 \right),
\end{eqnarray}
where $c_0$ is an arbitrary dimensionless constant and $c_1 = \pm 1$ (the numerical value of $c_1$ can be absorbed into the constant $\Lambda$). The coefficients inside the parentheses are chosen in such a way that the vacuum (\ref{eq:MG_vacuum}) is a solution to the
Einstein equations. Our example contains, therefore, a single continuous free parameter $c_0$. Two additional constraints should be imposed on $c_0$. The first one comes from the requirement that the graviton is non-tachyonic, $m_2^2 > 0$. This translates into the inequality
\begin{equation} \label{eq:BH:positive-mass}
c_0 - 6 c_1 \geq 0 .
\end{equation}
The second condition is necessary to ensure that scalar modes with pathological behavior do not reappear upon addition of higher-derivative terms (recall that the model (\ref{eq:MG_action}) is understood as the low-energy effective theory, so such terms are generically present). This condition is given by the inequality (\ref{eq:MG_ghost_UV}) which in the case at hand implies
\begin{equation} \label{eq:BH:no-tachyons}
c_0 > 0 .
\end{equation}

Before solving the Einstein equations for the particular model of massive gravity based on the function (\ref{eq:BH:fct-XW}), we should stress that the choice of this functional form is by no means unique. This particular form has been chosen in order to simplify the solution of the field equations, as will become clear below.

\subsection{The static spherically symmetric solutions}

Let us start with eq.~(\ref{eq:BH:equat4}) which has motivated our choice of the particular functional form (\ref{eq:BH:fct-XW}). This equation reads
\begin{eqnarray*}
0 = \dfrac{1}{X} \left[ - c_{0} + \phi^{\prime2} \left( 6 c_1 \dfrac{\phi^{4}}{r^4} + c_0 - 6 c_1  \right) \right] .
\end{eqnarray*}
Because of our choice of the function $\mathcal{F}$, this equation contains the variable $X$ as an overall factor only. Hence, it reduces
to a closed differential equation for $\phi$. The solution to this equation is
\begin{eqnarray*}
\phi = b r,
\end{eqnarray*}
where the constant $b$ satisfies the equation
\begin{eqnarray} \label{eq:BH:b}
0 = \left( b^2 - 1 \right) \left( 6 b^4 + 6 b^2 + \dfrac{c_0}{c_1} \right).
\end{eqnarray}
We are interested in real positive values\footnote{The case $b<0$ can be reduced to $b>0$ by the inversion of coordinates.} of $b$. For $c_0/c_1 > 0$ there is only one such solution,
\begin{eqnarray*}
b = 1,
\end{eqnarray*}
while for $c_0 / c_1 < 0$ there exists another one
\begin{eqnarray*}
b = \dfrac{1}{\sqrt{2}} \left( - 1 + \sqrt{1 - 2 c_0 / 3 c_1} \right)^{1/2}.
\end{eqnarray*}
Thus, at negative $c_0/c_1$ we have two different branches of solutions. Then, the remaining two equations \refp{eq:BH:equat2} and \refp{eq:BH:equat3} can be written as follows
\begin{eqnarray}
0 &=& \dfrac{\alpha^{\prime}}{r} + \dfrac{\alpha - 1}{r^2} - 3 \Lambda_{c} + m^2 c_0 \left( 1 - \dfrac{1}{X} \right) , \\
0 &=& \alpha^{\prime\prime} + \lambda \dfrac{\alpha - 1}{r^2} + \left( \dfrac{\alpha^{\prime}}{r} - 3 \Lambda_{c} \right) \left( 2 + \lambda \right) , \label{eq:BH:alpha(r)}
\end{eqnarray}
where
\begin{eqnarray*}
\lambda = - 12 b^6 \dfrac{c_1}{c_{0}} , && \Lambda_{c} = 2 m^2 c_1 \left( b^{6} - 1 \right).
\end{eqnarray*}
Recall that, according to our normalization, the constant $c_1$ only takes two values, $c_1 = \pm 1$. It is worth noting that $\Lambda_{c}$ corresponds to the cosmological constant present in the model.

Eq.~(\ref{eq:BH:alpha(r)}) is a linear inhomogeneous equation for $\alpha$. Its general solution can be found analytically. Making use
of this solution and integrating the remaining equations one finally obtains
\begin{eqnarray}
\dif s^2 &=& \alpha \, \dif t^2 - \alpha^{-1} \dif r^2 - r^2 \dif \Omega^2 , \nonumber \\
\phi^0 &=& t \pm \int \dfrac{\dif r}{\alpha} \left[ 1 - \alpha \left( \dfrac{S}{c_{0} m^2} \dfrac{\lambda-1}{r^{\lambda+2}} + 1 \right)^{-1} \right]^{1/2} , \nonumber \\
\phi^i &=& b x^i , \label{eq:BH:sol-XW}
\end{eqnarray}
where
\begin{eqnarray}  \label{eq:BH:sol_alpha}
\alpha \left( r \right) &=& 1 - \dfrac{r_{s}}{r} - \dfrac{S}{r^{\lambda}} + \Lambda_{c} r^2 .
\end{eqnarray}
Here $r_{s}$ and $S$ are two integration constants: $r_{s}$ is the usual Schwarzschild radius while $S$ is a scalar charge whose presence
reflects the modification of the gravitational interaction as compared to GR. At $S = 0$ this solution reduces to the conventional Schwarzschild solution describing a black hole of mass $M = r_s \left( 2 G \right)^{-1}$. The behavior of this class of solutions is determined by the two integration constants $r_s$ and $S$, and the value of the parameter $\lambda$. Let us discussed their properties.

\subsubsection{Asymptotic behavior}

The behavior of these solutions at spatial infinity depends on the constants $c_0$ and $c_1 = \pm 1$. Models with $c_1 = 1$ require $c_0 \geq 6$ to ensure that the graviton is non-tachyonic. Then, the only solution of eq.~(\ref{eq:BH:b}) is $b = 1$. Hence, there is no cosmological constant $\Lambda_c = 0$ and $\lambda < 0$ so that the metric is growing at spatial infinity as $S r^{|\lambda|}$. Such solutions do not describe asymptotically flat space-time and will therefore not be discussed further.

Models characterized by $c_1 = - 1$ are more interesting, provided that $c_0 > 0$ in order to satisfy eqs.~(\ref{eq:BH:positive-mass}) and (\ref{eq:BH:no-tachyons}). In that case $\lambda > 0$ and two branches of solutions exist. The first one posses a non-zero $\Lambda_c$, which can be positive or negative depending on the numerical value of $c_0$. These solutions asymptote to the generalization of the vacuum solution in presence of a non-zero cosmological constant. The second branch of solutions are characterized by $\Lambda_c = 0$ (and therefore by $b = 1$). They asymptote to Minkowski space-time. We will only consider solutions with $\Lambda_c = 0$ in what follows.

\subsubsection{The black-hole mass}

The total mass of the static, asymptotically flat space-time under consideration is given by the Komar mass \cite{Wald:1984rg} which takes the following form
\begin{eqnarray*}
M_{\tx{Komar}} \left( r \right) = \dfrac{1}{2 G} \left( r_s + \dfrac{\lambda S}{r^{\lambda-1}} \right) .
\end{eqnarray*}
\begin{itemize}
\item If $0 < \lambda < 1$, the third term on the right-hand side of eq.~(\ref{eq:BH:sol_alpha}) dominates at large distances with consequence that the Komar mass of these solutions grows with distances.
\item If $\lambda > 1$, the standard Schwarzschild term dominates at infinity implying that the Komar mass tends to $M \equiv r_s \left( 2 G \right)^{-1}$ when $r \rightarrow \infty$. The solutions with positive (negative) $M$ have attractive (repulsive) behavior at infinity.
\end{itemize}

\subsubsection{Behavior of the solutions}

At the origin $r=0$ both terms proportional to $r_s$ and $S$ are singular, so the metric always possesses a singularity unless $r_s = S = 0$. This singularity may or may not be hidden by the horizon depending on the signs and values of $r_s$ and $S$. The solutions possessing the horizon are candidates for modified black holes.

The horizon is always present if both $r_s$ and $S$ are positive. Such black holes have attractive gravitational potential at all distances,
which is stronger than for a conventional black hole of mass $M$. The horizon size of the modified black hole is larger than $r_s = 2 G M$. Solution where both $r_s$ and $S$ are negative imply to naked singularities and will therefore be disregarded. Let us study models with $\lambda > 1$ and $0 < \lambda < 1$ separately.
\begin{itemize}
\item For models characterized by $\lambda > 1$, the standard Schwarzschild term dominates at infinity. If $r_s > 0$ and $S < 0$, the presence of the horizon depends on the relative values of $S$ and $r_s$. It exists for sufficiently small $|S|$. Defining, on dimensional grounds, the mass parameter $s$ associated with $S$ by the relation $|S|=s^{-\lambda}$, the existence of an horizon requires that
\begin{equation} \label{eq:BH:horizon_1}
s \geq \dfrac{\lambda}{r_s} \left( \dfrac{1}{\lambda - 1} \right)^{\dfrac{\lambda - 1}{\lambda}} .
\end{equation}
The Newtonian potentials for solutions satisfying and not satisfying the condition (\ref{eq:BH:horizon_1}) are shown in Fig.~\ref{fg:BH:XW} - (a) and (b), respectively. When the horizon exists, the gravitational field is attractive all the way to the horizon. The attraction is weaker than in the case of the usual Schwarzschild black hole of mass $r_s$, and the horizon size is smaller. The behavior of the gravitational force with distance mimics that of the smaller-mass black hole plus a continuous distribution of \textquotedblleft dark matter\textquotedblright, with the total mass enclosed within the radius $r$ approaching $M$ as $r \rightarrow \infty$.

At $r_s < 0$ and $S > 0$, the modified black hole anti-gravitates at large distances and gravitates close to the horizon. The attraction changes to repulsion at
\begin{equation*}
r = r_{*} \equiv  \left| \dfrac{\lambda S}{r_s} \right|^{\dfrac{1}{\lambda - 1}}.
\end{equation*}
The corresponding Newtonian potential is shown in Fig.~\ref{fg:BH:XW} - (c).

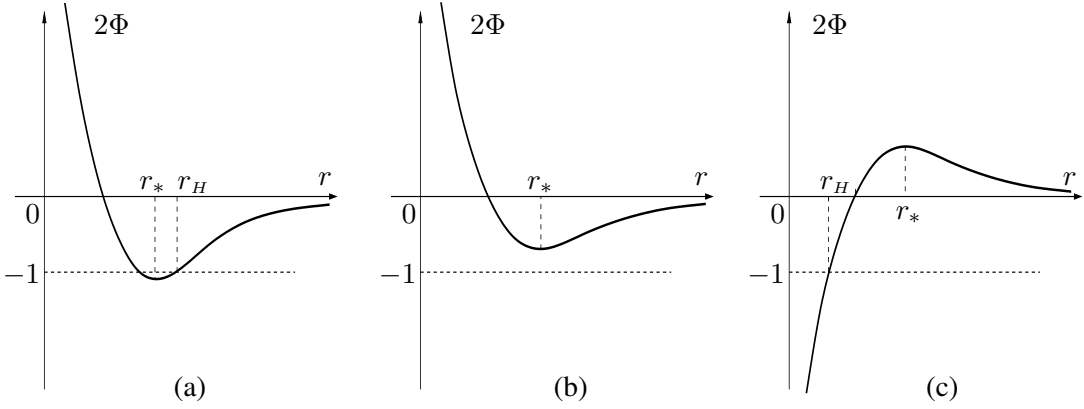
\begin{figure*}
\setlength{\unitlength}{1cm}
\begin{picture}(10,4.5)
\put(1.15,2.3){$0$}
\put(6.1,2.3){$0$}
\put(10.95,2.3){$0$}
\put(0.85,1.5){$-1$}
\put(5.8,1.5){$-1$}
\put(10.65,1.5){$-1$}
\put(2.05,4.8){$2 \Phi$}
\put(7,4.8){$2 \Phi$}
\put(11.5,4.8){$2 \Phi$}
\put(5,2.8){$r$}
\put(9.95,2.8){$r$}
\put(14.8,2.8){$r$}
\put(3.15,2.75){$r_{\scriptscriptstyle{H}}$}
\put(11.6,2.75){$r_{\scriptscriptstyle{H}}$}
\put(2.65,2.75){$r_{*}$}
\put(7.75,2.75){$r_{*}$}
\put(12.6,2.35){$r_{*}$}
\put(3.1,0){(a)}
\put(8.1,0){(b)}
\put(13,0){(c)}
\put(1,0){\includegraphics[angle=0,width=400pt,height=150pt]{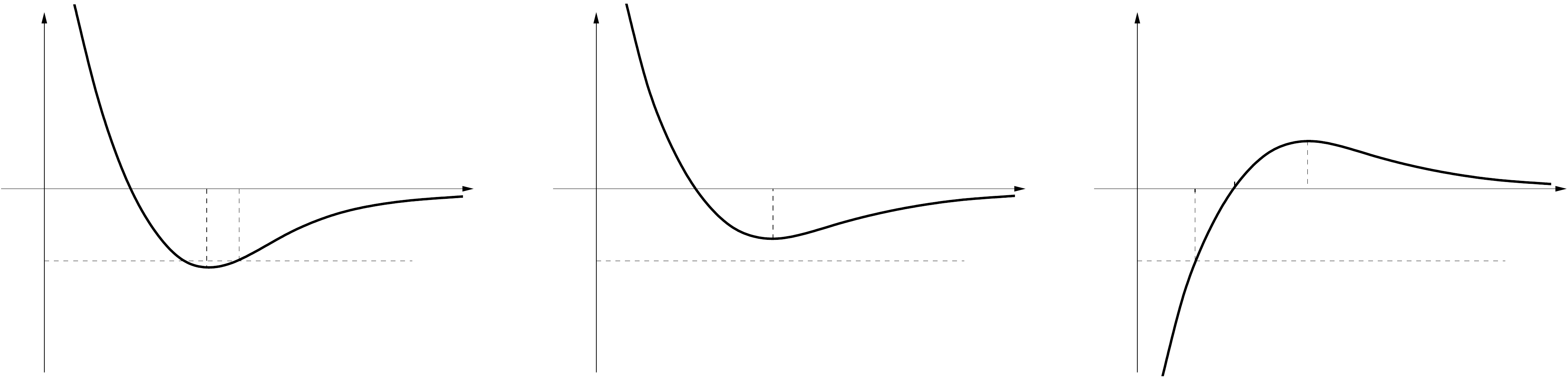}}
\end{picture}
\caption{The deviation of $g_{00}$ from one (proportional to the Newtonian potential $g_{00} - 1 = 2 \Phi$) for tree different situations. For models characterized by $\lambda > 1$, Figs.~(a) and (b) correspond to $r_s > 0$ and $S < 0$ with the numerical values satisfying (a) and not satisfying (b) eq.~(\ref{eq:BH:horizon_1}), while Fig.~(c) represents solution with $r_s < 0$ and $S > 0$. For models characterized by $0 < \lambda < 1$, Figs.~(a) and (b) correspond to $r_s < 0$ and $S > 0$ with the numerical values satisfying (a) and not satisfying (b) eq.~(\ref{eq:BH:horizon_2}), while Fig.~(c) represents solution with $r_s > 0$ and $S < 0$}
\label{fg:BH:XW}
\end{figure*}
\item For models characterized by $0 < \lambda < 1$, the new contribution proportional to the scalar charge $S$ dominates at infinity. If
$r_s > 0$ and $S < 0$, the modified black hole anti-gravitates at large distances and gravitates close to the horizon. As before, the attraction changes to repulsion at $r = r_{*}$ and the corresponding Newtonian potential is shown in Fig.~\ref{fg:BH:XW} - (c).

Finally, at $r_s < 0$ and $S > 0$, the presence of the horizon depends on the relative values of $S$ and $r_s$. The existence of an horizon requires that
\begin{equation} \label{eq:BH:horizon_2}
|r_s| \geq \dfrac{\lambda}{s} \left( \dfrac{1}{\lambda - 1} \right)^{\dfrac{\lambda - 1}{\lambda}} .
\end{equation}
The Newtonian potentials for solutions satisfying and not satisfying the condition (\ref{eq:BH:horizon_2}) are shown in Fig.~\ref{fg:BH:XW} - (a) and (b), respectively. As previously, when the horizon exists, the gravitational field is attractive all the way to the horizon.
\end{itemize}

A remark is in order at this point. In the conventional GR the constant $r_s$ or, equivalently, the black hole mass $M$, is also a free parameter which can, in principle, be positive or negative. In GR, however, only positive values make sense for the following reasons.  First, negative-mass Schwarzschild solutions possess naked singularity at the origin, which is physically unacceptable. Second, the conventional matter satisfies the null energy condition which ensures that any compact spherically-symmetric matter distribution has a positive mass \cite{Hawking:1973uf}. None of these arguments go through in the case of massive gravity. Fig.~\ref{fg:BH:XW} - (c) gives an example of solution with repulsive behavior at large distances and without naked singularity: as for a conventional black hole, the singularity of this solution is hidden behind the horizon.  The positivity of energy is also not expected in massive gravity. This is related to the fact that the background (\ref{eq:MG_vacuum}) breaks time translations, and only the combination of the time translations with the shifts of $\phi^0$ by a constant remains unbroken. In this respect the massive gravity model is exactly analogous to the Ghost Condensate model, where the negative-energy states have been constructed explicitely \cite{Arkani-Hamed:2005gu}.

\subsection{Correspondence with linear analysis}

The solutions found in the previous section have the asymptotic behavior different from that obtained in the linear perturbation theory discussed in chapter \ref{ch:weak_field}. In order to compare the exact solution with the results of the linear perturbation theory, let us discuss the exact solution in the gauge where $h \left( r \right) = 0$, $\delta \left( r \right) \neq 0$ (cf. section \ref{sc:BH_Ansatz}). In this gauge the perturbation theory corresponds to assuming that the variations of all metric components are of the same order. In other words, they are formally assigned a small parameter $\epsilon$ to the first power. The solutions described in chapter \ref{ch:weak_field} satisfy the Einstein equations expanded to the linear order in $\epsilon$.

The solution (\ref{eq:BH:sol-XW}) is not of this type. Transforming it into the gauge $h \left( r \right) = 0$ one finds that $g_{0r} = \delta \left( r \right)$ does not decay as fast as the perturbations of other components, for instance, as $\alpha - 1$. In fact, in the equations expanded in powers of perturbations the terms of order $\delta^2$ balance those linear in $\alpha - 1$. In other words, in the formal expansion of the solution (\ref{eq:BH:sol-XW}) in powers of the small parameter $\epsilon$ the perturbation $\delta$ should be assigned the order $\sqrt{\epsilon}$ rather than $\epsilon$. Hence, the solution (\ref{eq:BH:sol-XW}) is non-linear even at large distances from the center.  A similar phenomenon has been observed in the context of bigravity models in Ref.~\cite{Berezhiani:2008nr}.

Another difference between the solution (\ref{eq:BH:sol-XW}) and the solution to the linearized equations is that the former is static, while in the latter only metric components are static (in the gauge $g_{0i} = 0$). Indeed, the scalar fields $\Xi_0$ have time dependence which may be viewed as an accretion of a fluid with zero energy-momentum tensor (see eq.~(\ref{eq:MG_Xi0})).

\subsection{Gravitational field of a star} \label{sc:BH:grav-field-star}

In GR, one may relate the mass of a star to an integration constant of the vacuum solution in the exterior space by matching the interior and exterior solutions at the star surface (see, e.g., Ref.~\cite{Weinberg:1972}). In massive gravity, one may try to use the same approach to determine the scalar charge $S$ of an ordinary star. The analytical solution in the interior region is required for the matching procedure.

The star is described, to a good approximation, by a diagonal energy-momentum tensor $\mathcal{T}_{\mu}^{\nu} = ( \rho, -p, -p, -p )$, where $\rho$ and $p$ are the energy density and pressure inside the star, respectively. This energy-momentum tensor is assumed to be responsible for the external gravitational field described by eqs.~(\ref{eq:BH:sol-XW}). Since there is no direct coupling between the ordinary matter and the Goldstone fields, $\mathcal{T}_{\mu}^{\nu}$ must be conserved separately, $\nabla^\mu \mathcal{T}_{\mu}^{\nu} = 0$. For simplicity, we take the energy density to be constant at $r < R$, where $r = R$ is the surface of the star, and zero outside. The pressure $p$ cannot be chosen independently; it is determined by the conservation of $\mathcal{T}_{\mu}^{\nu}$.

Because of the spherical symmetry, the ansatz (\ref{eq:BH_ansatz}) holds. The Einstein equations in the interior of the star are obtained from eqs.~(\ref{eq:BH:equat1} - \ref{eq:BH:equat3}) by adding the contributions of the energy-momentum of the star, while eq.~(\ref{eq:BH:equat4}) remains unchanged. The resulting set of equations can be solved analytically. The solution reads
\begin{eqnarray*}
\alpha \left( r \right) &=& 1 - \dfrac{s_{1}}{r} - \dfrac{s_{2}}{r^{\lambda}} + \Lambda_{c} r^2 + \dfrac{\rho}{\Mpl} \left( \dfrac{r^2}{6} - \dfrac{R^2}{2} \right) + \mathcal{O} \left( \rho^2 \right) , \\
\beta \left( r \right) &=& \left[ 1 - \dfrac{s_{1}}{r} - \dfrac{s_{2}}{r^{\lambda}} + \Lambda_{c} r^2 - \dfrac{r^2 \rho}{3 \Mpl} \right]^{-1} , \nonumber \\
h \left( r \right) &=& \pm \int \dfrac{\dif r}{\alpha} \left[ 1 - \alpha \left( \dfrac{s_{2}}{c_{0} m^2} \dfrac{\lambda-1}{r^{\lambda+2}} + 1 \right)^{-1} \right]^{1/2} , \nonumber \\
\phi \left( r \right) &=& b r . \nonumber
\end{eqnarray*}
For simplicity, we have expanded the first equation in powers of $\rho$, while the other relations are exact. Since the geometry inside the star is regular, the integration constants $s_{1}$ and $s_{2}$ must be set to zero.

The interior solution has to be matched with the solution (\ref{eq:BH:sol-XW}) at $r=R$. It is convenient to match the variable $X$ which equals 1 in the interior region. In the gauge $h(r)=0$ this variable is nothing but the $g^{00}$ component of the metric. Hence, it must be continuous.  Making use of eqs.~(\ref{eq:BH:sol-XW}) one can see that the continuity of $X$ at $r=R$ requires that $S=0$. Therefore, the scalar charge of an ordinary star is zero.

It remains an open question how objects (e.g., black holes) with
$S\neq 0$ can be created. The argument given above does not apply to
time-dependent configurations, so it is possible that a non-zero
scalar charge may be acquired during the gravitational collapse.

\section[Numerical example]{$\mathcal{F}(Z^{ij})$ models}

As mentioned earlier, models characterized by the function $\func$ of a single variable $Z^{ij} = X^\gamma W^{ij}$ are of a particular interest. We discuss in this section the exact static spherically symmetric solutions in these models. Our goal is to demonstrate that the solutions found earlier are not specific to the particular form of the function (\ref{eq:BH:fct-XW}) and exist also in models obeying the dilatation symmetry (\ref{eq:dilatation_Z}).

In section \ref{sc:BH_XW} the analytical solutions of eqs.~(\ref{eq:BH:equat1} - \ref{eq:BH:equat4}) were obtained by choosing the function ${\cal F}$ in such a way that the dependence on $X$ factors out in eq.~(\ref{eq:BH:equat4}). Since now $\mathcal{F}$ has only one argument, the derivatives of $\mathcal{F}$ with respect to $X$ and $W^{ij}$ are no more independent. For this reason we did not succeed in constructing non-trivial examples where the Einstein equations are solvable analytically. Hence, to demonstrate the existence of unusual solutions we have to use numerical methods.

Consider the following function $\mathcal{F}$,
\begin{eqnarray} \label{eq:functZ}
\mathcal{F} = c_{0} \left( z_{1} + 2 + \dfrac{z_{1}^3 - 6 z_{1} z_{2} + 8 z_{3}}{3} \right) + 2 c_{1} \left( z_{1}^2 - 2 z_{2} - 1 - 2 \dfrac{z_{1}^3 - 6 z_{1} z_{2} + 8 z_{3}}{3} \right) ,
\end{eqnarray}
were $z_{n} \equiv \textrm{Tr} \left( Z^n \right)$ are tree independent scalars made out of $Z^{ij}$, with $z_{n} = \left( X^\gamma \right)^{n} w_{n}$. The coefficients in front of individual terms have been adjusted so that the flat metric and the scalar fields given by eq.~(\ref{eq:MG_vacuum}) solve the field equations. We are interested in solutions to the field equations that asymptote to this vacuum state.

In addition to the adjustments already made, the following inequality should be imposed on the coefficients $c_0$ and $c_1$ to ensure that the graviton is non-tachyonic,
\begin{equation} \label{eq:BH:positive-mass2}
c_0 - 2 c_1 \geq 0.
\end{equation}
This guarantees that the square of the graviton mass is non-negative. Moreover, this inequality is sufficient for the absence of pathological scalar modes which may appear upon addition of higher-derivative terms.  As in the previous example, the overall scale of the coefficients $c_0$ and $c_1$ can be absorbed in the parameter $\Lambda$, so without loss of generality we may set $c_1 = \pm 1$.

For this class of models, the field eqs.~(\ref{eq:BH:equat1} - \ref{eq:BH:equat4}) may be viewed as equations for
$\alpha(r)$, $\beta(r)$, $\vartheta(r) \equiv X^\gamma f_{1} - 1$ and $\varrho(r) \equiv X^\gamma f_{2} - 1$. Then eq.~\refp{eq:BH:equat1} gives $\beta$ in terms of $\alpha$, while eq.~\refp{eq:BH:equat4} enables to express $\varrho$ in terms of $\vartheta$. The two remaining equations form a coupled set of non-linear equations for $\alpha$ and $\vartheta$; they have to be solved numerically.

\begin{figure}
\begin{center}
\includegraphics[angle=0,width=10cm]{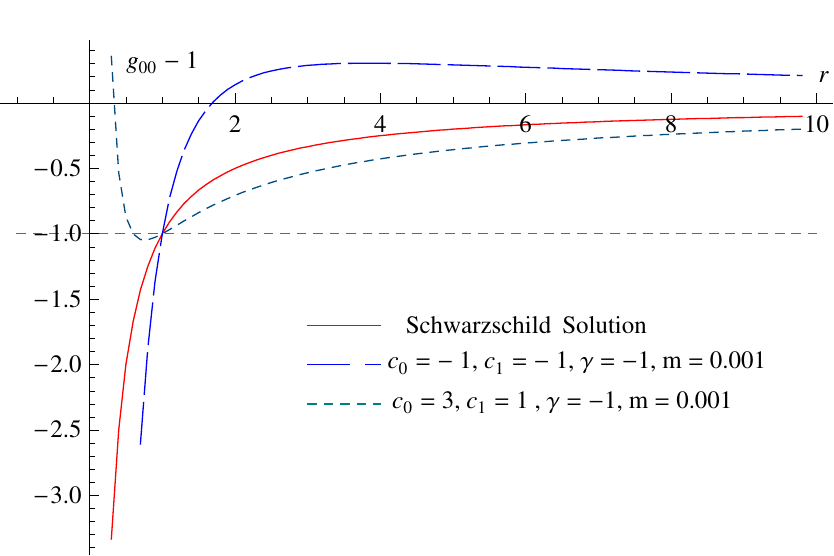}
\end{center}
\caption{The deviation of $g_{00}$ from one for three different cases: the usual Schwarzschild solution (solid line) and two solutions corresponding to different values of the parameters of the function \refp{eq:functZ} (long-dashed and short-dashed lines). The integration  constants of these solutions have been chosen such that the external
horizon is located at $r = 1$.} \label{fg:BH:Z}
\end{figure}

The numerical solutions are shown in Fig.~\ref{fg:BH:Z} for different value of the parameters $c_{0}$ and $c_{1}$. For all these graphs, we have assumed that the external horizon is located at $r = 1$ and that $\vartheta = 100$ at the horizon. The large value of $\vartheta$ is chosen in order to make the difference between the modified solution and the Schwarzschild solution visible on the plot (large values of $\vartheta$ correspond to large scalar charge $S$ of the previous section).

The plots show the behavior qualitatively similar to that discussed in section \ref{sc:BH_XW}. In particular:
\begin{itemize}
\item if $c_0 = c_1 = - 1$, the Newtonian potential $2 \Phi = \alpha - 1$ is attractive at short distances and becomes repulsive at larger distances;
\item if $c_0 = 3$ and $c_{1} = 1$, the Newtonian potential is attractive outside of the horizon, and becomes repulsive close to the singularity.
\end{itemize}
The deviations from the Schwarzschild metric are larger for larger values of the ``scalar charge'' (parameterized by the value of $\vartheta$ at the horizon). The Schwarzschild solution is recovered at $\vartheta\to 0$.

\section{Summary and prospects}

To summarize, there exist spherically symmetric vacuum solutions in massive gravity models which depend on two integration constants, the Schwarzschild radius $r_s$ and an extra parameter $S$ which can be called the \textquotedblleft scalar charge\textquotedblright. At zero value of the scalar charge the standard Schwarzschild solution is recovered, while at non-zero $S$ the metric is modified with respect to the Schwarzschild case.

The solutions having non-zero scalar charge exhibit much reacher behavior than the Schwarzschild solution in GR. As can be seen from the explicit example of section \ref{sc:BH_XW}, both the short and long distance behavior may be modified depending on the parameters of the model.

Unlike in GR, the solutions may have a negative Komar mass. Such solutions have repulsive gravitational interaction at large distances. At short distances the repulsion may change to attraction and give rise to the horizon, hiding the singularity at the origin. Such solutions represent anti-gravitating black holes.

In the case of a positive Komar mass, the $S$-dependent contributions may make the gravitational attraction weaker at short distances (cf. fig.~\ref{fg:BH:XW} - (a)). In this case the gravitational force decays with distance slower than $1/r^2$, thus mimicking the presence of dark matter. Interestingly, solutions with the same value of $M$ but different scalar charge $S$ have different behavior, which corresponds to different amount of the apparent \textquotedblleft dark matter\textquotedblright. This is in contrast with other models possessing modifications of the gravitational potential \cite{Milgrom:1983pn,Dvali:2000hr,Carroll:2003wy}, where the modification of the gravitational force is determined by the parameters of the model.

It is currently an open question how objects with non-zero scalar charge may be created. As has been argued in Sect.~\ref{sc:BH:grav-field-star}, the absence of direct coupling between the Goldstone fields and ordinary matter results in zero scalar charges of static matter distributions. Thus, the gravitational field of ordinary stars is described by the $S=0$ solutions, i.e., by the standard Schwarzschild metric. This may be not the case for black holes, especially the super-massive black holes in the centers of galaxies, which may be of primordial origin \cite{Carr:1974nx,Carr:2005bd}. In any case, this question requires further investigation.

Another open question is the stability of the modified black hole solutions. Several kinds of instabilities may be present. Among perturbations of the solutions there may exist unstable modes with the characteristic time scale of order of the horizon size; in this case the interpretation in terms of black holes is not possible. Second potential source of problems is generic presence of the higher-derivative terms not included in the action (\ref{eq:MG_action}). One has to check that the 2-parameter family of modified black holes survives their inclusion. By analogy with the ghost condensate case, one may expect that these terms produce at least a slow Jeans-type instability \cite{Arkani-Hamed:2003uy}, which is not, however, dangerous for the black hole interpretation (see discussion of section \ref{sc:Sym_Massive_Gravity}). Finally, the presence of negative mass solutions may lead to instabilities of the quantum-mechanical nature similar to those found in Ref.~\cite{Krotov:2004if}.

To conclude this list, let us mention also the solutions satisfying $h = 0$ which were not considered here. In this case the ten Einstein equations reduce to three equations for $\alpha$, $\beta$ and $\phi$ which form a (generically) well-defined system. It remains to be seen whether this system has asymptotically flat solutions. In any case, the Schwarzschild solution does not belong to this class which is characterized by $\alpha \beta \neq 1$.

\chapter{Structure formation} \label{ch:cosmo}

Given that massive gravity theories with spontaneous breaking of Lorentz-invariance pass the most obvious constraints, one may wonder if they reproduce correctly more subtle parts of modern cosmology, in  particular the formation of structures. In the standard cosmological model based on GR, the formation of the different structures observed in our Universe (galaxies, clusters of galaxies, super-clusters, voids, \ldots) is understood through the theory of cosmological perturbations (see \cite{Mukhanov:1990me,Uzan:2005,Bardeen:1980kt,Kolb:1994,Dodelson:2003,Mukhanov:2005} and references therein) which describes the evolution of the primordial inhomogeneities of the cosmic fluids filling the early Universe.

These primordial inhomogeneities grow with the expansion of the Universe because of the low-energy gravitational instability present in GR (the Jeans instability controlled by curvature). This instability implies that a small over-density will grow by attracting surrounding matter. The gravitational force exert by this over-density will then increase with consequence that it will attract more and more matter. In an expanding Universe, the growth of the over-density is partially counteracted by the expansion which tends to dilute it. Still, with some assumptions about cold dark matter \cite{Davis:1985rj,Liddle:1993fq,Hu:1998kj}, the general picture which emerges from this seems to be in agreement with the most recent cosmological observations \cite{Liddle:1993fq,Bertschinger:1998tv}.

It is not obvious \emph{a priori} that massive gravity theories could reproduce correctly the mechanism of structure formation since the background contains the condensates of four Goldstone fields whose perturbations mix with the matter density perturbations. Still, we will show in this chapter that cosmological perturbation theory in massive gravity models have predictions similar to those of GR \cite{Bebronne:2007qh}.

This chapter is organized as follows. First, we review the cosmological solutions in massive gravity with Lorentz-symmetry breaking. The aim of this discussion is to show that models invariant under the dilatation symmetry (\ref{eq:dilatation_Z}) are attractors of the cosmological evolution \cite{Dubovsky:2005dw}. Then, we move to the original part and study perturbations of the cosmological solutions found in such models.

\section{Cosmological models} \label{sc:FLRW}

GR has provided a complete framework for cosmologists to work with. Hence, it is of common belief that the laws of gravity rule the Universe's dynamic. Consequently, cosmological solutions have played a central role in any theory of gravity since the formulation of GR in the beginning of the twenty century. Massive gravity models with spontaneous breaking of the Lorentz symmetry are no exceptions. Indeed, cosmological solutions were the first exact solutions of the massive gravitational field equations to be studied (after the vacuum solution, of course).

The Standard Model of cosmology, known as the $\Lambda\tx{CDM}$ model, is based on a few assumptions. One of them, the cosmological principle, states that the Universe is homogeneous and isotropic on cosmological distances. This assumption fixes the geometry of the Universe to be described by the Friedmann-Lema\^{\i }tre-Robertson-Walker (FLRW) metric
\begin{eqnarray} \label{eq:FLRW_metric}
\dif s^2 = a(\eta)^2 \left( \dif \eta^2 - \dif x_i \dif x^i \right) &\tx{with}& \dif x_i \dif x^i = \dfrac{1}{1 - k r^2} \dif r^2 + r^2 \dif \Omega^2 ,
\end{eqnarray}
where $k = -1, 0, 1$ for an open, flat or closed Universe, respectively. Hence, the cosmological principle reduces the dynamics of the Universe to the dynamics of a single variable, the scale factor $a \left( \eta \right)$. The cosmological principle also imposes constraints on the energy-momentum tensor of any cosmological fluid, since only perfect fluids are homogeneous and isotropic. Hence, the energy-momentum tensor (\ref{eq:En_Mo_tens}) of any cosmological fluid reads
\begin{eqnarray*}
\mathcal{T}_{\mu\nu} = \left( \rho + p \right) v_\mu v_\nu - p g_{\mu\nu} ,
\end{eqnarray*}
where $\rho = \mathcal{T}_{\mu\nu} v^\mu v^\nu$ and $p = \mathcal{T}_{\mu\nu} j^{\mu\nu} / 3$ are the energy and pressure densities measured by an observer of four-velocity $v_\mu$ which is comoving with the fluid, respectively. This four-velocity obeys the geodesic equation $v^\nu \nabla_\nu v^\mu = 0$ with $v^\mu v_\mu = 1$, implying that the affine parameter of the observer can be chosen such that $v_\mu = \left( a, 0, 0, 0 \right)$.

Although the cosmological principle was first based on philosophical considerations, its is now supported by several high-precisions cosmological observations, such as Large Sky Survey \cite{AdelmanMcCarthy:2007wu,Cole:2005sx} or studies of the CMB anisotropies \cite{Dunkley:2008ie}. Consequently, theories of massive gravity have to contain this cosmological principle in their cosmological solutions. For this reason, let us assume the following spatially-flat homogeneous and isotropic ansatz \cite{Dubovsky:2004sg} for the Goldstone scalar fields,
\begin{eqnarray}
\phi^0 = \phi \left( \eta \right) , & \phi^i = \tau \left( \eta \right) x^i ,
\end{eqnarray}
together with the spatially-flat FLRW metric ($k = 0$ in (\ref{eq:FLRW_metric})). This ansatz cannot be directly generalized to the case of an open or closed Universe, and it is not known whether massive gravity models based on (\ref{eq:MG_action}) admit solutions of these types. With this ansatz, the Einstein equations for the massive gravitational field reduce to the usual Friedmann and Raychaudhuri equations of GR with extra contributions consisting in an energy and pressure densities for the Goldstone fields:
\begin{eqnarray} \label{eq:FLRW_equations_GR}
3 \mathcal{H}^2 = \dfrac{a^2}{\Mpl} \left( \rho + \rho_{\phi} \right) , & 2 \mathcal{H}^\prime + \mathcal{H}^2 = - \dfrac{a^2}{\Mpl} \left( p + p_{\phi} \right),
\end{eqnarray}
where prime denotes the derivative with respect to the conformal time $\eta$. Here $\mathcal{H} = a^\prime / a$ is the conformal Hubble parameter so that $H = \mathcal{H} / a$ is the physical Hubble parameter, $\rho$ and $p$ are the total energy and pressure densities of usual matter fields (baryons, photons, \ldots ) respectively, while $\rho_{\phi}$ and $p_{\phi}$ are the energy and pressure densities for the Goldstone scalar fields (section \ref{sc:ii_origin}). Note that the Goldstone anisotropic stress tensor $\pi_{\mu\nu}^\phi$ is identically zero for this ansatz in agreement with the homogeneity and isotropy of the background.

For an arbitrary function $\func$ of $X$ and $W^{ij}$, the four Goldstone equations obtained by varying the action with respect to $\phi^\mu$ reduce to
\begin{eqnarray} \label{eq:FLRW_Golds_V1}
0 = X^{1/2} \partial_\eta \left( a^3 X^{1/2} \mathcal{F}_X \right) + 3 a^3 \dfrac{\tau^\prime}{\tau} W \mathcal{F}_W .
\end{eqnarray}
As before, $\partial \func / \partial W^{ij} = \delta_{ij} \mathcal{F}_W$ while $W \equiv - \delta_{ij} W^{ij} / 3 = a^{-2} \tau^2$ and $X = a^{-2} \phi^{\prime2}$. Note that the Goldstone equations are equivalent to the equations of energy-momentum conservation for the Goldstone fluid, $\nabla_\mu t^{\mu\nu} = 0$. Consequently, those equations can be formulated in the same way as the conservation equations of usual matter fields. For any cosmological fluid including the Goldstone fluid, the equations for energy and momentum conservation read
\begin{eqnarray*}
\rho^\prime_i + 3 \mathcal{H} \left( \rho_i + p_i \right) , &\tx{where}& i = \tx{baryons}, \tx{photons}, \phi^\mu, \ldots \, .
\end{eqnarray*}
Since the conservation of total energy-momentum is a consequence of the Einstein equations, if there is $N$ non-interacting usual matter fields in the Universe, there is $N+2$ independent equations describing the Universe's dynamic which may be chosen as
\begin{eqnarray} \label{eq:FLRW_dyn}
3 \mathcal{H}^2 = \dfrac{a^2}{\Mpl} \left( \rho + \rho_{\phi} \right) , & 0 = \rho^\prime_{\phi} + 3 \mathcal{H} \left( \rho_{\phi} + p_{\phi} \right) , & 0 = \rho^\prime_n + 3 \mathcal{H} \left( \rho_n + p_n \right) ,
\end{eqnarray}
where the indices $n = 1, \ldots, N$ labels the usual matter fields and $\rho = \sum_{n} \rho_n$.

Given $N$ equations of state $p_n = w_n \rho_n$, one can determined the energy densities of those fields as functions of the scale factor $\rho_n \propto a^{-3(1+w_n)}$. Similarly, the Goldstone equation gives $\rho_\phi$ as a function of the scale factor. Finally, the Friedmann equation gives the scale factor as a function of the conformal time $\eta$. Yet, there are two unknown functions in the Goldstone ansatz, $\phi(\eta)$ and $\tau(\eta)$, with only one Goldstone equation to constrain them. Hence, there is a freedom in choosing one of those two functions. This freedom is a direct consequence of the fact that the action is supposed to be invariant under the symmetry (\ref{eq:MG_Sym_phi}). Indeed, the ansatz for $\phi^i$ is still homogeneous as a shift of the spatial coordinates $x^i$ by a constant can be compensate by a $\phi^0$ dependent shift of the fields $\phi^i$.

Note however that this ambiguity can be fixed by specifying the boundary conditions for the fields $\phi^i$ at spatial infinity \cite{Dubovsky:2005dw}. For instance, imagine that the space is a torus of size $L$. Then, $\phi^i$ would have to satisfy some periodicity condition,
\begin{eqnarray*}
\phi^i \left( \eta, x^i \right) = \phi^i \left( \eta, x^i + L^i \right) - L^i ,
\end{eqnarray*}
implying $\tau = \tx{const}$. Other boundary conditions may lead to time-depend $\tau \left( \eta \right)$. Therefore, the ambiguity in choosing different functions $\tau \left( \eta \right)$ is analogous to the ambiguity in choosing the vacuum in theories with flat directions.

\subsection{Cosmological attractors}

One choice which seems quite natural, as it looks like the vacuum solution, consist in requiring that only $\phi^0$ varies with time, so that $\tau = 1$. Then, in cosmological background the function $\func$ depends only on $X = a^{-2} \phi^{\prime2}$ and $W = a^{-2}$, and the Goldstone equation (\ref{eq:FLRW_Golds_V1}) reduces to
\begin{eqnarray} \label{eq:FLRW_Golds_X}
X^{1/2} \mathcal{F}_X = \dfrac{\tx{const.}}{a^{3}} .
\end{eqnarray}
This equation implies that at late time either $X$ or $\mathcal{F}_X$ goes to zero. Since the mass parameters $m_i^2$ are linear combination of the derivatives of $\func$, one may wonder whether they go to zero or remain finite in the limit $a \rightarrow \infty$. Indeed, mass parameters going to zero will imply that the cutoff scale decreases with the scale factor. Similarly, if $X$ goes to infinity, the validity of the low energy effective field theory becomes questionable.

There is a broad class of massive gravity theories for which the mass parameters are finite in the limit $a \rightarrow \infty$, so that the effective field theory description remains valid \cite{Dubovsky:2005dw}. Let us assume that $X$ asymptotes to some power of $a$ at late time. This requirement is not that restrictive since eq.~(\ref{eq:FLRW_Golds_X}) implies that any algebraic function $\func \left( X, a^{-2} \right)$ satisfies it. Then, there is a constant $\gamma$ such that the combination $a^{-2} X^\gamma$ goes to a non-zero constant $Z_0$ in the limit of growing scale factor $a \rightarrow \infty$. Let us replace $X$ by a new variable $Z \equiv a^{-2} X^\gamma$. Then, $\func \left( X , a^{-2} \right)$ becomes a function of $Z$ and $a$
\begin{eqnarray*}
\mathcal{F} \left( X, a^{-2} \right) = \mathcal{F} \left( a^{2/\gamma} Z^{1/\gamma}, a^{-2} \right) \equiv \tilde{\func} \left( Z, a^{-2} \right) .
\end{eqnarray*}
If one assumes further that the function $\tilde{\func} \left( Z, a^{-2} \right)$ is regular in the limit of growing scale factor, one finally find that
\begin{eqnarray*}
\tilde{\func} \left( Z, a^{-2} \right) \rightarrow \tilde{\func}_0 \left( Z_0 \right) &\tx{when}& a \rightarrow \infty.
\end{eqnarray*}
This discussion implies that a wide class of functions $\func \left( X , W^{ij} \right)$ depend only on the combination $Z^{ij} \equiv X^\gamma W^{ij}$ in the limit when the scale factor goes to infinity. Hence, there is another motivation (beside the correct Newtonian limit) for studying theories which depend on a single argument $Z^{ij}$ and satisfy the dilation symmetry (\ref{eq:dilatation_Z}).

Let us discuss the cosmological solution of massive gravity possessing this dilatation symmetry. For these models, the derivatives of $\func$ with respect to $X$ and $W^{ij}$ are proportional to one another,
\begin{eqnarray*}
\mathcal{F}_X = \dfrac{- 3 \gamma Z}{X} \mathcal{F}_Z , & \mathcal{F}_W = \dfrac{Z}{W} \mathcal{F}_Z ,
\end{eqnarray*}
where $\partial \mathcal{F} / \partial Z^{ij} \equiv \delta_{ij} \mathcal{F}_Z$ and $Z \equiv - \delta_{ij} Z^{ij} / 3 = \phi^{\prime2\gamma} / a^{2\gamma+2}$. The energy and pressure densities for the Goldstone fluid read
\begin{eqnarray*}
\rho_\phi = - \Lambda^4 \left( \dfrac{1}{2} \func + 3 \gamma Z \func_ Z \right) , & p_\phi = w_\phi \rho_\phi ,
\end{eqnarray*}
with
\begin{eqnarray} \label{eq:EoS_Z}
w_\phi = - 1 + 2 \left( 3 \gamma - 1 \right) \dfrac{Z \func_Z}{\func + 6 \gamma Z \func_ Z} .
\end{eqnarray}
The Goldstone equation (\ref{eq:FLRW_Golds_X}) reduces to
\begin{eqnarray*} \label{eq:FLRW_Golds_Z}
Z^{1-1/2\gamma} \mathcal{F}_Z = \dfrac{\tx{const.}}{a^{3-1/\gamma}} .
\end{eqnarray*}
For a given function $\func$, this equation determines the dependence of the variable $Z^{ij}$ on the scale factor. One has to specify $\func$ in order to solve it. However, this relation gives some hints about the Goldstone dynamics without need of an exact solution. For example, models characterized by $\gamma = 1 / 3$ are particular. The Goldstone equation implies then that $Z^{ij}$ is constant without constraints on $\mathcal{F}_Z$, while the equation of state (\ref{eq:EoS_Z}) implies that $\rho_\phi$ behaves as a cosmological constant whose value is determined by the initial conditions.

One can also determined some features of models characterized by $\gamma > 1/3$ or $\gamma < 0$ without need of an exact solution. Indeed, for those value of $\gamma$ the Goldstone equation implies that
\begin{eqnarray*}
Z^{1-1/2\gamma} \mathcal{F}_Z \rightarrow 0 &\tx{when}& a \rightarrow \infty .
\end{eqnarray*}
Hence, either $Z$ goes to zero so that the mass parameters decrease with the scale factor, or $\mathcal{F}_Z$ goes to zero. The case of interest is when $Z$ goes to a constant $Z_0$ in this limit such that $\func_Z \left( Z_0 \right) = 0$. In this case, the graviton mass remains finite at $a \to \infty$ and the effective field theory description remains valid. Moreover, for models satisfying $\func_Z \left( Z_0 \right) = 0$, $Z = Z_0$ is a natural solution to the Goldstone equation for which $\phi^\prime \propto a^{1+1/\gamma}$. For those models, the Goldstone fluid also corresponds to a cosmological constant whose value is fixed by the initial conditions.

Finally, it is worth noting that for $0 < \gamma < 1/ 3$ and regular function $\mathcal{F}$, $Z$ grows with the scale factor implying that the effective field theory description breaks at late time.

\subsection{Other example}

Other massive gravity theories may imply different cosmological models. For the sake of illustration, one may consider solutions with $\tau \neq 1$ before going to the original part of this chapter. Let us have a look the following function of $X$ and $W^{ij}$
\begin{eqnarray*}
\mathcal{F} = \left( W^{ii} - W^{ij} W^{ij} \right) P \left( X \right) ,
\end{eqnarray*}
where $P \left( X \right)$ is a function of $X$ only. For this particular model, the mass of the graviton is given by $m_2^2 = 2 m^2 W \left( 1 + 4 W \right) P \left( X \right)$. Let us choose the ansatz $\tau = a$ such that $W = 1$. Then the homogeneous and isotropic Goldstone fluid is pressure-less, $p_{\phi} = 0$, so that the Goldstone equation (\ref{eq:FLRW_Golds_V1}) reduces to
\begin{eqnarray} \label{eq:Cos_Mod_1_Gold}
\rho^\prime_{\phi} = - 3 \mathcal{H} \rho_{\phi} &\rightarrow& \rho_{\phi} = \dfrac{\rho_{dm}}{a^{3}} ,
\end{eqnarray}
where $\rho_{dm}$ is a constant corresponding to the initial Goldstone energy density. Hence, depending on the particular model considered, massive gravity theories could provide a cold dark matter candidate in the right hand side of the Friedmann equation
\begin{eqnarray*}
3 \mathcal{H}^2 = \dfrac{a^2}{\Mpl} \left( \rho + \dfrac{\rho_{dm}}{a^{3}} \right) .
\end{eqnarray*}
Moreover, if $P \left( X \right)$ is simply given by
\begin{eqnarray*}
P \left( X \right) = 1 + X ,
\end{eqnarray*}
then eq.~(\ref{eq:Cos_Mod_1_Gold}) implies that $X = 1 - \rho_{dm} / ( 3 a^{3})$ so that $P \rightarrow 2$ in the limit $a \rightarrow \infty$. Hence, the graviton mass tends to a non-zero constant in the limit of growing scale factor.

\section{Cosmological perturbations} \label{sc:cosmo_pert}

To address the question of structure formation, we focus on massive gravity theories possessing the dilatation symmetry (\ref{eq:dilatation_Z}) and neglect possible deviations from the point $Z = Z_0$ by considering models for which $\func_Z \left( Z_0 \right) = 0$. For those models, $Z^{ij} = - Z_0 \delta^{ij}$ is a solution of the Goldstone equations so that the mass parameters $m_i^2$ are constants while $\phi^\prime = Z_0^{1/2\gamma} a^{1+1/\gamma}$. Then, $\rho_\phi$ corresponds to a cosmological constant and the EoM (\ref{eq:FLRW_dyn}) for the background reduce to
\begin{eqnarray} \label{eq:FLRW_dyn_primord}
3 \mathcal{H}^2 = \dfrac{a^2}{\Mpl} \left( \rho + \rho_{\phi} \right) , & 0 = \rho^\prime_n + 3 \mathcal{H} \left( \rho_n + p_n \right) .
\end{eqnarray}
We also assume that the total cosmological constant $\rho_\Lambda$, which includes $\rho_\phi$, is of the order of the present-day cosmological constant. Its contribution to the Friedmann equation at the epoch of structure formation is therefore negligible. With all these considerations in mind, the solution to eqs.~(\ref{eq:FLRW_dyn_primord}) can be parameterized as
\begin{eqnarray} \label{eq:FLRW_Solution}
a = \left( \dfrac{\eta}{\eta_0} \right)^{2 / ( 1 + 3 w )} , & \rho = \rho_0 a^{- 3 ( 1 + w )} ,
\end{eqnarray}
where $\rho_0$ and $\eta_0$ are constants related through
\begin{eqnarray*}
\eta_0 \equiv \dfrac{2}{H_0 \left( 1 + 3 w \right)} , & H_0 \equiv \sqrt{\dfrac{\rho_0}{3 \Mpl}} .
\end{eqnarray*}

\subsubsection{Transverse and longitudinal perturbations}

In order to describe perturbations of a flat FLRW space-time, one has to generalize the decomposition (\ref{eq:Pert_Mk}) into transverse and longitudinal fields \cite{Mukhanov:1990me}, which then reads
\begin{eqnarray} \label{eq:Pert_FLRW}
\left. \begin{array}{l}
h_{00} = 2 a^2 \varphi, \\
h_{0i} = a^2 \left( S_{i} + \partial_{i} B \right) , \\
h_{ij} = a^2 \left( 2 \psi \delta_{ij} - 2 \partial_{i} \partial_{j} E - \partial_{i} F_{j} - \partial_{j} F_{i} + H_{ij} \right) ,
\\
\end{array} \right| &&
\begin{array}{l}
\xi^{0} = \xi_{0} , \\
\xi^{i} = \xi_{i}^{T} + \partial_{i} \xi ,
\end{array}
\end{eqnarray}
where, as in Minkowski space-time, the vector perturbations $S_i$, $F_i$ and $\xi_{i}^{T}$ are transverse, while the tensor perturbation $H_{ij}$ is transverse and traceless. Since FLRW background is time-dependent, the gauge-invariant scalar perturbations introduced in section \ref{sc:Pert_Mink} have to be generalized as follows
\begin{eqnarray*}
\begin{array}{ll}
\Phi = \varphi - a^{-1} \left[ a \left( E^\prime + B \right) \right]', & \Psi = \psi + \mathcal{H} \left( E^\prime + B \right) ,  \\
\Xi^0 = \xi^{0} - \phi^{\prime} \left( B + E^\prime \right) , & \Xi = \xi - E .
\end{array}
\end{eqnarray*}
The two gauge-invariant vector fields are still given by the same combinations as in flat space-time,
\begin{eqnarray*}
\varpi_{i} = S_{i} + F_{i}^{\prime} , & & \sigma_{i} = \xi_{i}^{T} - F_{i} ,
\end{eqnarray*}
and $H_{ij}$ is still gauge-invariant as expected from its tensorial structure.

\subsubsection{Perturbations of the energy-momentum tensor}

The notations introduced in section \ref{sc:Weak_Field} in order to describe the energy-momentum perturbations above Minkowski space-time need also to be generalized to FLRW background. The main difference as compared to the flat space-time originates in the non-zero value of the energy and pressure densities of the background, so that the most general perturbations of the energy-momentum tensor of ordinary matter fields are parameterized in the following way,
\begin{eqnarray*}
\delta \mathcal{T}_{\mu\nu} &=& \left( \delta \rho + \delta p \right) v_{\mu} v_{\nu} - g_{\mu\nu} \delta p + \left( \rho + p \right) \left( v_{\nu} \delta v_{\mu} + v_{\mu} \delta v_{\nu} \right) - p h_{\mu\nu} \nonumber \\
& & + \left( v_{\mu} \delta q_{\nu} + v_{\nu} \delta q_{\mu} \right) + \delta \pi_{\mu\nu} .
\end{eqnarray*}
As previously, $\delta \rho$ and $\delta p$ are the perturbations of the matter and pressure densities measured by a comoving observer, $\delta q_\mu$ is the perturbation of the energy flux perpendicular to $v_\mu$ and $\delta \pi_{\mu\nu}$ is the perturbation of the anisotropic stress tensor. As in Minkowski space-time, the fact that $v_\mu = \left( a, 0, 0, 0 \right)$ implies that $\delta q_0 = \delta \pi_{0\nu} = 0$. The observer's four-velocity is a time-like vector. Then, it satisfies $v_\mu v^\mu = 1$ which at the linearized level implies that
\begin{eqnarray*}
\delta v_0 = a \varphi .
\end{eqnarray*}
Therefore, the linearized energy-momentum tensor of any cosmological fluid is given by
\begin{eqnarray} \label{eq:FLR_En_mom_tensor_matt}
\delta \mathcal{T}_{00} &=& a^2 \left( \delta \rho + 2 \rho \varphi \right) , \\
\delta \mathcal{T}_{0i} &=& a \left( \rho + p \right) \delta v_i - a^2 p \left( S_i + \partial_i B \right) + a \delta q_i \nonumber , \\
\delta \mathcal{T}_{ij} &=& a^2 \left[ \delta_{ij} \delta p - p \left( 2 \psi \delta_{ij} - 2 \partial_i \partial_j E - \partial_i F_j - \partial_j F_i + H_{ij} \right) \right] + \delta \pi_{ij} \nonumber .
\end{eqnarray}
Let us introduce the following parametrization
\begin{eqnarray*}
a \left( \rho + p \right) \delta v_i + a \delta q_i &\equiv& a^2 \left( \zeta_i + \partial_i \zeta \right) , \\
\delta \pi_{ij} &\equiv& a^2 \left[ \left( 3 \partial_i \partial_j - \delta_{ij} \partial_k^2 \right) \pi + \partial_i \pi_j + \partial_j \pi_i + \pi_{ij} \right] ,
\end{eqnarray*}
where the vector perturbations $\zeta_i$ and $\pi_i$ are transverse while the tensor perturbation $\pi_{ij}$ is transverse and traceless. At this point, let us comment about the interpretation of $\zeta_i + \partial_i \zeta$. In this three-dimensional vector field are hidden both the three-dimensional velocity perturbation $\delta v_i$ and the perturbation of the energy flux $\delta q_i$. The equations for the gravitational field are not able to differentiate between these two three-dimensional vector fields, with consequence that only $\zeta_i + \partial_i \zeta$ will be determined by these equations. To differentiate between $\delta v_i$ and $\delta q_i$, one must had an equation for $\delta q_i$. In other words, one has to suppose some physics which will give a non-zero $\delta q_i$. The discussion of such physics is beyond this work. Therefore, we will assume $\delta q_i = 0$.

In Minkowski space-time, the perturbations of the energy-momentum tensor are all gauge - invariant perturbations. This is a direct consequence of the fact that this tensor is identically zero in the vacuum. Since the energy-momentum tensor of usual matter is different from zero in the FLRW background, some of the perturbations introduced above are no more gauge-invariant. Indeed, the perturbations $\pi_{ij}$, $\pi_i$ and $\pi$ of the anisotropic stress tensor are still gauge invariant as well as the transverse vector $\zeta_i$. The other perturbations, namely $\delta \rho$, $\delta p$ and $\zeta$ are not gauge-invariant. Hence, one needs to introduce three gauge-invariant scalars
\begin{eqnarray*}
\delta_\rho = \left[ \delta\rho - \rho^\prime \left( E^\prime + B \right) \right] / \rho, & \delta_p = \left[ \delta p - p^\prime
\left( E^\prime + B \right) \right] / p, & \delta_\zeta = \zeta - \left( \rho + p \right) \left( E^\prime + B \right) .
\end{eqnarray*}
With all these notations, the equations for the conservation of energy and momentum read
\begin{eqnarray} \label{eq:FLRW_conservation}
0 &=& \left( \rho \delta_\rho \right)^\prime + 3 \mathcal{H} \left( \rho \delta_\rho + p \delta_p \right) - 3 \left( \rho + p \right) \Psi^\prime - \partial_i^2 \delta_\zeta , \nonumber \\
0 &=& \delta_\zeta^\prime + 4 \mathcal{H} \delta_\zeta - p \delta_p - 2 \partial_k^2 \pi - \Phi \left( \rho + p \right) , \nonumber \\
0 &=& \zeta_i^\prime + 4 \mathcal{H} \zeta_i - \partial_k^2 \pi_i .
\end{eqnarray}

\subsection{Jeans instability}

Contrary to what have been done in chapter \ref{ch:weak_field}, the goal here is not to determine the gravitational interaction created by an arbitrary source, but rather to understand how the inhomogeneities described by $\delta \mathcal{T}_{\mu\nu}$ evolve under their own gravitational interaction.

As a warming up, let us go back to Minkowski space-time and restrict ourselves to Newtonian gravity. In the Euclidian space, the usual equations of hydrodynamics \cite{Landau:1971} for a fluid of density $\rho$ and pressure $p$ are the conservation and Euler equations, which read
\begin{eqnarray*}
0 &=& \partial_t \rho + \partial_i \left( \rho v_i \right), \\
0 &=& \partial_t v_i + v_j \partial_j v_i + \dfrac{1}{\rho} \partial_i p + \partial_i \Phi,
\end{eqnarray*}
where $v_i$ is the velocity of the fluid, while $\Phi$ is Newton's potential satisfying the usual Poisson equation. A static and uniform fluid is described by the following solution to these two equations:
\begin{eqnarray*}
\begin{array}{cccc}
\rho = \tx{constant} , & p = \tx{constant} , & v_i = 0 , & \Phi_0 = 0 .
\end{array}
\end{eqnarray*}
Any perturbation of the fluid is characterized by an energy $\delta \rho$, a pressure $\delta p$, a velocity $\delta v_i$ and a potential $\Phi$ which are solutions of the following three equations
\begin{eqnarray*}
0 &=& \partial_t \delta \rho + \rho \partial_i \delta v_i , \\
0 &=& \partial_t \delta v_i + \dfrac{1}{\rho} \partial_i \delta p + \partial_i \Phi, \\
0 &=& \partial_i^2 \Phi - 4 \pi G \delta \rho .
\end{eqnarray*}
Combining these three equations gives a differential equation for $\delta \rho$ known as the sound equation in the presence of gravity
\begin{eqnarray*}
0 = \left( \partial_t^2 - c_s^2 \partial_i^2 - 4 \pi G \rho \right) \delta \rho , & & c_s^2 \equiv \left( \dfrac{\partial p}{\partial \rho} \right)_{S} = \dfrac{\delta p}{\delta \rho} ,
\end{eqnarray*}
where $c_s$ is the speed of sound. The solution of this wave equation is simply given by a plane wave whose time dependence is
\begin{eqnarray*}
\delta \rho \propto \tx{exp.} \left[ - i \sqrt{c_s^2 k_{J}^2 \left( \dfrac{k^2}{k_{J}^2} - 1 \right)} t \right] ,
\end{eqnarray*}
where $k_J = \sqrt{4 \pi G \rho / c_s^{2}}$ is the Jeans wave number. One concludes from this calculation that there are two opposite effects in competition here: while gravity tends to increase over-densities, random thermal motion described by $\delta p$ acts as to dilute the over-dense regions. The amplitude of the Jeans wave number signals the transition between small perturbations $k < k_J$ for which gravity dominates and large perturbations $k > k_J$ for which gravity is negligible. In the last case, the perturbations are just sound waves. Indeed, for $k < k_J$ the time dependence of $\delta \rho$ is given by an exponential which characterizes the gravitational instability, while for $k > k_J$ its time dependence is just an oscillating function. In an expending Universe both contributions are diluted by the expansion so that matter over-densities grow as a power of the scale factor.

\subsection{Evolution of the inhomogeneities}

Theories of structure formation involve two distinct parts. The first one is a theory for the production of the primordial inhomogeneities while the second is a model to describe the evolution of these primordial inhomogeneities in the expanding Universe. Models of inflation (see \cite{Uzan:2005,Kolb:1994,Dodelson:2003,Mukhanov:2005,Linde:1990,Lyth:1998xn} and references therein) are perhaps the most popular candidates as theory for the primordial inhomogeneities, while their evolution is understood in the Standard Model of Cosmology through GR \cite{Uzan:2005,Kolb:1994,Dodelson:2003,Mukhanov:2005,Liddle:1993fq,Bertschinger:1998tv,Bernardeau:2001qr}.

The present discussion will only concern the evolution of the primordial inhomogeneities. For this purpose, one has to consider a system of fourteen independent equations made of ten linearized Einstein equations and four linearized Goldstone equations\footnote{We could also discuss the perturbations by considering the ten linearized Einstein equations along with the four equations (\ref{eq:FLRW_conservation}) which describe the conservation of the energy-momentum tensor of usual matter. Indeed, the Einstein equations imply the conservation of the total energy-momentum tensor with consequence that there is effectively fourteen independent equations.}. In massive gravity models, these equations reduce to those of GR in the limit $m \rightarrow 0$. As in the flat space-time, the equations for the tensor, vector and scalar perturbations decouple in the linear regime allowing for a separate study of these three sectors. The equations governing each of these sectors are derived in appendix \ref{app:sc:line_equ}.

\subsubsection{Tensor perturbations}

Although tensor perturbations are not generated by density fluctuations, they are produced in many theories of the primordial inhomogeneities in addition to scalar perturbations. For example, gravitational waves are generated in most models of inflation \cite{Starobinsky:1979ty,Rubakov:1982df}. If they were produced in the early Universe, tensor modes would have left an imprint in the CMB polarization \cite{Kosowsky:1994cy,Zaldarriaga:2003bb,Keating:2006zy}\footnote{For a discussion about the influence of the graviton mass on the CMB polarization, see \cite{Dubovsky:2009xk,Bessada:2009np}.}. Hence, it is important to understand how they evolve during the expansion of the Universe.

\begin{figure}
\begin{center}
\includegraphics[width=7cm]{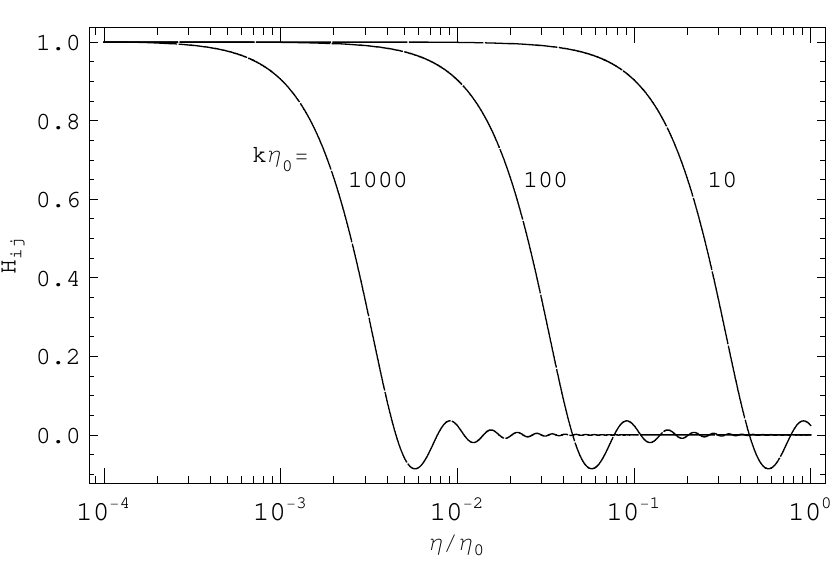}
\includegraphics[width=7cm]{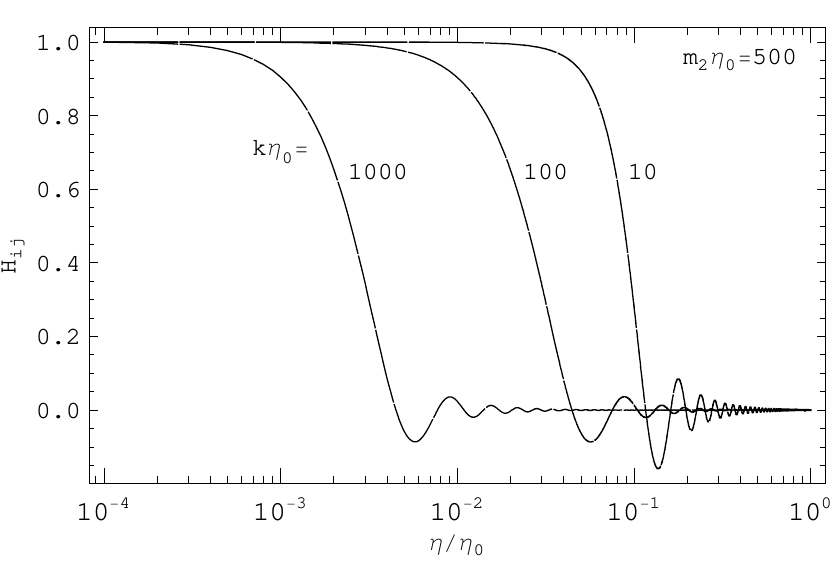}
\end{center}
\caption{Evolution of gravitational waves in a flat FLRW Universe dominated by usual matter with $w = 0$ as a function of conformal time. On the left, the predictions of GR for three different modes labeled by their comoving momenta. On the right, the same modes in massive gravity theories. The graviton mass has a negligible influence on small scale while it strongly affects large scale modes.} \label{fg:Tensor_modes}
\end{figure}

The tensor modes are given by an inhomogeneous equation which reads
\begin{eqnarray} \label{eq:FLRW_Tens}
0 = H_{ij}^{\prime\prime} + 2 \mathcal{H} H_{ij}^{\prime} - \partial_i^2 H_{ij} + a^2 \left( m_2^2 H_{ij} + \dfrac{2 \pi_{ij}}{\Mpl} \right) .
\end{eqnarray}
If the mass of the graviton is larger than the Hubble constant, $m_2 \gg \mathcal{H}/a$, which we assume to be the case in what follows, eq.~(\ref{eq:FLRW_Tens}) describes massive gravitational waves. For simplicity, let us neglect the anisotropic stress $\pi_{ij} = 0$\footnote{We then ignore, among other effects, the damping provided by cosmological neutrinos \cite{Weinberg:2003ur}.}. Then this equation is identical to the equation for a free massive scalar field in a flat FLRW background. In the three-dimensional Fourier space, the previous equation reduces to
\begin{eqnarray*}
0 = H_{ij}^{\prime\prime} + 2 \mathcal{H} H_{ij}^{\prime} + k^2 H_{ij} , & p^2 \gg m_2^2 ,
\end{eqnarray*}
or
\begin{eqnarray*}
0 = H_{ij}^{\prime\prime} + 2 \mathcal{H} H_{ij}^{\prime} + a^2 m_2^2 H_{ij} , & p^2 \ll m_2^2 ,
\end{eqnarray*}
in the relativistic and non-relativistic limits, respectively. Note that $p^2 \equiv a^{-2} k^2$ is the physical momentum. These equations state that the amplitude of the gravitational waves scales like $H_{ij} \propto a^{-1}$ and $H_{ij} \propto a^{-3/2}$ in the relativistic and non-relativistic limits, respectively. The exact solution of eq.~(\ref{eq:FLRW_Tens}) for a matter dominated Universe are plotted in Fig.~\ref{fg:Tensor_modes} for both GR and massive gravity models. As expected, the graviton mass strongly affects the dynamics of large scale modes (recall that $\lambda = 2 \pi / k$) which are the latest to enter the cosmological horizon, while small scale modes which enter the horizon and decay first are practically not affected by the graviton mass.

Before concluding this discussion about tensor modes, let us mention the possibility of non - relativistic gravitational waves to be a dark matter candidate \cite{Dubovsky:2004ud}, although this possibility seems strongly constrained by pulsar observations \cite{Pshirkov:2008nr}.

\subsubsection{Vector perturbations}

Primordial vector modes describe vortical fluid perturbations in the early Universe. For example, primordial magnetic fields may generate vector perturbations through their anisotropic stress \cite{Subramanian:2003sh,Lewis:2004ef} \footnote{For a discussion about the influence of primordial vector modes on the CMB polarization, see \cite{Lewis:2004kg}.}. The three gauge-invariant equations describing vector perturbations are
\begin{eqnarray*}
\partial_j^2 \varpi_i - a^2 \dfrac{2 \zeta_i}{\Mpl} = 0 , & \varpi_i' + 2 \mathcal{H} \varpi_i - a^2 \left( m_2^2 \sigma_i + \dfrac{2 \pi_i}{\Mpl} \right) = 0 , & m_2^2 \sigma_{i} = 0.
\end{eqnarray*}
The first of these equations allows to express $\zeta^i$ in terms of $\varpi_i$, while the third equation gives $\sigma_i = 0$. Therefore,
the only non-trivial equation is the second one. It differs from the conventional one by the term proportional to the graviton mass $m_2^2$ which cancels at $\sigma_i = 0$. Thus, this equation is the conventional one
\begin{eqnarray} \label{eq:FLRW_Vect}
\varpi_i^\prime + 2 \mathcal{H} \varpi_i - a^2 \dfrac{2 \pi_i}{\Mpl} = 0 ,
\end{eqnarray}
and there are no differences in the vector sector as compared to GR. If one neglects the anisotropic stress $\pi_i$, the previous equation describes a field with the amplitude decreasing as $\varpi_i \propto a^{-2}$.

\subsubsection{Scalar perturbations}

Observations of the CMB anisotropy spectrum shows that primordial scalar perturbations were most certainly dominated by adiabatic perturbations \cite{Dunkley:2008ie,Crotty:2003rz}. For such perturbations, $\delta_p$ can be expressed in terms of $\delta_\rho$ by means of the matter equation of state which reads
\begin{eqnarray*}
\delta p = c_s^2 \, \delta \rho &\rightarrow& \delta_p = \dfrac{c_s^2}{w} \, \delta_\rho ,
\end{eqnarray*}
where $c_s$ is the sound velocity ($c_s^2 = w$ for a perfect fluid). The scalar sector consists of 8 scalar perturbations: $\Phi$, $\Psi$, $\Xi$, $\Xi^0$, $\delta_\rho$, $\delta_p$, $\delta_\zeta$ and $\pi$. The behavior of these perturbations is governed by the previous equation of state along with four Einstein equations and two Goldstone equations
\begin{eqnarray}
0 &=& - 2 \partial_j^2 \Psi + 6 \mathcal{H} \left( \mathcal{H} \Phi + \Psi^\prime \right) + a^2 \left[ \dfrac{\rho \delta_\rho}{\Mpl} + m_4^2 \left( \partial_j^2 \Xi + 3 \Psi - 3 \gamma \Phi + 3 \gamma \dfrac{\Xi^{0\prime}}{\phi^{\prime}} \right) \right], \label{eq:FLRW_Scalar_eq_1} \\
0 &=& 2 \left( \Psi^\prime + \mathcal{H} \Phi \right) - a^2 \dfrac{\delta_\zeta}{\Mpl} , \label{eq:FLRW_Scalar_eq_2} \\
0 &=& \Phi - \Psi + a^2 \left( m_2^2 \Xi + \dfrac{3 \pi}{\Mpl} \right) , \label{eq:FLRW_Scalar_eq_3} \\
0 &=& - 2 \Psi^{\prime\prime} - 2 \Phi \left( \mathcal{H}^2 + 2 \mathcal{H}^\prime \right) + \partial_j^2 \left( \Psi - \Phi \right) - 2 \mathcal{H} \left( 2 \Psi + \Phi \right)^\prime + a^2 \dfrac{p \delta_p - \partial_i^2 \pi}{\Mpl} \nonumber \\
& & - \dfrac{a^2 m_2^2}{3} \partial_j^2 \Xi + a^2 m_4^2 \left( \Phi - \dfrac{\Xi^{0\prime}}{\phi^{\prime}} - \dfrac{\Psi}{\gamma} - \dfrac{1}{3 \gamma} \partial_j^2 \Xi \right) \label{eq:FLRW_Scalar_eq_4} , \\
0 &=& m_4^2 \partial_0 \left[ \dfrac{a^4}{\phi^{\prime}} \left( 3 \gamma \left( \Phi - \dfrac{\Xi^{0\prime}}{\phi^{\prime}} \right) - 3 \Psi - \partial_i^2 \Xi \right) \right] , \label{eq:FLRW_Scalar_eq_5} \\
0 &=& m_4^2 \left( \dfrac{1}{3\gamma} \partial_j^2 \Xi + \dfrac{\Psi}{\gamma} + \dfrac{\Xi^{0\prime}}{\phi^{\prime}} - \Phi \right) - \dfrac{2}{3} m_2^2 \partial_j^2 \Xi . \label{eq:FLRW_Scalar_eq_6}
\end{eqnarray}
There is therefore 7 equations for 8 scalar perturbations. To close this system, we must either add an equation describing the physics behind $\pi$, or we must set $\pi = 0$. We choose the second option here.

Since we are interested in massive gravity models possessing the dilatation symmetry (\ref{eq:dilatation_Z}), the relations $m_0^2 = 3 \gamma m_4^2$ and $m_4^2 = \gamma \left( 3 m_3^2 - m_2^2 \right)$ between the mass parameters have been used in the previous system of equations. This system can be solved as follows. At $m_4^2\neq 0$, eq.~(\ref{eq:FLRW_Scalar_eq_6}) can be used to express $\Phi - \Xi^{0\prime} / \phi^{\prime}$ in terms of the other fields so that eq.~(\ref{eq:FLRW_Scalar_eq_5}) becomes a closed equation for $\Xi$
\begin{eqnarray*}
0 = \partial_i^2 \left[ \Xi^\prime + \left( 3 - \dfrac{1}{\gamma} \right) \mathcal{H} \Xi \right] .
\end{eqnarray*}
The solution of this equation which does not grow at spatial infinity 
reads
\begin{eqnarray} \label{eq:FLRW_XI}
m_2^2 \Xi = \dfrac{1}{\gamma} a^{1/\gamma - 3} \Psi_0 \left( x^i \right) ,
\end{eqnarray}
where $\Psi_0 \left( x^i \right)$ is the function of the space coordinates introduced in section \ref{sc:Weak_Field}. As in Minkowski space-time, this function arises as an integration constant which is related to in the presence of a mode with the dispersion relation $\omega^2 = 0$. Then, eqs.~(\ref{eq:FLRW_Scalar_eq_1}), (\ref{eq:FLRW_Scalar_eq_2}) and (\ref{eq:FLRW_Scalar_eq_6}) can be used to express $\delta_\rho$, $\delta_\zeta$ and $\Xi^0$ in terms of $\Phi$ and $\Psi$, while eq.~(\ref{eq:FLRW_Scalar_eq_3}) reads
\begin{eqnarray}
\Psi - \Phi = \dfrac{1}{\gamma} a^{1/\gamma - 1} \Psi_0 .
\label{eq:FLRW_PHI}
\end{eqnarray}
With the account of all these relations, the remaining equation (\ref{eq:FLRW_Scalar_eq_4}) becomes a closed inhomogeneous equation for $\Psi$,
\begin{eqnarray*}
0 &=& \Psi^{\prime\prime} + 3 \mathcal{H} \left( 1 + c_s^2 \right) \Psi^\prime + \left[ \left( 1 + 3 c_s^2 \right) \mathcal{H}^2 + 2 \mathcal{H}^\prime - c_s^2 \partial_j^2 \right] \Psi \\
& & - \dfrac{1}{\gamma} a^{1/\gamma - 1} \left[ \left( 3 c_s^2 + \dfrac{1}{\gamma} \right) \mathcal{H}^2 + 2 \mathcal{H}^\prime - \gamma c_s^2 \partial_j^2 \right] \Psi_0 .
\end{eqnarray*}
This relation is known as the Bardeen equation. Inverting the relation $a = a \left( \eta \right)$ enables one to write this equations as
\begin{eqnarray} \label{eq:FLRW_PSI}
0 &=& \partial_a^2 \Psi + \dfrac{1}{a} \left( 4 + 3 c_s^2 + \dfrac{\mathcal{H}^\prime}{\mathcal{H}^2} \right) \partial_a \Psi + \dfrac{1}{a^2} \left[ 1 + 3 c_s^2 + 2 \dfrac{\mathcal{H}^\prime}{\mathcal{H}^2} - \dfrac{c_s^2 \partial_i^2}{\mathcal{H}^2} \right] \Psi \nonumber \\
& & + \dfrac{1}{\gamma} a^{1/\gamma - 3} \left[ \dfrac{\gamma c_s^2 \partial_i^2}{\mathcal{H}^2} - \left( 3 c_s^2 + \dfrac{1}{\gamma} + 2 \dfrac{\mathcal{H}^\prime}{\mathcal{H}^2} \right) \right] \Psi_0 .
\end{eqnarray}
Once the solution to this equation is found, the other variables are determined by eqs.~(\ref{eq:FLRW_Scalar_eq_1}),
(\ref{eq:FLRW_Scalar_eq_2}), (\ref{eq:FLRW_Scalar_eq_6}) and (\ref{eq:FLRW_PHI}). In particular, if $\Psi$ is a solution to eq.~(\ref{eq:FLRW_PSI}), the density contrast is given by
\begin{eqnarray} \label{eq:FLRW_rho}
\delta_\rho = \dfrac{2 \Mpl}{a^2 \rho} \left[ \left( \partial_j^2 - 3 a \mathcal{H}^2 \partial_a - 3 \mathcal{H}^2 \right) \Psi + a^{1/\gamma - 1} \left( 3 \mathcal{H}^2 \dfrac{1}{\gamma} - \partial_j^2 \right) \Psi_0 \right] .
\end{eqnarray}

The conventional cosmological perturbations are recovered by setting the mass parameters to zero, $m_i^2=0$. In this case, eq.~(\ref{eq:FLRW_Scalar_eq_3}) gives $\Phi - \Psi = 0$ which implies $\Psi_0 = 0$ (cf. eqs.~(\ref{eq:FLRW_XI}) and (\ref{eq:FLRW_PHI})). Then both eqs.~(\ref{eq:FLRW_PSI}) and (\ref{eq:FLRW_rho}) reduce to the standard equations describing cosmological perturbations in the Einstein theory. Note that the value of $\Psi_0$ is determined essentially by the initial conditions. Setting $\Psi_0 = 0$ would eliminate the $\Psi_0$-dependent terms in eqs.~(\ref{eq:FLRW_PSI}) and (\ref{eq:FLRW_rho}) and bring these relations to the conventional form even in the case $m_2^2 \neq 0$.

In order to stress the differences between massive gravity models and GR, let us consider different situations.

\subsubsection{Matter perturbations}

For matter perturbations,  $c_s^2 = 0$ and our previous discussion of the Jeans instability suggest that no random thermal motion will be able to dilute the over-dense regions since $k_J = \infty$. This naive expectation is confirmed by the exact solution to eq.~(\ref{eq:FLRW_PSI}). Indeed, the equation for $\Psi$ in a matter-dominated Universe ($w = 0$) reduces to
\begin{eqnarray} \label{eq:FLRW_PSI_matter}
\partial_a^2 \Psi + \dfrac{7}{2 a} \partial_a \Psi + \dfrac{1}{\gamma} a^{1/\gamma - 3} \left( 1 - \dfrac{1}{\gamma} \right) \Psi_0 = 0,
\end{eqnarray}
which differs from the standard one by the presence of an inhomogeneous term proportional to $\Psi_0$. The solution to this
equation reads
\begin{eqnarray*} \label{043}
\Psi = \dfrac{2}{2 + 3 \gamma} a^{1/\gamma - 1} \Psi_0 \left( x^i \right) + a^{-5/2} c_1 \left( x^i \right) + c_2 \left( x^i \right) ,
\end{eqnarray*}
where $c_i(x^i)$ are two integration constants. Substituting this solution into eq.~(\ref{eq:FLRW_rho}) one finds the density contrast
\begin{eqnarray} \label{eq:FLRW_rho_matter}
\delta_\rho &=& \left( \dfrac{2 a}{3 H_0^2} \partial_i^2 + 3 \right) \dfrac{c_1 \left( x^i \right)}{a^{5/2}} + \left( \dfrac{2 a}{3 H_0^2} \partial_i^2 - 2 \right) c_2 \left( x^i \right) \nonumber \\
& & - \dfrac{1}{2 + 3 \gamma} a^{1/\gamma - 1} \left( \dfrac{2 a \gamma}{H_0^2} \partial_i^2 - 6 \right) \Psi_0 \left( x^i \right) .
\end{eqnarray}
The first two terms in this equation are precisely the ones which appear in the standard Einstein theory, the second term describing the linear growth of the perturbations, $\delta_\rho \propto a$.  The difference with the conventional case consists in the third term on the right hand side of eq.~(\ref{eq:FLRW_rho_matter}). The perturbations corresponding to this term grow proportionally to $a^{1/\gamma}$. For $\gamma > 1$ or $\gamma < 0$ these \textquotedblleft anomalous\textquotedblright perturbations grow slower than the standard ones. Another case of interest is given by $\gamma = 1$. Indeed, in this case the $\Psi_0$-term in eq.~(\ref{eq:FLRW_rho_matter}) can be absorbed by a redefinition of $c_2(x^i)$ so that the density contrast has the same value as in GR.

\subsubsection{Radiation perturbations}

The situation is similar for a relativistic fluid $c_s^2 = 1 / 3$, except for the presence of pressure to compensate the gravitational instability. This has for consequence that the equations for $\Psi$ is more difficult to solve analytically. Hence, we will only consider radiation perturbations in the radiation epoch $w = 1 / 3$. In that case, eq.~(\ref{eq:FLRW_PSI}) reduces to the following one,
\begin{eqnarray} \label{eq:FLRW_PSI_radiation1}
0 = \partial_a^2 \Psi + \dfrac{4}{a} \partial_a \Psi - \dfrac{\partial_i^2}{3 H_0^2} \Psi + \dfrac{1}{\gamma} a^{1/\gamma - 3} \left[ \dfrac{a^2 \gamma \partial_i^2}{3 H_0^2} - \left( \dfrac{1}{\gamma} - 1 \right) \right] \Psi_0 .
\end{eqnarray}
In the three-dimensional momentum space, the solution to this equation is given by
\begin{eqnarray*}
\Psi = \dfrac{c_1 \left( k^i \right)}{\mathbf{n}^2} \left[ \dfrac{\sin \mathbf{n}}{\mathbf{n}} - \cos \mathbf{n} \right] + \dfrac{c_2 \left( k^i \right)}{\mathbf{n}^2} \left[ \dfrac{\cos \mathbf{n}}{\mathbf{n}} + \sin \mathbf{n} \right] + f \left( \mathbf{n} , \gamma \right) ,
\end{eqnarray*}
where $\mathbf{n} = \eta k / \sqrt{3}$ is proportional to the scale factor, $c_i(k^i)$ are two integration constants and $f(\mathbf{n} , \gamma)$ is given by
\begin{eqnarray*}
f \left( \mathbf{n} , \gamma \right) &\equiv& \dfrac{\Psi_0}{\gamma} \left( \dfrac{\sqrt{3}}{k \eta_0} \right)^{1/\gamma-1} \int \dif y \, \Theta \left( \mathbf{n} - y \right) \left[ \sin \left( \mathbf{n} - y \right) + \dfrac{\cos \left( \mathbf{n} - y \right)}{\mathbf{n}} \right. \\
& & \left. - \dfrac{\cos \left( \mathbf{n} - y \right)}{y} + \dfrac{\sin \left( \mathbf{n} - y \right)}{\mathbf{n} y} \right] \dfrac{y^{1/\gamma-1}}{\mathbf{n}^2} \left( \gamma y^2 + \dfrac{1}{\gamma} - 1 \right) .
\end{eqnarray*}
The density contrast is obtained by substituting this solution into eq.~(\ref{eq:FLRW_rho})
\begin{eqnarray*}
\delta_\rho &=& - 2 c_1 \left( k^i \right) \left( \dfrac{2 \cos \mathbf{n}}{\mathbf{n}^2} - \cos \mathbf{n} + \dfrac{2 \sin \mathbf{n}}{\mathbf{n}} - \dfrac{2 \sin \mathbf{n}}{\mathbf{n}^3} \right) \\
& & - 2 c_2 \left( k^i \right) \left( \dfrac{2 \cos \mathbf{n}}{\mathbf{n}} - \dfrac{2 \cos \mathbf{n}}{\mathbf{n}^3} + \sin \mathbf{n} - \dfrac{2 \sin \mathbf{n}}{\mathbf{n}^2} \right) \\
& & + 2 \left( \dfrac{\sqrt{3}}{k \eta_0} \mathbf{n} \right)^{1/\gamma-1} \left( \dfrac{1}{\gamma} + \mathbf{n}^2  \right) \Psi_0 - 2 \left( \mathbf{n}^2 + \mathbf{n} \partial_\mathbf{n} + 1 \right) f \left( \mathbf{n} , \gamma \right) .
\end{eqnarray*}
For generic values of $\gamma$, this expression is cumbersome. Yet, one may suspect that for $\gamma < - 1$ and for $0 < \gamma$ the density contrast grows faster than in GR because of the third term in the right hand side of this expression.

For simplicity let us concentrate on modes much smaller than the Hubble scale, $k^2 \gg \mathcal{H}^2$. For those modes, $\mathbf{n} \gg 1$ and the density contrast reduce to
\begin{eqnarray*}
\delta_\rho &\sim& 2 c_1 \left( k^i \right) \cos \mathbf{n} - 2 c_2 \left( k^i \right) \sin \mathbf{n} \nonumber \\
& & + 2 \left( \dfrac{\sqrt{3}}{k \eta_0} \right)^{1/\gamma-1} \left[ \mathbf{n}^{1 + 1 / \gamma} - \int_0^\mathbf{n} \dif{y} \, y^{1 + 1 / \gamma} \sin \left( \mathbf{n} - y \right) \right] \Psi_0 .
\end{eqnarray*}
As one may see from this expression, for $-1\leq \gamma<0$ the $\Psi_0$-dependent contribution to the density contrast decays with the scale factor so that only the standard contribution remains. Thus, in this range of $\gamma$ the perturbations behave just as predicted by GR in both matter and radiation-dominated epochs.

The case $\gamma = 1$ is also special here. Indeed, it is straightforward to show that $f(\mathbf{n},1) = \Psi_0$ plus a term which is absorbed by a redefinition of $c_1(k^i)$. Thus, the dependence on $\Psi_0$ cancels out in the density contrast so that only the standard part remains. At other values of $\gamma$, the $\Psi_0$-dependent contributions to the density contrast grow in the radiation - dominated Universe.

\subsubsection{Dark energy perturbations}

Before going to the conclusions of this chapter, let us consider perturbations of a dark energy fluid characterized by $c_s^2 = - 1$ in a dark energy-dominated Universe $w = -1$. The equation for $\Psi$ reduces to
\begin{eqnarray*}
0 &=& \partial_a^2 \Psi + \dfrac{2}{a} \partial_a \Psi + \dfrac{\eta_0^2 \partial_i^2}{a^4} \Psi + a^{1/\gamma - 3} \left[ \dfrac{1}{\gamma} - \dfrac{1}{\gamma^2} - \dfrac{\eta_0^2 \partial_i^2}{a^2} \right] \Psi_0 .
\end{eqnarray*}
In the three-dimensional momentum space, the solution to this equation is given by
\begin{eqnarray} \label{eq:Psi_DE_pert}
\Psi = c_1 \left( k^i \right) \cosh \left( \dfrac{k \eta_0}{a} \right) + c_2 \left( k^i \right) \sinh \left( \dfrac{k \eta_0}{a} \right) + \Psi_0 a^{1/\gamma-1} ,
\end{eqnarray}
where as before $c_i(k^i)$ are two integration constants. The density contrast is obtained by substituting this solution into eq.~(\ref{eq:FLRW_rho})
\begin{eqnarray*}
\delta_\rho &=& 2 \left( - \left( \dfrac{k^2 \eta_0^2}{3 a^2} + 1 \right) \cosh \left( \dfrac{k \eta_0}{a} \right) + \dfrac{k \eta_0}{a} \sinh \left( \dfrac{k \eta_0}{a} \right) \right) c_1 \left( k^i \right) \\
& & + 2 \left( - \left( \dfrac{k^2 \eta_0^2}{3 a^2} + 1 \right) \sinh \left( \dfrac{k \eta_0}{a} \right) + \dfrac{k \eta_0}{a} \cosh \left( \dfrac{k \eta_0}{a} \right) \right) c_2 \left( k^i \right) .
\end{eqnarray*}
The density contrast as given by this relation should be the correct solution of our problem. However, things are more subtle here. For both matter and radiation perturbations, equation (\ref{eq:FLRW_Scalar_eq_2}) gives $\delta_\zeta$ once the potentials $\Psi$ and $\Phi$ are determined. For dark energy perturbations, $\delta_\zeta \propto \rho \left( 1 + w \right)$ vanishes identically so that equation (\ref{eq:FLRW_Scalar_eq_2}) reduces to $\Psi^\prime = - \mathcal{H} \Phi$. Therefore, this is an extra equation which has to be satisfied by (\ref{eq:Psi_DE_pert}), and which implies that $c_1 = c_2 = \Psi_0 = 0$. Thus, the density contrast vanishes identically $\delta_\rho = 0$ as in the conventional GR. It is worth noting that these results do also apply in an inflationary epoch.

\section{Summary and prospects}

Let us conclude this chapter by summarizing the previous discussion. After having discussed cosmological models in massive gravity theories, we have studied the perturbations in one specific cosmological model based on the sub-class of theories invariant under the dilatation symmetry (\ref{eq:dilatation_Z}). We have argued that cosmological perturbations contain two contributions, the \textquotedblleft normal\textquotedblright and the \textquotedblleft anomalous\textquotedblright one. The first, normal contribution has the behavior identical to that found in the conventional GR. It is therefore in agreement with observations to the same extent as the latter. In particular, the \textquotedblleft normal\textquotedblright contribution can describe successfully at least the linear stage of the structure formation.

The second, \textquotedblleft anomalous\textquotedblright contribution of the cosmological perturbations is specific to the model of massive gravity with the dilatation symmetry (\ref{eq:dilatation_Z}). These perturbations originate from the condensation of the four scalar fields present in the model and depend on the unknown function $\Psi_0(x^i)$ already introduced in section \ref{sc:Weak_Field}. This function enters the solutions of the EoM as an integration constant. Hence, its value cannot be determined within the model (\ref{eq:MG_action_Z}) and has to be specified as an initial conditions.

The behavior of the \textquotedblleft anomalous\textquotedblright perturbations at different stages of the evolution of the Universe depends on the value of $\gamma$. At the matter-dominated stage the anomalous perturbations grow not faster than the standard ones for $\gamma < 0$ and $1 \leq \gamma$. In the radiation-dominated epoch this occurs at $- 1 \leq \gamma < 0$ and $\gamma=1$. Thus, at $- 1 \leq \gamma < 0$ and $\gamma = 1$ the normal perturbations dominate at both radiation and matter-dominated stages. During an inflationary stage or in a dark energy-dominated Universe, the \textquotedblleft anomalous\textquotedblright perturbations cancels out in the density contrast.

The appearance of this time-independent arbitrary function is not surprising since the same function $\Psi_0$ already enters the expression for the gravitational potential of an isolated massive body (\ref{eq:Scalar_F(Z)_Pot}). Its origin may be traced back to the existence of the scalar mode with the dispersion relation $\omega^2 = 0$. This mode is not dynamical in the models considered. However, the action (\ref{eq:MG_action_Z}) is a low-energy effective action, so one should expect corrections containing higher-derivative terms to be present. In general, these corrections make $\Psi_0$ a dynamical variable with the dispersion relation $\omega^2 = \alpha k^4$, where $\alpha$ is a small coefficient (see section \ref{sc:Sym_Massive_Gravity}). Therefore, $\Psi_0$ becomes a slowly varying function of time. The slow evolution may drive $\Psi_0$ to a particular value at the inflationary epoch and thus prepare the initial conditions for the radiation-dominated stage. If this initial value of $\Psi_0$ is small, then the growth of the \textquotedblleft anomalous\textquotedblright part of perturbations may become irrelevant and corresponding values of $\gamma$ phenomenologically acceptable. This question deserves furthers studies.

Finally, it is worth noting that in the Standard Model of Cosmology, perturbations had to grown since recombination by a factor greater than the one predicted by GR alone. Indeed, GR fails at getting the baryons perturbations anywhere near the amplitude required to generate the observed structures of the Universe. For this reason, it is generally assumed that dark matter perturbations start growing before the emission of the CMB \cite{Liddle:1993fq}. Things are different in massive gravity models since for $\gamma < - 1$ or $0 < \gamma < 1$ or $1 < \gamma$, the density contrast grows faster than in GR. It would be interesting to see if this could eventually eliminate the need for dark matter in the formation of structures.

\chapter{Conclusions \& Final comments} \label{ch:conclusion}

In the first part of this thesis, we have reviewed the problems and issues of theories with massive gravitons, before introducing a particular class of theories for which the Lorentz symmetry is spontaneously broken by the vacuum expectation value of four scalar fields. This discussion had two aims. The first one was to illustrate the difficulties which arise when trying to give a mass to gravitons. The second was to show how theories with Lorentz symmetry breaking go through all these difficulties and describe consistent effective field theories which reduce to GR in the limit of vanishing graviton mass. In this first part, we have also introduced a minimal class of theories possessing only two massive propagating modes with helicities $\pm 2$, and have shown that for the sub-class of models invariant under the dilatation symmetry (\ref{eq:dilatation_Z}) there is no modification of the Newtonian potential.

In the second part of this thesis, we have discussed the original contributions to the study of this minimal class of massive gravity theories. Let us summarize here the conclusions of these works.

\vspace{0.5cm}

It has been argued in chapter \ref{ch:inst_int} that physical instantaneous interactions are present in this minimal class of massive gravity theories because of the spontaneous breaking of the Lorentz-invariance. This symmetry breaking allows for independent modifications of the dispersion relations of the tensor, vector and scalar sectors so that the instantaneous contributions to the potentials do not cancel in the graviton propagator, unlike in GR. A concrete example of instantaneous interaction is given for theories invariant under the dilatation symmetry (\ref{eq:dilatation_Z}): we demonstrated by direct calculations that a gravitational source localized in space is responsible for an instantaneous frequency shift of a distant light beam. This interaction originates in an anisotropic sound wave which propagates at an infinite velocity in the Goldstone fluid. It is supposed to be responsible for the violation of the black hole \textquotedblleft no-hair\textquotedblright \, theorem which motivates the work summarized in chapter \ref{ch:bh}.

\vspace{0.5cm}

In chapter \ref{ch:bh}, the static vacuum spherically symmetric solutions in massive gravity have been obtained both analytically and numerically. The solutions depend on two integration constants, instead of one in GR: the Schwarzschild radius $r_s$ and an additional \textquotedblleft scalar charge\textquotedblright $S$. At zero value of $S$ and positive $r_s$ the standard Schwarzschild black hole solutions are recovered. Depending on the parameters of the model and the signs of $r_s$ and $S$, the solutions may or may not have horizon. Those with the horizon describe modified black holes provided they are stable against small perturbations. In the analytically solvable example, the modified black hole solutions may have both attractive and repulsive (anti-gravitating) behavior at large distances.  At intermediate distances the gravitational potential of a modified black hole may mimic the presence of dark matter. Modified black hole solutions are also found numerically in more realistic massive gravity models which are attractors of the cosmological evolution.

\vspace{0.7cm}

Finally, in chapter \ref{ch:cosmo} we have discussed more subtle parts of modern cosmology, namely the mechanism of structure formation.
After having introduced cosmological models in the minimal class of massive gravity theories, we have studied the perturbations in one specific cosmological model based on the sub-class of theories invariant under the dilatation symmetry (\ref{eq:dilatation_Z}). The Friedman equation in these models acquires an unconventional term due to the Lorentz-breaking condensates which has the equation of state $w=-1/(3\gamma)$ with $\gamma$ being a free parameter taking values outside of the range $[0,1/3]$. Apart from the standard contributions, the perturbations above the Friedmann background contain an extra piece which is proportional to an arbitrary function $\Psi(x^i)$ of the space coordinates. This function appears as an integration constant and corresponds to a non-propagating scalar mode which may, however,
become dynamical with the account of the higher-derivative corrections. For $-1<\gamma<0$ and $\gamma=1$ the \textquotedblleft anomalous\textquotedblright perturbations grow slower than the standard ones and thus the model is compatible with observations. Whether the model is experimentally acceptable at other values of $\gamma$ depends on the value of the function $\Psi(x^i)$ at the beginning of the radiation-dominated epoch.

\vspace{0.5cm}

In this thesis, we have addressed only a few questions related to massive gravitons and possible modifications of GR at large scales. Many of the interesting questions still remain open. Among all possible further investigations, let us summerize some which may seem promising in light of what have been learned in this thesis.
\begin{itemize}
\item It is still an open question how objects (stars, black holes, \ldots) with a non-zero scalar charge $S$ may be created. Since a non-zero scalar charge will imply a modification of the Schwarzschild metric, this question surely deserves further investigations.
\item Another open question concern the stability of the spherically symmetric solutions. These solutions describe modified black hole provided they are stable against small perturbations. It is therefore important to study this issue in more detail.
\item The presence of a physical instantaneous interaction should allow to look behind the horizon of a black hole or behind the cosmological horizon. It would therefore be interesting to understand how information is carried through these horizons.
\item It would be interesting to compare the predictions of massive gravity theories concerning cosmological perturbations directly to observations, in order to determine which values of $\gamma$ and $\Psi_0$ correspond to the best fit to observations.
\end{itemize}
The work presented in this thesis correspond to a minimal class of models, which does not claim to be the only viable candidates for theories with massive gravitons. Even if some aspects of these models have been studied in this thesis and elsewhere, an intensive theoretical as well as experimental work has still to be done before one is able to conclude with certainty about their ability to describe correctly the gravitational interaction. Finally, we would like to stress that if GR correctly describes the gravitational interaction from cosmological scales up to the Planck scale, the study of massive gravity models enables one to comprehend which deviations from the standard laws of gravity are possible and how we may constrain them.

\appendix

\part[Appendix]{\usefont{OT1}{pzc}{m}{n}\selectfont Appendix}

\chapter{The gravitational field equations} \label{app:ch:equations}

The action for the massive gravitational field reads
\begin{eqnarray*}
\act = \act_{\tx{GR}} + \act_{\phi} &\tx{with}& \act_{\phi} = \int \textrm{d}^4 x \sqrt{- g} \Lambda^4 \mathcal{F} \left( X , V^i , W^{ij} \right) ,
\end{eqnarray*}
where $X$, $V^i$ and $W^{ij}$ are given by
\begin{eqnarray*}
\begin{array}{ccc}
X = g^{\mu\nu} \partial_\mu \phi^0 \partial_\nu \phi^0 , & V^{i} = g^{\mu\nu} \partial_\mu \phi^0 \partial_\nu \phi^i , & W^{ij} = g^{\mu\nu} \partial_\mu \phi^i \partial_\nu \phi^j - \dfrac{V^i V^j}{X} ,
\end{array}
\end{eqnarray*}
and where $\phi^\mu$ are four scalar fields. The variation of this action with respect to the metric gives the Einstein equations
\begin{eqnarray*}
\mathcal{G}_{\mu\nu} = \dfrac{1}{\Mpl} \left( \mathcal{T}_{\mu\nu} + t_{\mu\nu} \right),
\end{eqnarray*}
which are ten equations determining the metric. $\mathcal{T}_{\mu\nu}$ is the energy-momentum tensor of standard matter fields while $t_{\mu\nu}$ is the energy momentum tensor of the scalar fields, obtained by varying $\act_{\phi}$ with respect to the metric
\begin{eqnarray}
t_{\mu\nu} &=& \Lambda^{4} \left[ - \dfrac{1}{2} g_{\mu\nu} \mathcal{F} + \dfrac{1}{2}
\dfrac{\partial \mathcal{F}}{\partial V^{i}} \left( \partial_{\mu} \phi^{i} \partial_{\nu} \phi^{0} + \partial_{\mu} \phi^{0}
\partial_{\nu} \phi^{i} \right) + \left( \partial_{\mu} \phi^{i} \partial_{\nu} \phi^{j} + \dfrac{V^i V^j}{X^2} \partial_{\mu} \phi^0 \partial_{\nu} \phi^0 \right. \right. \nonumber \\
& & \left. \left. - \dfrac{V^j}{X} \left( \partial_{\mu} \phi^{i} \partial_{\nu} \phi^{0} + \partial_{\mu} \phi^0 \partial_{\nu} \phi^i \right) \right) \dfrac{\partial \mathcal{F}}{\partial W^{ij}} + \dfrac{\partial \mathcal{F}}{\partial X} \partial_{\mu} \phi^{0} \partial_{\nu} \phi^{0} \right] \label{eq:Ann_En_Mom_Gold} .
\end{eqnarray}
The variation of the action with respect to the scalar fields give four Goldstone equations which read
\begin{eqnarray}
0 &=& \partial_\beta \left\lbrace \sqrt{- g} g^{\alpha\beta} \left[ \left( \dfrac{\partial \mathcal{F}}{\partial X} + \dfrac{\partial \mathcal{F}}{\partial W^{ij}} \dfrac{V^{i} V^{j}}{X^2} \right) \delta^0_\mu \partial_\alpha
\phi^{0} + \dfrac{\partial \mathcal{F}}{\partial W^{ij}} \delta^i_\mu \partial_\alpha \phi^{j} \right. \right. \nonumber\\
& & \left. \left. + \left( \dfrac{1}{2} \dfrac{\partial \mathcal{F}}{\partial V^{i}} - \dfrac{\partial \mathcal{F}}{\partial W^{ij}} \dfrac{V^{j}}{X} \right) \partial_\alpha \left( \phi^{0} \delta^i_\mu + \phi^{i} \delta^0_\mu
\right) \right] \right\rbrace . \label{eq:Goldstone_equation}
\end{eqnarray}

It is worth noting that the Goldstone equations implies the conservation of energy-momentum tensor of the scalar fields. This is easily demonstrated after the introduction of more compact notations. Let $\chi^{\mu\nu}$ be
\begin{eqnarray*}
\chi^{\mu\nu} = g^{\alpha\beta} \partial_\alpha \phi^\mu \partial_\beta \phi^\nu.
\end{eqnarray*}
Then, the energy-momentum tensor of the scalar fields reads
\begin{eqnarray*}
t_{\mu\nu} = \Lambda^{4} \left[ - \dfrac{1}{2} g_{\mu\nu} \mathcal{F} + \mathcal{F}_{\alpha\beta} \partial_\mu \phi^\alpha \partial_\nu \phi^\beta \right] , && \mathcal{F}_{\alpha\beta} \equiv \dfrac{\partial \mathcal{F}}{\partial \chi^{\alpha\beta}} ,
\end{eqnarray*}
while the Goldstone equations are given by
\begin{eqnarray*}
0 = \partial_\mu \left( \sqrt{- g} \mathcal{F}_{\alpha\beta} g^{\mu\nu} \partial_\nu \phi^\beta \right) .
\end{eqnarray*}
With these new notations, it is straightforward to show that
\begin{eqnarray*}
\nabla_{\lambda} t_{\mu}^{\lambda} = \dfrac{\partial_\mu \phi^\alpha}{\sqrt{- g}} \partial_{\lambda} \left( \sqrt{- g} \mathcal{F}_{\alpha\beta} g^{\lambda\gamma} \partial_\gamma \phi^\beta \right) ,
\end{eqnarray*}
where $\nabla_{\mu}$ ($\nabla^{\mu}$) is the covariant derivative with respect to $x^\mu$ ($x_\mu$). Consequently, the Goldstone equations and the conservation equations for the scalar fields are identical provided that $\partial_\mu \phi^\alpha \neq 0$. This discussion enable to identify the number of independent equations of the massive gravitational field. If there is no usual matter field in the right hand side of the Einstein equations, then the Goldstone equations are not independent equations since they are a consequence of the fact that the Einstein tensor is divergence-less. On the contrary, if one consider space-time with matter field in it, then the Einstein equations implies the conservation of the total energy-momentum tensor $\mathcal{T}_{\mu\nu} + t_{\mu\nu}$ with consequences that the Goldstone equations are independent equations. In these situations, one has to consider either the conservation equations for the matter fields or the Goldstone equations along with the Einstein equations.

\section{Energy-momentum tensor} \label{app:sc:energy_momentum}

The energy-momentum tensor associated to a matter field Lagrangian $\lag$ is given by the variation of its Lagrangian with respect to the metric
\begin{eqnarray*}
\delta \left( \int \textrm{d}^4 x \sqrt{- g} \lag \right) = \int \textrm{d}^4 x \sqrt{- g} \mathcal{T}_{\mu\nu} \delta g^{\mu\nu},
\end{eqnarray*}
with
\begin{eqnarray*}
\mathcal{T}_{\mu\nu} = \dfrac{1}{\sqrt{- g}} \dfrac{\delta \left( \sqrt{- g} \mathcal{L}_\omega \right)}{\delta g^{\mu\nu}}
\end{eqnarray*}
The tensor defined by this last relation is a symmetric tensor of rank two. It contains all the information concerning the energy and momentum of the matter fields described by this Lagrangian.

Lets $v^\mu$ with $v_\mu v^\mu = 1$ be the velocity of an observer comoving with the fluid described by the energy-momentum tensor. Then, the two following tensors are projection tensor along and perpendicular to $v^\mu$
\begin{eqnarray*}
v^\mu v^\nu , & j_{\mu\nu} \equiv v_\mu v_\nu - g_{\mu\nu} ,
\end{eqnarray*}
respectively. The energy-momentum tensor of the fluid can be decomposed as
\begin{eqnarray*}
\mathcal{T}_{\mu\nu} = \rho v_{\mu} v_{\nu} + q_\mu v_\nu + q_\nu v_\mu - p j_{\mu\nu} + \pi_{\mu\nu} ,
\end{eqnarray*}
where $\rho = \mathcal{T}_{\mu\nu} v^\mu v^\nu$ and $p = \mathcal{T}_{\mu\nu} j^{\mu\nu} / 3$ are respectively the energy and pressure densities measured by the comoving observer, $q^\mu = - \mathcal{T}_{\gamma\lambda} v^\lambda j^{\gamma\mu}$ is the energy flux perpendicular to $v^\mu$ and $\pi^{\mu\nu} = \mathcal{T}_{\lambda\gamma} j^{\lambda\mu} j^{\gamma\nu} + j^{\mu\nu} \mathcal{T}_{\lambda\gamma} j^{\lambda\gamma} / 3 $ is the anisotropic pressure tensor
\begin{eqnarray*}
q_\mu v^\mu = 0 ,& \pi_{\mu\nu} v^\nu = 0 , & \pi_\mu^\mu = 0
\end{eqnarray*}

With these notations, it is possible to determine the energy and pressure densities of the Goldstone scalar fields. Indeed, one has that
\begin{eqnarray*}
\rho_{\phi} &=& - \dfrac{1}{2} \Lambda^{4} \mathcal{F} + v^\mu v^\nu \Lambda^{4} \left[ \dfrac{1}{2}
\dfrac{\partial \mathcal{F}}{\partial V^{i}} \left( \partial_{\mu} \phi^{i} \partial_{\nu} \phi^{0} + \partial_{\mu} \phi^{0}
\partial_{\nu} \phi^{i} \right) + \left( \partial_{\mu} \phi^{i} \partial_{\nu} \phi^{j} + \dfrac{V^i V^j}{X^2} \partial_{\mu} \phi^0 \partial_{\nu} \phi^0 \right. \right. \\
& & \left. \left. - \dfrac{V^j}{X} \left( \partial_{\mu} \phi^{i} \partial_{\nu} \phi^{0} + \partial_{\mu} \phi^0 \partial_{\nu} \phi^i \right) \right) \dfrac{\partial \mathcal{F}}{\partial W^{ij}} + \dfrac{\partial \mathcal{F}}{\partial X} \partial_{\mu} \phi^{0} \partial_{\nu} \phi^{0} \right] ,
\end{eqnarray*}
\begin{eqnarray*}
p_{\phi} &=& \dfrac{1}{2} \Lambda^{4} \mathcal{F} + \dfrac{j^{\mu\nu}}{3} \Lambda^{4} \left[ \dfrac{1}{2}
\dfrac{\partial \mathcal{F}}{\partial V^{i}} \left( \partial_{\mu} \phi^{i} \partial_{\nu} \phi^{0} + \partial_{\mu} \phi^{0}
\partial_{\nu} \phi^{i} \right) + \left( \partial_{\mu} \phi^{i} \partial_{\nu} \phi^{j} + \dfrac{V^i V^j}{X^2} \partial_{\mu} \phi^0 \partial_{\nu} \phi^0 \right. \right. \\
& & \left. \left. - \dfrac{V^j}{X} \left( \partial_{\mu} \phi^{i} \partial_{\nu} \phi^{0} + \partial_{\mu} \phi^0 \partial_{\nu} \phi^i \right) \right) \dfrac{\partial \mathcal{F}}{\partial W^{ij}} + \dfrac{\partial \mathcal{F}}{\partial X} \partial_{\mu} \phi^{0} \partial_{\nu} \phi^{0} \right] .
\end{eqnarray*}

\section{The mass parameters} \label{app:sc:mass_param}

In backgrounds characterized by flat three-dimensional space $g_{ij} \propto \delta_{ij}$, there are five mass parameters $m_i^2$, $i = 0 \ldots 4$, defined by the following relations
\begin{eqnarray*}
m_0^2 &=& \dfrac{\Lambda^{4}}{\Mpl} \left( X \mathcal{F}_{X} + 2 X^2 \mathcal{F}_{XX} \right) , \\
m_1^2 &=& \dfrac{2 \Lambda^{4}}{\Mpl} \left( - X \mathcal{F}_{X} - W \mathcal{F}_{W} + \dfrac{1}{2} X W \mathcal{F}_{VV} \right) , \\
m_2^2 &=& \dfrac{2 \Lambda^{4}}{\Mpl} \left( W \mathcal{F}_{W} - 2 W^2 \mathcal{F}_{WW2} \right) , \\
m_3^2 &=& \dfrac{\Lambda^{4}}{\Mpl} \left( W \mathcal{F}_{W} + 2 W^2 \mathcal{F}_{WW1} \right) , \\
m_4^2 &=& - \dfrac{\Lambda^{4}}{\Mpl} \left( X \mathcal{F}_{X} + 2 X W \mathcal{F}_{XW} \right) ,
\end{eqnarray*}
where $W = -1 / 3 \delta_{ij} W^{ij}$ and where the first and second non-zero derivatives of the function $\mathcal{F}$ are denoted as follows
\begin{eqnarray*}
\dfrac{\partial \mathcal{F}}{\partial X} \equiv \mathcal{F}_{X}, && \dfrac{\partial^2 \mathcal{F}}{\partial X^2} \equiv \mathcal{F}_{XX} , \\
\dfrac{\partial^2 \mathcal{F}}{\partial V^{i}V^{j}} \equiv \mathcal{F}_{VV} \delta_{ij}, && \dfrac{\partial \mathcal{F}}{\partial W^{ij}} \equiv \mathcal{F}_{W} \delta_{ij} , \\
\dfrac{\partial^2 \mathcal{F}}{\partial X W^{ij}} \equiv \mathcal{F}_{XW} \delta_{ij}, && \dfrac{\partial^2 \mathcal{F}}{\partial W^{ij} W^{kl}} \equiv \mathcal{F}_{WW1} \delta_{ij} \delta_{kl} + \mathcal{F}_{WW2} \left( \delta_{ik} \delta_{jl} + \delta_{il} \delta_{jk} \right) .
\end{eqnarray*}
For function independent of $V^i$ as the ones considered in this thesis, the mass $m_1^2$ is proportional to a combination of the energy-momentum tensor of the scalar fields
\begin{eqnarray*}
m_1^2 = - 2 \left( \rho_\phi + p_\phi \right) ,
\end{eqnarray*}
with consequence that this mass vanishes in Minkowski space-time. For the class of models characterized by a function $\mathcal{F} = \mathcal{F} \left( Z^{ij} \right)$ with \mbox{$Z^{ij} = X^\gamma W^{ij}$}, it is straightforward to show that the five mass parameters are
\begin{eqnarray*}
m_0^2 = \dfrac{\Lambda^{4}}{\mathcal{M}^2_{pl}} \gamma [ 3 ( 1 - 2 \gamma ) Z \mathcal{F}_{Z} + 6 \gamma Z^2 \left( 3 \mathcal{F}_{ZZ1} + 2 \mathcal{F}_{ZZ2} \right) ] , & & m_1^2 = \dfrac{2 \Lambda^{4}}{M^2_{pl}} \left( 3 \gamma - 1 \right) Z \mathcal{F}_{Z} , \\
m_2^2 = \dfrac{2 \Lambda^{4}}{\mathcal{M}^2_{pl}} \left( Z \mathcal{F}_{Z} - 2 Z^2 \mathcal{F}_{ZZ2} \right) , & & m_3^2 = \dfrac{\Lambda^{4}}{\mathcal{M}^2_{pl}} \left( Z \mathcal{F}_{Z} + 2 Z^2 \mathcal{F}_{ZZ1} \right) , \\
m_4^2 = \dfrac{\Lambda^{4}}{\mathcal{M}^2_{pl}} \gamma \left[ Z \mathcal{F}_{Z} + 2 Z^2 \left( 3 \mathcal{F}_{ZZ1} + 2 \mathcal{F}_{ZZ2} \right) \right] . &&
\end{eqnarray*}
where
\begin{eqnarray*}
\dfrac{\partial \mathcal{F}}{\partial Z^{ij}} &\equiv& \mathcal{F}_{Z} \delta_{ij}, \\
\dfrac{\partial^2 \mathcal{F}}{\partial Z^{ij} Z^{kl}} &\equiv& \mathcal{F}_{ZZ1} \delta_{ij} \delta_{kl} + \mathcal{F}_{ZZ2} \left( \delta_{ik} \delta_{jl} + \delta_{il} \delta_{jk} \right) .
\end{eqnarray*}
For those particular models, one has
\begin{eqnarray*}
m_0^2 = 3 \gamma \left( m_4^2 - m_1^2 / 2 \right) , & m_4^2 = \gamma \left( 3 m_3^2 - m_2^2 \right) .
\end{eqnarray*}

\section{Linearized gravity} \label{app:sc:line_equ}

Since metric theories of gravity are generally non-linear, perturbation theory plays an important role in exploring such theories. Consider small fluctuations about a fixed background with metric $\gamma_{\mu\nu}$
\begin{eqnarray*}
g_{\mu\nu} = \gamma_{\mu\nu} + h_{\mu\nu} &,& | h_{\mu\nu} | \sim \epsilon \ll 1 .
\end{eqnarray*}
With this parametrization, and with $\gamma = \det \left( \gamma_{\mu\nu} \right)$ and $h = h_{\mu\nu} \gamma^{\mu\nu}$, the inverse of the metric and its determinant $g = \det \left( g_{\mu\nu} \right)$ are given by
\begin{eqnarray*}
g^{\mu\nu} &=& \gamma^{\mu\nu} - h^{\mu\nu} + h^{\mu\alpha} \gamma_{\alpha\beta} h^{\beta\nu} + \mathcal{O} \left( \epsilon^3 \right) , \\
\sqrt{- g} &=& \sqrt{- \gamma} \left[ 1 + \dfrac{1}{2} h + \dfrac{1}{4} \left( \dfrac{h^2}{2} - h_{\mu\nu} h^{\mu\nu} \right) + \mathcal{O} \left( \epsilon^3 \right) \right] .
\end{eqnarray*}
With these relations, the Ricci tensor is given up to quadratic order in the fluctuations by
\begin{eqnarray*}
\ricci_{\mu\nu} &=& \ricci_{\mu\nu} {\big |}_{h_{\mu\nu} = 0} + \dfrac{1}{2} \left[ \nabla_{\mu} \nabla^{\alpha} h_{\nu\alpha} + \nabla_{\nu} \nabla^{\alpha} h_{\mu\alpha} - \nabla_{\alpha} \nabla^{\alpha} h_{\mu\nu} - \nabla_{\mu} \nabla_{\nu} h \right] \\
& & + \dfrac{1}{2} \nabla_{\nu} \left( h^{\alpha\beta} \nabla_{\mu} h_{\alpha\beta} \right) + \dfrac{1}{4} \left( \nabla_{\mu} h_{\nu\alpha} + \nabla_{\nu} h_{\mu\alpha} - \nabla_{\alpha} h_{\mu\nu} \right) \nabla^{\alpha} h \\
& & + \dfrac{1}{2} \gamma^{\alpha\beta} \nabla^{\lambda} h_{\mu\alpha} \left( \nabla_{\lambda} h_{\nu\beta} - \nabla_{\beta} h_{\nu\lambda} \right) - \dfrac{1}{4} \nabla_{\mu} h^{\alpha\beta} \nabla_{\nu} h_{\alpha\beta} \\
& & - \dfrac{1}{2} \nabla_{\alpha} \left[ h^{\alpha\beta} \left( \nabla_{\mu} h_{\nu\beta} + \nabla_{\nu} h_{\mu\beta} \right) - h^{\alpha\beta} \nabla_{\beta} h_{\mu\nu} \right] + \mathcal{O} \left( \epsilon^3 \right) ,
\end{eqnarray*}
where $\nabla_{\mu}$ ($\nabla^{\mu}$) is the covariant derivative with respect to $x^\mu$ ($x_\mu$). These relations allows to express the Einstein-Hilbert action and the gravitational field equations up to the second order in the perturbations.

\subsection{The gravity sector} \label{app:sc:line_EH}

If the Ricci tensor is decomposed according to
\begin{eqnarray*}
\ricci_{\mu\nu} + \delta \ricci_{\mu\nu} + \delta^2 \ricci_{\mu\nu} + \ldots \, ,
\end{eqnarray*}
where $\delta \ricci_{\mu\nu}$ and $\delta^2 \ricci_{\mu\nu}$ are respectively the linearized and quadratic contributions to the Ricci tensor, then the Einstein-Hilbert action describing gravitational perturbations above a fixed space-time is given up to the quadratic order by the following relation
\begin{eqnarray*}
\act_{\tx{EH}} &=& - \Mpl \int \dif^4 x \, \sqrt{- \gamma} \left[ \ricci + \delta \ricci_{\mu\nu} \gamma^{\mu\nu} - \left( h^{\mu\nu} - \dfrac{1}{2} h \gamma^{\mu\nu} \right) \ricci_{\mu\nu} - \left( h^{\mu\nu} - \dfrac{1}{2} h \gamma^{\mu\nu} \right) \delta \ricci_{\mu\nu} \right. \\
& & \left. + \delta^2 \ricci_{\mu\nu} \gamma^{\mu\nu} + \left( \dfrac{1}{8} \gamma^{\mu\nu} h^{2} - \dfrac{1}{4} \gamma^{\mu\nu}  h_{\lambda\gamma} h^{\lambda\gamma} + h^{\mu\lambda} h_\lambda^\nu - \dfrac{1}{2} h h^{\mu\nu} \right) \ricci_{\mu\nu} \right] .
\end{eqnarray*}
In this action, the zero order contribution fixes the background space-time. The first order contributions are proportional to the EoM of the background and therefore vanish. The only relevant terms are the quadratic ones, which in Minkowski space-time reduce to
\begin{eqnarray*}
\mathcal{L}_{\tx{EH}} {\Big |}_{\tx{Mink.}} &=& \dfrac{1}{4} \left[ \partial_{\alpha} h^{\mu\nu} \partial^{\alpha} h_{\mu\nu} - 2 \partial_{\mu} h^{\mu\nu} \partial^{\alpha} h_{\nu\alpha} + h \left( \partial^{\mu} \partial_{\mu} h - 2 \partial^{\mu} \partial^{\nu} h_{\mu\nu} \right) \right].
\end{eqnarray*}

There is two equivalent ways to get the linearized Einstein equations for the gravitational field. The first possibility is to linearize the action up to the second order, then varying this action with respect to the perturbations. This approach is used in the chapter \ref{ch:mg} for Minkowski background. The second possibility is to linearize directly the Einstein equations. This second option is often easier since linear equations are simpler than quadratic actions. For this reason, this second procedure is used to derive the linearized EoM in FLRW background needed in chapter \ref{ch:cosmo}.

The left-hand-side of Einstein's equations is given by the Einstein tensor
\begin{eqnarray*}
\mathcal{G}_{\mu\nu} = \mathcal{R}_{\mu\nu} - \dfrac{1}{2} g_{\mu\nu} \mathcal{R}
\end{eqnarray*}
In order to linearize this tensor, let us concentrate on the flat Friedmann-Lema\^{\i}tre-Robertson-Walker solution described by the metric (\ref{eq:FLRW_metric}) with $k = 0$. Then, with the notations introduced in section \ref{sc:cosmo_pert}, the first order Einstein tensor  reads
\begin{eqnarray*}
\delta \mathcal{G}_{00} &=& 2 \partial_i^2 \Psi - 6 \mathcal{H} \psi^\prime , \\
\delta \mathcal{G}_{i0} &=& \partial_{i} \left[ 2 \left( \mathcal{H} \varphi + \psi^\prime \right) + \left( 2 \mathcal{H}^{\prime} + \mathcal{H}^2 \right) B \right] + \dfrac{1}{2} \partial_i^2 \varpi_{i} + S_i \left( 2 \mathcal{H}^{\prime} + \mathcal{H}^2 \right) , \\
\delta \mathcal{G}_{ij} &=& \delta_{ij} \left[ 2 \mathcal{H} \left( 2 \psi + \varphi \right)^\prime + 2 \psi^{\prime\prime} - \partial_k^2 \left( \Psi - \Phi \right) + 2 \left( \psi + \varphi \right) \left( \mathcal{H}^2 + 2 \mathcal{H}^\prime \right) \right] \\
& & + \partial_i \partial_j \left[ \Psi - \Phi - 2 \left( 2 \mathcal{H}^{\prime} + \mathcal{H}^2 \right) E \right] - \mathcal{H} H_{ij}^{\prime} + \dfrac{1}{2} \left( \partial_k^2 H_{ij} - H_{ij}^{\prime\prime} \right) + \left( 2 \mathcal{H}^{\prime} + \mathcal{H}^2 \right) H_{ij} \\
& & + \mathcal{H} \left( \partial_i \varpi_j + \partial_j \varpi_i \right) + \dfrac{1}{2} \left( \partial_i \varpi^\prime_j + \partial_j \varpi^\prime_i \right) - \left( \mathcal{H}^2 + 2 \mathcal{H}^\prime \right) \left( \partial_i F_j + \partial_j F_i \right) .
\end{eqnarray*}
where $\mathcal{H}$ is the conformal Hubble parameter and the prime denotes the derivative with respect to the conformal time $\eta$. The Einstein tensor in Minkowski space-time is deduced from these relations by taking $a = 1$, $\mathcal{H} = 0$, $\psi = \Psi$ and by replacing the prime by a dot which correspond to the derivative with respect to the time $t$.

\subsection{The scalar fields sector}

As discussed in section \ref{sc:FLRW}, the homogeneous and isotropic ansatz for the Goldstone scalar fields reads
\begin{eqnarray*}
\phi^0 = \phi \left( \eta \right) , & \phi^i = x^i .
\end{eqnarray*}
With this ansatz, it is possible to show that the Goldstone equation (\ref{eq:FLRW_Golds_V1}) can be put in the following form
\begin{eqnarray} \label{eq:FLRW_Golds_ma}
m_0^2 \left( \dfrac{\phi^{\prime\prime}}{\phi^{\prime}} - \mathcal{H} \right) = 3 \mathcal{H} m_4^2 .
\end{eqnarray}
Hence, this equation relates the masses $m_0^2$ and $m_4^2$. This relation will be used to express the linearized field equations in a gauge invariant formulation. But before writing the EoM, one need to determine the linearized energy-momentum tensor of the Goldstone fields. This tensor is given by
\begin{eqnarray*}
\delta t_{00} &=& a^2 \left[ 2 \rho_\phi \varphi + m_0^2 \Mpl \left( \xi^{0\prime} / \phi^{\prime} - \varphi \right) + \left( \rho_\phi + p_\phi + m_4^2 \Mpl \right) \left( \partial_i^2 \Xi + 3 \psi \right) \right] , \\
\delta t_{0i} &=& a^2 \left[ \left( \rho_\phi + p_\phi \right) \left( \xi_i^T + \partial_i \xi \right)^{\prime} + \rho_\phi \left( S_i + \partial_i B \right) + \dfrac{m_1^2 \Mpl}{2} \left( \varpi_i + \sigma_i^{\prime} - \dfrac{\partial_i \Xi^{0}}{\phi^\prime} + \partial_i \Xi^{\prime} \right) \right] , \\
\delta t_{ij} &=& a^2 \Mpl \left[ \dfrac{1}{2} m_2^2 \left( \partial_j \xi_i^T + \partial_i \xi_j^T + 2 \partial_i \partial_j \xi \right) - m_3^2 \delta_{ij} \left( 3 \psi + \partial_i^2 \Xi \right) + \delta_{ij} m_4^2 \left( \varphi - \xi^{0\prime} / \phi^{\prime} \right) \right. \\
& & \left. + \left( \dfrac{1}{2} m_2^2 - \dfrac{p_\phi}{\Mpl} \right) \left( 2 \psi \delta_{ij} - \partial_i F_j - \partial_j F_i - 2 \partial_i \partial_j E + H_{ij} \right) \right] .
\end{eqnarray*}

\subsection{The Einstein equations} \label{app:sc:line_equa}

The energy-momentum tensor of usual matter fields have been discussed in section \ref{sc:cosmo_pert} (see relations \ref{eq:FLR_En_mom_tensor_matt}). Therefore, with the relations introduced above one can write the Einstein equations
\begin{eqnarray*}
0 &=& 2 \partial_i^2 \Psi - 6 \mathcal{H} \psi^\prime \\
& & - a^2 \left[ \dfrac{\delta \rho + 2 \left( \rho + \rho_\phi \right) \varphi}{\Mpl} + m_0^2 \left( \dfrac{\xi^{0\prime}}{\phi^{\prime}} - \varphi \right) + \left( \dfrac{\rho_\phi + p_\phi}{\Mpl} + m_4^2 \right) \left( \partial_i^2 \Xi + 3 \psi \right) \right] , \\
0 &=& \partial_{i} \left[ 2 \left( \mathcal{H} \varphi + \psi^\prime \right) + \left( 2 \mathcal{H}^{\prime} + \mathcal{H}^2 \right) B \right] + \dfrac{1}{2} \partial_i^2 \varpi_{i} + S_i \left( 2 \mathcal{H}^{\prime} + \mathcal{H}^2 \right) - \dfrac{a}{\Mpl} \left[ \left( \rho + p \right) \delta v_i + \delta q_i \right. \\
& & \left. - a p \left( S_i + \partial_i B \right) \right] - a^2 \left[ \dfrac{\rho_\phi + p_\phi}{\Mpl} \left( \xi_i^T + \partial_i \xi \right)^{\prime} + \dfrac{m_1^2}{2} \left( \varpi_i + \sigma_i^{\prime} + \partial_i \left( \Xi^{\prime} - \dfrac{\Xi^{0}}{\phi^\prime} \right) \right) \right. \\
& & \left. + \dfrac{\rho_\phi}{\Mpl} \left( S_i + \partial_i B \right) \right] ,
\end{eqnarray*}\begin{eqnarray*}
0 &=& \delta_{ij} \left[ 2 \mathcal{H} \left( 2 \psi + \varphi \right)^\prime - 2 \left( \varphi + \psi \right) \left( 2 \mathcal{H}^2 + \mathcal{H}^\prime \right) + 2 \psi^{\prime\prime} - \partial_k^2 \left( \Psi - \Phi \right) + 6 \left( \psi + \varphi \right) \left( \mathcal{H}^2 + \mathcal{H}^\prime \right) \right] \\
& & + \partial_i \partial_j \left[ \Psi - \Phi - 2 \left( 2 \mathcal{H}^{\prime} + \mathcal{H}^2 \right) E \right] - \mathcal{H} H_{ij}^{\prime} + \dfrac{1}{2} \left( \partial_k^2 H_{ij} - H_{ij}^{\prime\prime} \right) + \left( 2 \mathcal{H}^{\prime} + \mathcal{H}^2 \right) H_{ij} \\
& & + \mathcal{H} \left( \partial_i \varpi_j + \partial_j \varpi_i \right) + \dfrac{1}{2} \left( \partial_i \varpi^\prime_j + \partial_j \varpi^\prime_i \right) - \left( \mathcal{H}^2 + 2 \mathcal{H}^\prime \right) \left( \partial_i F_j + \partial_j F_i \right) - \dfrac{a^2}{\Mpl} \left[ \delta_{ij} \delta p \right. \\
& & \left. - p \left( 2 \psi \delta_{ij} - 2 \partial_i \partial_j E - \partial_i F_j + \partial_j F_i + H_{ij} \right) + \left( 3 \partial_i \partial_j - \delta_{ij} \partial_k^2 \right) \pi + \partial_i \pi_j + \partial_j \pi_i + \pi_{ij} \right] \\
& & - a^2 \left[ \dfrac{1}{2} m_2^2 \left( \partial_j \xi_i^T + \partial_i \xi_j^T + 2 \partial_i \partial_j \xi \right) - m_3^2 \delta_{ij} \left( 3 \psi + \partial_i^2 \Xi \right) + \delta_{ij} m_4^2 \left( \varphi - \xi^{0\prime} / \phi^{\prime} \right) \right. \\
& & \left. + \left( \dfrac{1}{2} m_2^2 - \dfrac{p_\phi}{\Mpl} \right) \left( 2 \psi \delta_{ij} - \partial_i F_j - \partial_j F_i - 2 \partial_i \partial_j E + H_{ij} \right) \right]
\end{eqnarray*}
Those equations correspond respectively to the $00$, $0i$ and $ij$ components of the Einstein equations. By making use of the Friedmann and Raychaudhuri equations (\ref{eq:FLRW_equations_GR}) for the background along with the Goldstone equation in the form (\ref{eq:FLRW_Golds_ma}), and with the gauge-invariant fields introduced in section \ref{sc:cosmo_pert}, the linearized Einstein equations for the massive gravitational field consist of one tensor equation
\begin{eqnarray*}
0 = H_{ij}^{\prime\prime} - \partial_i^2 H_{ij} + 2 \mathcal{H} H_{ij}^{\prime} + a^2 \left( m_2^2 H_{ij} + \dfrac{2 \pi_{ij}}{\Mpl} \right),
\end{eqnarray*}
two vector equations
\begin{eqnarray*}
0 &=& a^{-2} \partial_i^2 \varpi_{i} - \dfrac{2 \zeta_i}{\Mpl} - \left( m_1^2 + 2 \dfrac{\rho_\phi + p_\phi}{\Mpl} \right) \left( \varpi_i + \sigma_i^\prime \right) , \\
0 &=& \varpi^\prime_i + 2 \mathcal{H} \varpi_i - a^2 \left( m_2^2 \sigma_i + \dfrac{2 \pi_i}{\Mpl} \right) ,
\end{eqnarray*}
and four scalar equations
\begin{eqnarray*}
0 &=& 2 \partial_j^2 \Psi - 6 \mathcal{H} \left( \mathcal{H} \Phi + \Psi^\prime \right) \\
& & - a^2 \left[ \dfrac{\rho \delta_\rho}{\Mpl} - m_0^2 \left( \Phi - \dfrac{\Xi^{0\prime}}{\phi^{\prime}} \right) + \left( \dfrac{\rho_\phi + p_\phi}{\Mpl} + m_4^2 \right) \left( \partial_j^2 \Xi + 3 \Psi \right) \right], \\
0 &=& 2 \partial_i \left( \Psi^\prime + \mathcal{H} \Phi \right) + a^2 \partial_i \left[ \dfrac{m_1^2}{2} \left( \dfrac{\Xi^{0}}{\phi^{\prime}} - \Xi^\prime \right) - \dfrac{\rho_\phi + p_\phi}{\Mpl} \Xi^\prime - \dfrac{\delta_\zeta}{\Mpl} \right] , \\
0 &=& \partial_i \partial_j \left[ \Phi - \Psi + a^2 \left( m_2^2 \Xi + \dfrac{3 \pi}{\Mpl} \right) \right] , \\
0 &=& - 2 \Psi^{\prime\prime} - 2 \Phi \left( \mathcal{H}^2 + 2 \mathcal{H}^\prime \right) + \partial_j^2 \left( \Psi - \Phi \right) - 2 \mathcal{H} \left( 2 \Psi + \Phi \right)^\prime + a^2 \dfrac{p \delta_p - \partial_i^2 \pi}{\Mpl} - a^2 m_3^2 \partial_j^2 \Xi \nonumber \\
& & + a^2 m_4^2 \left( \Phi - \dfrac{\Xi^{0\prime}}{\phi^{\prime}} \right) + a^2 \left( m_2^2 - 3 m_3^2 \right) \Psi .
\end{eqnarray*}

\subsection{The Goldstone equations}

The variation of the action with respect to the scalar fields give four Goldstone equations (\ref{eq:Goldstone_equation}). At the linearized level, those four equations give one vector equation
\begin{eqnarray*}
0 &=& \partial_\eta \left[ a^4 \left( m_1^2 + 2 \dfrac{\rho_\phi + p_\phi}{\Mpl} \right) \left( \varpi_i + \sigma_i^\prime \right) \right] - a^{4} m_2^2 \partial_j^2 \sigma_i ,
\end{eqnarray*}
and two scalar equations
\begin{eqnarray*}
0 &=& \partial_0 \left[ \dfrac{a^4}{\phi^{\prime}} \left( m_0^2 \left( \Phi - \dfrac{\Xi^{0\prime}}{\phi^{\prime}} \right) - m_4^2 \left( 3 \Psi + \partial_i^2 \Xi \right) \right) \right] + \dfrac{a^4}{2 \phi^{\prime}} m_1^2 \partial_i^2 \left( \Xi^\prime - \dfrac{\Xi^{0}}{\phi^{\prime}}\right) , \\
0 &=& \partial_i \left\{ a^{-4} \partial_0 \left[ a^4 \left( \dfrac{m_1^2}{2} + \dfrac{\rho_\phi}{\Mpl} \right) \left( \Xi^\prime - \dfrac{\Xi^{0}}{\phi^{\prime}} \right) + a^4 \dfrac{ p_\phi}{\Mpl} \Xi^{\prime} \right] + \left( m_3^2 - m_2^2 \right) \partial_j^2 \Xi \right. \\
& & \left. - \left( m_4^2 + \dfrac{\rho_\phi + p_\phi}{\Mpl} \right) \Phi + \left( 3 m_3^2 - m_2^2 \right) \Psi + \left( m_4^2 + \dfrac{\rho_\phi}{\Mpl} \right) \dfrac{\Xi^{0\prime}}{\phi^{\prime}} \right\} .
\end{eqnarray*}

\section{Static spherically symmetric ansatz} \label{app:sc:bh}

For the ansatz (\ref{eq:BH_ansatz}) discussed in section \ref{sc:BH_Ansatz}, the non-zero components of the Einstein tensor are given by the
following relations
\begin{eqnarray*}
\mathcal{G}_{0}^{0} &=& \dfrac{1}{r^2} \left[ 1 - \left( \dfrac{r}{\beta} \right)^{\prime} \right], \\
\mathcal{G}_{r}^{r} &=& \dfrac{1}{r^2} \left( 1 - \dfrac{\alpha + r \alpha^{\prime}}{\alpha \beta} \right), \\
\mathcal{G}_{\theta}^{\theta} &=& \mathcal{G}_{\varphi}^{\varphi} = - \dfrac{1}{4 r} \left[ \dfrac{\alpha^{\prime} + r \alpha^{\prime\prime}}{\alpha \beta} + \left( \dfrac{2 \alpha + r \alpha^{\prime}}{\alpha \beta} \right)^{\prime} \right].
\end{eqnarray*}
For functions $\mathcal{F}$ which are invariant under rotations of the Goldstone fields $\phi^i$ internal space, the derivatives of $\mathcal{F}$ with respect to $W^{ij}$ are given by
\begin{eqnarray*}
\dfrac{\partial \mathcal{F}}{\partial W^{ij}} &=& \mathcal{F}_{1} \delta_{ij} + 2 \mathcal{F}_{2} W^{ij} + 3 \mathcal{F}_{3} W^{ik} W^{kj},
\end{eqnarray*}
where $\mathcal{F}_{i} \equiv \partial \mathcal{F} / \partial w_{i}$. Therefore, the components of the energy-momentum tensor (\ref{eq:Ann_En_Mom_Gold}) of the four Goldstone fields which are not identically zero are given by
\begin{eqnarray*}
t_{0}^{0} &=& \Lambda^{4} \left[ - \dfrac{1}{2} \mathcal{F} + \dfrac{1}{\alpha} \left( \mathcal{F}_{X} + \dfrac{\partial \mathcal{F}}{\partial W^{ij}} \dfrac{V^{i} V^{j}}{X^{2}} \right) \right] , \\
t_{r}^{r} &=& \Lambda^{4} \left[ - \dfrac{1}{2} \mathcal{F} - \left( \mathcal{F}_{X} + \dfrac{\partial \mathcal{F}}{\partial W^{ij}} \dfrac{V^{i} V^{j}}{X^{2}} \right) \dfrac{h^{\prime2}}{\beta} + \dfrac{1}{\beta} \dfrac{\partial \mathcal{F}}{\partial W^{ij}} \left( - \partial_r \phi^i \partial_r \phi^j + \dfrac{2 V^{j}}{X} \partial_r \phi^i \partial_r \phi^0 \right) \right] , \\
t_{\theta}^{\theta} &=& t_{\varphi}^{\varphi} = \Lambda^{4} \left[ - \dfrac{1}{2} \mathcal{F} + \dfrac{\partial \mathcal{F}}{\partial W^{ij}} \partial^\theta \phi^i \partial_\theta \phi^j \right] , \\
t_{0}^{r} &=& - \dfrac{\Lambda^{4} h^{\prime}}{\beta} \left[ \mathcal{F}_{X} + \dfrac{\partial \mathcal{F}}{\partial W^{ij}} \left( \dfrac{V^{i} V^{j}}{X^{2}} + \dfrac{\partial_r \phi^i \partial_r \phi^{j}}{\beta X} \right) \right].
\end{eqnarray*}

\chapter{Instantaneous interaction} \label{app:ch:ii}

This appendix provided some details about the calculations presented in section \ref{ch:inst_int} devoted to a physical instantaneous interactions in massive gravity models.

\section{Transverse and longitudinal projections operators} \label{app:sc:Trans_Long_Op}

The key of the understanding of the instantaneous interaction present in Lorentz-violating electrodynamics and in Lorentz-violating massive gravity models lies in the concept of transverse and longitudinal sources. Defining the transverse projection operator by \cite{brill:832}
\begin{eqnarray} \label{eq:an_proj_T}
T_{ij} \left( x , x^\prime \right) = \left[ \delta_{ij} \delta^3 \left( x - x^\prime \right) - \dfrac{1}{4 \pi} \dfrac{\partial^2}{\partial x^i \partial x^{\prime j}} \left( \dfrac{1}{| x - x^\prime |} \right) \right] ,
\end{eqnarray}
and the longitudinal projections operator by
\begin{eqnarray} \label{eq:an_proj_L}
L_{ij} \left( x , x^\prime \right) = \dfrac{1}{4 \pi} \dfrac{\partial^2}{\partial x^i \partial x^{\prime j}} \left( \dfrac{1}{| x - x^\prime |} \right) ,
\end{eqnarray}
one can show that
\begin{eqnarray*}
\dfrac{\partial}{\partial x^i} T_{ij} \left( x , x^\prime \right) = \dfrac{\partial}{\partial x^{\prime j}} T_{ij} \left( x , x^\prime \right) = \cod{\varepsilon}{i}{jk} \dfrac{\partial}{\partial x^j} L_{kl} \left( x , x^\prime \right) = 0 .
\end{eqnarray*}
The transverse and longitudinal components of any three-dimensional vector field $V^i \left( x \right)$ are given by
\begin{eqnarray*}
V^{T}_i \left( x \right) = \int \textrm{d}^3 x^\prime T_{ij} \left( x , x^\prime \right) V_j \left( x^\prime \right) , & & V^{L}_i \left( x \right) = \int \textrm{d}^3 x^\prime L_{ij} \left( x , x^\prime \right) V_j \left( x^\prime \right) .
\end{eqnarray*}
One can easily verify the following identities
\begin{eqnarray*}
V^{T}_i \left( x \right) + V^{L}_i \left( x \right) = V_i \left( x \right) , & \dfrac{\partial}{\partial x^i} V^{T}_i \left( x \right) = 0 , & \left( \nabla \times V^{L} \right)_i \left( x \right) = 0 .
\end{eqnarray*}
The longitudinal vector field can be expressed through
\begin{eqnarray*}
V^{L}_i = \partial_i v (x) &\textrm{with}& v \left( x \right) = \dfrac{1}{4 \pi} \int \textrm{d}^3 x^\prime \dfrac{\partial}{\partial x^{\prime j}} \left( \dfrac{1}{| x - x^\prime |} \right) V_j \left( x^\prime \right) .
\end{eqnarray*}

\section{First order geodesics}

Let us first consider the geodesic motion of an observer of four-velocity $v^\alpha$. This observer follows a time-like geodesic $x^\alpha \left( \tau_{ob} \right) = \left( t \left( \tau_{ob} \right) , x^i \left( \tau_{ob} \right) \right)$ with an affine parameter $\tau_{ob}$. It obeys the following equation
\begin{eqnarray*}
\dfrac{\dif v^\alpha}{\dif \tau_{ob}} + \C{\mu}{\nu}{\alpha} v^\mu v^\nu = 0 &\tx{with}& v^\alpha \equiv \dfrac{\dif x^\alpha}{\dif \tau_{ob}} .
\end{eqnarray*}
The four velocity of the observer is such that $v^\mu v_\mu = 1$. A natural solution to the previous equation is Minkowski space-time is given by $v^\mu = \left( 1 , 0 , 0 , 0 \right)$. Therefore, the geodesic followed by the observer reads $x^\alpha \left( \tau_{ob} \right) = \left( \tau_{ob} , x^i \right)$.

Perturbations of the metric $\eta_{\mu\nu} + h_{\mu\nu}$ imply perturbations of the geodesic $x^\alpha \left( \tau_{ob} \right) + \delta x^\alpha \left( \tau_{ob} \right)$ which in turn implies perturbations of the four velocity $v^\alpha \left( \tau_{ob} \right) + \delta v^\alpha \left( \tau_{ob} \right)$. Since the observer is supposed to be massive, its perturbed time-like geodesic satisfies
\begin{eqnarray*}
\left( v^\mu + \delta v^\mu \right) \left( \eta_{\mu\nu} + h_{\mu\nu} \right) \left( v^\nu + \delta v^\nu \right) = 1 &\rightarrow&
\delta v^0 = - \varphi,
\end{eqnarray*}
implying that the zero component of the velocity perturbation is given by $h_{00} / 2$. The linearized geodesic equation for the massive observer becomes
\begin{eqnarray*}
\dfrac{\dif \delta v^\alpha}{\dif \tau_{ob}} + \delta \C{0}{0}{\alpha} = 0 .
\end{eqnarray*}
Under a infinitesimal gauge transformation $x^\mu \rightarrow x^{\prime\mu} = x^\mu - y^\mu$ of the coordinates, the geodesic perturbations transform as
\begin{eqnarray*}
\delta x^\mu \longrightarrow \delta x^{\prime\mu} = \delta x^\mu - y^\mu .
\end{eqnarray*}
Hence, the following quantities are gauge-invariant fields
\begin{eqnarray*}
\delta\mathbf{n}^0 = \delta x^0 + B + \dot{E} , & \delta\mathbf{n}^i = \delta x^i - F^i + \partial^i E .
\end{eqnarray*}
It is then straightforward to show that the gauge-invariant four-velocity of the observer is fixed by the following equation
\begin{eqnarray*}
\dfrac{\dif \delta \mathbf{v}^\mu}{\dif \tau_{ob}} = \delta_i^\mu \dot{\varpi}_i - \delta^{\mu\nu} \partial_\nu \Phi,
\end{eqnarray*}
with
\begin{eqnarray*}
\delta \mathbf{v}^0 \equiv \delta v^0 + \dot{B} + \ddot{E} , & \delta \mathbf{v}^i \equiv \delta v^i - \dot{F}^i + \partial^i \dot{E} .
\end{eqnarray*}
\vspace{0.1cm}

Light follows null geodesics $x^\alpha \left( \tau_{ph} \right) = \left( t \left( \tau_{ph} \right) , x^i \left( \tau_{ph} \right) \right)$ characterized by an affine parameter $\tau_{ph}$ and satisfying the following geodesic equation
\begin{eqnarray*}
\dfrac{\dif u^\alpha}{\dif \tau_{ph}} + \C{\mu}{\nu}{\alpha} u^\mu u^\nu = 0 , & u^\alpha \equiv \dfrac{\dif x^\alpha}{\dif \tau_{ph}} ,
\end{eqnarray*}
with $u^\mu u_\mu = 0$. A solution to this equation in flat space-time can always be formulated as $u^\mu = \left( 1, n^i \right)$ with $n_i^2 = 1$. Hence, the geodesic can be parameterized as $x^\alpha \left( \tau_{ph} \right) = c^\alpha + \tau_{ph} u^\alpha$ where $c^\alpha$ is a constant four vector. As previously, perturbations of the metric implies perturbations of the geodesic $x^\alpha \left( \tau_{ph} \right) + \delta x^\alpha \left( \tau_{ph} \right)$ which in turn imply perturbations of the four-vector tangent to the geodesic of the photons $u^\alpha \left( \tau_{ph} \right) + \delta u^\alpha \left( \tau_{ph} \right)$. Therefore, one has
\begin{eqnarray*}
\left( u^\mu + \delta u^\mu \right) \left( \eta_{\mu\nu} + h_{\mu\nu} \right) \left( u^\nu + \delta u^\nu \right) = 0
&\rightarrow&
2 \delta u^0 + h_{00} - 2 n^i \delta u^i + 2 n^i h_{0i} + n^j n^i h_{ij} = 0.
\end{eqnarray*}
This relation implies that $\delta u^0$ is determined by $\delta u^i$ and the metric perturbations. Since $\delta u^\alpha$ transforms as $\delta v^\alpha$ under a four dimensional coordinates reparametrization and since
\begin{eqnarray*}
\dfrac{\dif f}{\dif \tau_{ph}} = \dot{f} + n^i \partial_i f
\end{eqnarray*}
the gauge-invariant four-velocity of light reads
\begin{eqnarray*}
\delta \mathbf{u}^0 \equiv \delta u^0 + \dot{B} + \ddot{E} + n^i \partial_i \left( B + \dot{E} \right) , &
\delta \mathbf{u}^i \equiv \delta u^i - \dot{F}^i + \partial^i \dot{E} + n^j \partial_j \left( F^i + \partial^i E \right) .
\end{eqnarray*}
Hence, the geodesic of light is given to the first order by
\begin{eqnarray*}
\dfrac{\dif \delta \mathbf{u}^\alpha}{\dif \tau_{ph}} &=& - \left[ \dot{\Phi} + \dot{\Psi} + 2 n^i \partial_i \Phi + \dfrac{n^i n^j}{2} \left( \partial_{i} \varpi_j + \partial_{j} \varpi_i - \dot{H}_{ij} \right) \right] \delta_0^\alpha \\
& & + \delta^{\alpha i} \left[ \partial_{i} \left( \Psi - \Phi \right) + \dot{\varpi}_i \right] - 2 n^i \left[ \dot{\Psi} \delta^\alpha_i - \dfrac{1}{2} \delta^{\alpha j} \left( \dot{H}_{ij} + \partial_{i} \varpi_j - \partial_{j} \varpi_i \right) \right] \\
& & - n^i n^j \left[ \left( \delta_i^\alpha \partial_j + \delta_j^\alpha \partial_i \right) \Psi - \dfrac{1}{2} \delta^{\alpha l} \left( \partial_j H_{il} + \partial_i H_{jl} - \partial_l H_{ij} \right) \right]
\end{eqnarray*}

\section{Frequency shift}

If an observer of four-velocity $v^\mu$ sees photons with a four-momentum $p^\mu = \omega_0 u^\mu$, he will conclude that they have a frequency fixed by the following relation
\begin{eqnarray*}
\omega = v_\nu p^\nu = \omega_0 .
\end{eqnarray*}
Perturbations of the metric imply a Doppler frequency shift because of the previous relation. Indeed, the frequence of the light beam is given at the linearized level by
\begin{eqnarray*}
\dfrac{\delta \omega}{\omega} = \delta \mathbf{u}^0 - \delta \mathbf{v}^0 + n_i \left( \delta \mathbf{v}^i + \varpi^i \right).
\end{eqnarray*}
After substituting in this relation the solutions to the geodesic equations
\begin{eqnarray*}
\delta \mathbf{v}^0 &=& - \Phi , \\
\delta \mathbf{v}^i &=& \varpi_i - \int \dif \tau_{ob} \, \partial_i \Phi  ,\\
\delta \mathbf{u}^0 &=& - 2 \Phi + \int \dif \tau_{ph} \, \left[ \dot{\Phi} - \dot{\Psi} + \dfrac{n^i n^j}{2} \left( \dot{H}_{ij} - \partial_{i} \varpi_j - \partial_{j} \varpi_i \right) \right] ,
\end{eqnarray*}
one finally found that the spectral shift is given by
\begin{eqnarray*}
\dfrac{\delta \omega}{\omega} &=& \int \dif \tau_{ph} \left[ \dot{\Psi} - n^i \partial_i \Phi + n^i n^j \left( \dfrac{1}{2} \dot{H}_{ij} - \partial_{i} \varpi_j \right) \right] + \int \dif \tau_{ob} \, n^i \partial_i \Phi .
\end{eqnarray*}
Substituting the solution of the linearized EoM of GR into this relation gives eq.~(\ref{eq:ShiftGR}) after some manipulations.

In Lorentz-violating massive gravity models based on the action (\ref{eq:MG_action_Z}), there is an extra contribution as compared to GR due to the presence of a mass term in eq.~(\ref{eq:Gravitons}) for the gravitational waves. For simplicity, we will determined the derivative of the spectral shift with respect to the affine parameters $\tau_{ph}$ of the light geodesic
\begin{eqnarray*}
\textbf{a}_\omega \equiv \dfrac{\dif}{\dif \tau_{ph}} \left( \dfrac{\delta \omega}{\omega} - \dfrac{\delta \omega^{\tx{GR}}}{\omega} \right) ,
\end{eqnarray*}
and demonstrate that this term give rise to a physical instantaneous interaction. Since
\begin{eqnarray*}
\textbf{a}_\omega = \dfrac{n^i n^j}{2} \dot{H}^{\Delta}_{ij} &\tx{with}& H^{\Delta}_{ij} = H_{ij} - H_{ij} |_{\tx{GR}},
\end{eqnarray*}
one has to solve the two following equations
\begin{eqnarray*}
\left( \Box + m_2^2 \right) \dot{H}_{ij} = - \dfrac{2 \dot{\pi}_{ij}}{\Mpl} , & \Box \dot{H}_{ij} |_{\tx{GR}} = - \dfrac{2 \dot{\pi}_{ij}}{\Mpl} .
\end{eqnarray*}

\subsubsection{Sources of tensorial modes} \label{app:sc:ii_source_tens}

The transverse (\ref{eq:an_proj_T}) and longitudinal (\ref{eq:an_proj_L}) projections operators can be used to determined the transverse and traceless anisotropic stress tensor $\pi_{ij}$. Indeed, one can show that for an arbitrary energy-momentum tensor $\delta \mathcal{T}_{\mu\nu}$, the transverse and traceless anisotropic stress tensor is given by
\begin{eqnarray*}
\pi_{ij} = \left[ \delta_{ik} \delta_{jl} - \delta_{ik} \dfrac{\partial_j \partial_l}{\partial_n^2} - \dfrac{\partial_i \partial_k}{\partial_n^2} \delta_{jl} + \dfrac{\delta_{ij}}{2} \dfrac{\partial_k \partial_l}{\partial_n^2} + \dfrac{1}{2} \dfrac{\partial_i \partial_k}{\partial_n^2} \dfrac{\partial_j \partial_l}{\partial_n^2} \right] \delta \mathcal{T}_{kl} + \dfrac{1}{2} \left[ \dfrac{\partial_i \partial_j}{\partial_k^2} - \delta_{ij} \right] \delta \mathcal{T}_{nn} ,
\end{eqnarray*}
with for an arbitrary functions $f$
\begin{eqnarray*}
\dfrac{1}{\partial_n^2} f \left( x \right) = - \dfrac{1}{4 \pi} \int \dif^3 x^\prime \dfrac{1}{| x - x^\prime |} f \left( x^\prime \right) .
\end{eqnarray*}
For the example of section \ref{sc:II_ex} consisting in a source which appears at $t = 0$ and described by
\begin{eqnarray*}
\delta \mathcal{T}_{00} &=& 2 \mu_{ij} t^2 \Theta \left( t \right) \partial_i \partial_j \delta^3 \left( x - x_s \right) , \\
\delta \mathcal{T}_{0i} &=& 4 \mu_{ij} t \Theta \left( t \right) \partial_j \delta^3 \left( x - x_s \right) , \\
\delta \mathcal{T}_{ij} &=& 4 \mu_{ij} \Theta \left( t \right) \delta^3 \left( x - x_s \right) ,
\end{eqnarray*}
with $x_s = \left( 0, 0, d \right)$ and $\mu_{ij} = \delta_{2i} \delta_{2j}$, the only component of the transverse and traceless anisotropic stress tensor $\pi_{ij}$ which enters the equation of the frequency shift through the gravitational waves $H_{ij}$ is given by
\begin{eqnarray*}
\pi_{11} = 2 \Theta \left( t \right) \left( \dfrac{\partial_x^2 \partial_y^2}{\partial_i^2 \partial_j^2} - \dfrac{\partial_z^2}{\partial_i^2} \right) \delta^3 \left( x - x_s \right) ,
\end{eqnarray*}
because the light's geodesic is supposed to be along the $x$ direction with $n^i = - \delta^i_1$.

\subsubsection{Derivative of the spectral shift}

The derivative of the spectral shift with respect to the affine parameters $\tau_{ph}$ of the light geodesic is given by the differences between the solutions of the tensor equations in GR and in the class of massive gravity models considered here. For the example of section \ref{sc:II_ex} summarized above, on has that
\begin{eqnarray*}
\dot{H}_{11} &=& - \dfrac{1}{\Mpl} \left( \dfrac{\partial_x^2 \partial_y^2}{\partial_i^2 \partial_j^2} - \dfrac{\partial_z^2}{\partial_i^2} \right) G_m^{+} \left( t, 0, x, x_s \right) , \\ \\
\dot{H}_{11} |_{\tx{GR}} &=& - \dfrac{1}{\Mpl} \left( \dfrac{\partial_x^2 \partial_y^2}{\partial_i^2 \partial_j^2} - \dfrac{\partial_z^2}{\partial_i^2} \right) G^{+} \left( t, 0, x, x_s \right) ,
\end{eqnarray*}
implying that
\begin{eqnarray*}
\textbf{a}_\omega = - \dfrac{1}{2 \Mpl} \left( \dfrac{\partial_x^2 \partial_y^2}{\partial_i^2 \partial_j^2} - \dfrac{\partial_z^2}{\partial_i^2} \right) \Delta G^{+} \left( t, 0, x, x_s \right)
\end{eqnarray*}
Note that if $\chi = t^2 - \left( x - x_s \right)_i^2$, the difference $\Delta G^{+}$ between the Green functions of the d'Alembert and Klein-Gordon equations is given by
\begin{eqnarray*}
\Delta G^{+} \left( t, 0, x, x_s \right) \equiv G_m^{+} \left( t, 0, x, x_s \right) - G^{+} \left( t, 0, x, x_s \right) = - \Theta \left( \chi \right) \dfrac{m_2 \mathcal{J}_{1} \left( m_2 \sqrt{\chi} \right)}{4 \pi \sqrt{\chi}} ,
\end{eqnarray*}
where $\mathcal{J}_{1}$ is a Bessel functions of the first kind. For $m_2^2 \chi$ very small, the following Taylor expansion gives a good approximation of $\Delta G^{+}$
\begin{eqnarray} \label{eq:an_taylor_bess}
\dfrac{\mathcal{J}_{1} \left( m_2 \sqrt{\chi} \right)}{m_2 \sqrt{\chi}} = \dfrac{1}{2} - \dfrac{m_2^2 \chi}{16} + \mathcal{O} \left( m_2^3 \right) .
\end{eqnarray}

After some manipulations, the derivative of the spectral shift with respect to the affine parameters $\tau_{ph}$ of the light geodesic reads
\begin{eqnarray}
\textbf{a}_\omega &=& \dfrac{1}{6 \Mpl} \Delta G^{+} \left( t, 0, x, x_s \right) + \dfrac{1}{8 \pi \Mpl} \int \dif^3 x^{\prime} \left( \dfrac{1}{| x - x^\prime |^3} - \dfrac{3 \left( z - z^{\prime} \right)^2}{| x - x^\prime |^5} \right) \Delta G^{+} \left( t, 0, x^\prime, x_s \right) \nonumber \\
& & - \dfrac{9}{32 \pi^2 \Mpl} \int \dif^3 x^{\prime} \dif^3 x^{\prime\prime} \dfrac{\left( x - x^{\prime} \right) \left( y - y^{\prime} \right)}{| x - x^{\prime} |^5} \dfrac{\left( x^\prime - x^{\prime\prime} \right) \left( y^\prime - y^{\prime\prime} \right)}{| x^\prime - x^{\prime\prime} |^5} \Delta G^{+} \left( t, 0, x^{\prime\prime}, x_s \right) . \label{eq:an_aomega_ex}
\end{eqnarray}
The first term in the right-hand-side of this equation vanishes outside of the light cone of the source since it involves the Heaviside function $\Theta \left( t^2 - d^2 \right)$. Let us then concentrate on the second term. With the Taylor expansion (\ref{eq:an_taylor_bess}), it is possible to show that the second contribution to $\textbf{a}_\omega$ can be approximate at the origin $x = 0$ by
\begin{eqnarray*}
\dfrac{m_2^2 d}{128} \left[ \int_{0}^{+\infty} \dif r \, \Theta \left( t^2 - r^2 - d^2 - 2 r d \right) - \int_{0}^{+\infty} \dif r \, \Theta \left( t^2 - r^2 - d^2 + 2 r d \right) \right] + \mathcal{O} \left( m_2^3 \right) .
\end{eqnarray*}
For $0 < t < d$ which is outside of the light cone of the source, this term clearly becomes an instantaneous contribution to the derivative of the spectral shift $\textbf{a}_\omega$
\begin{eqnarray*}
- \dfrac{m_2^4 d}{64} t + \mathcal{O} \left( m_2^5 \right) .
\end{eqnarray*}
Since the third term in (\ref{eq:an_aomega_ex}) is of order $\mathcal{O} \left( m_2^5 \right)$, the conclusion to this calculation is that the derivative of the spectral shift with respect to the affine parameters $\tau_{ph}$ of the light geodesic
\begin{eqnarray*}
\textbf{a}_\omega \left( t , 0 \right) = - \dfrac{m_2^4 d}{64} t + \mathcal{O} \left( m_2^5 \right) ,
\end{eqnarray*}
is instantaneously affected by the appearance of the source $\delta T_{\mu\nu}$ characterized by a non-trivial tensorial anisotropic stress.


\newpage
\renewcommand{\bibname}{\usefont{OT1}{pzc}{m}{n}\selectfont Bibliography}

\end{document}